\documentclass[12pt,a4paper,oneside,footinclude,BCOR5mm,]{article}

\usepackage[utf8]{inputenc}
\usepackage[english]{babel}
\usepackage[T1]{fontenc}
\usepackage{courier}

\usepackage[autostyle,italian=guillemets,]{csquotes}
\usepackage[style=authoryear,citestyle=authoryear,natbib,hyperref,dashed=false,maxbibnames=10,minbibnames=1,maxcitenames=2,giveninits=true,   uniquename=false,sorting=nyt]{biblatex}
\usepackage{url} 
\usepackage{hyperref}  

\usepackage{pdfsync}
\usepackage{epsfig}
\usepackage{amsmath}
\usepackage{amsfonts}
\usepackage{fancyhdr}
\usepackage{authblk}
\usepackage{subfig}
\usepackage{graphicx}
\usepackage{rotating}
\usepackage{booktabs}
\usepackage{multirow}
\usepackage{array,longtable}

\usepackage{trfsigns}
\usepackage{fdsymbol}
\usepackage{courier} 

\usepackage{xcolor}
\definecolor{princetonorange}{rgb}{1.0, 0.56, 0.0}

\usepackage{adjustbox}

\usepackage{soul}

\begin{document}

\title{GMM-lev estimation and individual heterogeneity: Monte Carlo evidence and empirical applications\thanks{We are grateful to the participants at the International Panel Data Conference for their helpful comments. Special thanks to Alok Bhargava, Maurice Bun, Jacques Mairesse, Patrick Sevestre. We thank Annapaola Zenato for her initial help with the research.}} 
\author{Maria Elena Bontempi\thanks{Corresponding author, Department of Economics, University of Bologna, Italy.
\newline
Email: mariaelena.bontempi@unibo.it} \\ 
Jan Ditzen\thanks{Department of Economics, University of Bozen, Italy.
\newline
Email: jan.ditzen@unibz.it}}
\date{\today}

\maketitle
\thispagestyle{empty}

\begin{abstract}
We introduce a new estimator, CRE-GMM, which exploits the correlated random effects (CRE) approach within the generalised method of moments (GMM), specifically applied to level equations, GMM-lev. It has the advantage of estimating the effect of measurable time-invariant covariates using all available information. This is not possible with GMM-dif, applied to the equations of each period transformed into first differences, while GMM-sys uses little information as it adds the equation in levels for only one period. The GMM-lev, by implying a two-component error term containing individual heterogeneity and shock, exposes the explanatory variables to possible double endogeneity. For example, the estimation of actual persistence could suffer from bias if instruments were correlated with the unit-specific error component. The CRE-GMM deals with double endogeneity, captures initial conditions and enhance inference. Monte Carlo simulations for different panel types and under different double endogeneity assumptions show the advantage of our approach. The empirical applications on production and R\&D contribute to clarify the advantages of using CRE-GMM. 
\newline
\end{abstract}

\section{Introduction}\label{sec:intro}
The generalized method of moments (GMM) is widely used in applied economic research to estimate linear dynamic panel data models, mainly as GMM-dif (model in first differences instrumented by lagged levels, \cite{Arellano1991}) and GMM-sys (which adds model in levels to the model in first differences and uses lagged first differences to instrument, \cite{Arellano1995,Blundell1998}).
    
We investigate the effect of exploiting uniquely the model in levels explicitly combined with the \citet{Mundlak1978}'s idea of "Correlated Random Effects" (CRE) in a unified framework. Our approach addresses unobserved heterogeneity and dynamic panel estimation simultaneously. Instruments will be defined as lagged first differences or levels according to the presence or absence of correlation between the explanatory variables and individual heterogeneity; lags' selection will depend on the classification of the explanatory variables as exogenous, predetermined and endogenous in terms of correlation with the idiosyncratic shock. We show that the \citet{Mundlak1978}'s approach is useful to handle this ``double'' endogeneity of the explanatory variables arising from two sources. The first source of endogeneity (\textit{endogeneity because of heterogeneity}) emerges from the correlation between the covariates and unobservable individual-specific characteristics, while the second source (\textit{standard endogeneity}) depends on the correlation of the covariates with the idiosyncratic shocks. Hence, we propose the Correlated Random Effects generalized method of moments on levels, the CRE-GMM estimation, which combines two methods to deal with both unobserved individual-specific effects (handled by the \citet{Mundlak1978}'s approach) and dynamic panel data model estimation challenges (handled by the GMM-lev method). We consider alternative settings: macro panels (small N and long T, for example N=25 and T=40), multilevel panels (the number of groups is high relative to the number of observations per group, for example N=100 and T=20), and longitudinal panels (N greatly larger than T, for example, N=1000 and T=10). 

The first aim of our CRE-GMM method is to maintain the levels of the equation of interest as they allow for estimating the effects of measurable time-invariant explanatory variables while considering the role of unmeasurable individual heterogeneity. Indeed, time invariant variables are lost when first differencing the data. For example, on macro panels, we cannot estimate the effects on $CO^2$ emissions of measurable institutional features driving time-invariant heterogeneity while controlling for country-specific unobserved heterogeneity with first differences to avoid omitted variable bias \citep{Haber2011}. \citet{Kropko2020} discusses the interpretations and identification of within-unit and cross-sectional variation in panel data models and shows that using only within variation, as in \citet{Acemoglu2008}, leads to incorrect conclusions about null effects of democracy on GDP. 
On longitudinal panels we cannot study innovative investments as a function of measurable individual characteristics (e.g., size, sector, market power) that are of great interest to researchers. To obtain unbiased effects, it is necessary to consider unobservable factors like management quality, ownership motivation, and cost of capital (further discussion is \cite{Gormley2014}).
In fields like education, psychologyl, sociology, and political science, standard approaches involve multilevel, multidimensional, or hierarchical models. While the fixed effects specification controls for correlations between lower-level predictors and higher-level residuals through within or first-differencing transformations, it cannot estimate the effects of higher-level, time-invariant, variables (time-invariant variables). Unfortunately, this approach models out higher-level variables, rendering any correlations with covariates irrelevant, and it does not address the source of endogeneity, which can itself be of interest \citep{Gelman2006,Colli2020,Hill2020,Imai2019}. A recent discussion from the economic perspective is in \citet{Matyas2017,Yang2021}.

Our second research aim tackles a common challenge in applied empirical studies. When using the GMM-lev estimator for dynamic panel data models, we often observe a high estimated autoregressive parameter, akin to the upward-biased pooled OLS estimation, ignoring individual heterogeneity. This implies that neglecting or failing to capture individual heterogeneity results in a mix of ``spurious'' persistence (due to unobserved unit-specific permanent characteristics) and ``true" persistence (causal effect of past realizations on the current realization of the dependent variable) \citep{Heckman1991,Page2006,Bednar2012}. An illustrative example from innovation literature highlights the difficulty of estimating the causal effect of past R\&D activities on current R\&D investment due to the path-dependent nature of technical changes \citep{Atkinson1969}. Spurious persistence, a product of uncontrolled individual heterogeneity, significantly impacts the innovation process \citep{Peters2009}.

An important issue in the estimation of level dynamic panel data models with unit-effects is whether it is possible to remove the unobservable individual heterogeneity from the model, or from the regressors, or from any variable that may be used as an instrument \citep{Kiviet2007}. Our CRE-GMM approach explicitly extends the estimated equation of interest to capture the initial endowment of each unit which is measured by pre-sample realisations of the variables of the model. Hence, it is able to deal with a mix of units whose differences are not captured uniquely by the initial observation of the dependent variable, but by a set of observations of the variables concurring to the dynamic process. 

In our extended level regressions, we treat individual heterogeneity as random. In the words of \citet{Nerlove2008}, while fixed effects models may be appropriate in cases in which a population is sampled exhaustively (e.g., data from geographic regions over time) or in which it is desired to predict individual behaviour (e.g., the probability that a given individual in a sample will default on a loan), random effects models are more consistent with \citet{Haavelmo1944}’s view that the ``population'' consists of an infinity of decisions made by individuals who are different from each other and who may change their behaviours over time. 

Our approach can be related to the static model using an instrumental variables approach to deal with endogeneity due to the correlation of the individual heterogeneity with some of time-varying and time-invariant explanatory variables \citep{Hausman1981}. \citet{Kripfganz2019} extend \citet{Hausman1981} to dynamic models by proposing a two-stage GMM estimation in which at the first stage a GMM-dif is used to estimate the role of time-varying covariates, while at the second stage a GMM-lev can be used to estimate the effects of measurable time-invariant covariates. In both approaches, all the explanatory variables are assumed to be strictly exogenous with respect to the idiosyncratic error term. \citet{Arellano1995} consider a dynamic panel data model with individual effects assumed to be random and equations in levels. \citet{Riju2019} extend the correlated random effects approach to linear static panel data models in which both the individual heterogeneity and the idiosyncratic shocks are correlated with time-varying explanatory variables, and implement instrumental variables approaches. Indeed, our CRE-GMM approach could also be used to estimate static models that need instrumental variables estimation for endogeneity together with consideration of the correlation of regressors with individual heterogeneity;\footnote{Internal instruments have the advantage that they are readily available while external instruments need to be carefully selected.}  Fundamental to our paper is also \citet{Bun2006} investigating GMM-dif, GMM-lev, GMM-sys supposing predetermined $x_{it}$ possibly correlated with individual heterogeneity. For GMM-lev and GMM-sys the leading term of the bias is strongly affected by the magnitude of the individual effects as well by any correlation of the regressor and the effect. 

Our work is also related to \citet{Phillips2010,Phillips2015,Hsiao2018,Hsiao2020,Alvarez2022} who use quasi-likelihood estimator of the conditional mean approach \citep{Chamberlain1980,Mundlak1978} applied to dynamic panel data models. 
\citet{Phillips2010,Phillips2015} assume random cross-sectional effects and have regressors that include lagged values of the dependent variable and may include other explanatory variables that are exogenous to the time-varying error component but may be correlated with unobserved time-invariant components.  Initial conditions do not matter if the model is first augmented with an appropriate control function. The feasible GLS iteration produces a quasi maximum-likelihood (QML) estimator of the augmented model that is consistent and asymptotically normal as $N$ increases while $T$ is fixed, regardless of initial conditions, log-likelihood misspecification or conditional heteroskedasticity. \citet{Hsiao2018} show that a properly specified QML estimator that uses the approaches from \citet{Mundlak1978,Chamberlain1980} to condition the unobserved effects and initial values on the observed strictly exogenous covariates is asymptotically unbiased if $N \rightarrow \infty$ whether $T$ is fixed or $\rightarrow \infty$. \citet{Hsiao2020} extends the analysis by comparing the feasible GLS for the conditional mean approach of \citet{Mundlak1978} and the linear difference approach treating initial values as random variables and as fixed constants. \citet{Alvarez2022} show that a correlated random effects estimation in levels may achieve substantial efficiency gains relative to estimation from data in differences. Among sociologists, \citet{Allison2017,Moral2013,Moral2019} consider likelihood-based estimation of panel data under fixed-$T$ and large-$N$ settings.  

Other related works are \citet{Wooldridge2000,Wooldridge2005a,Rabe2013} discussing how the {Mundlak1978}'s approach can be used to handle the initial conditions problem and the likelihood function in dynamic non-linear unobserved effects panel data models. 
\citet{Bai2009,Bai2013} uses the \citet{Mundlak1978}'s device in panels where there is a factor error structure correlated with the regressors. Levels are used instead of differencing, as differencing the data tends to remove useful information and is ineffective in removing the individual effects. 

In contrast to the studies cited above, we would like to relax the exogeneity assumption of the explanatory variables other than the lagged dependent variable. In other words, we would like to better understand the role of endogenenity due to heterogeneity (the correlation between the explanatory variables and individual characteristics) and the role of endogeneity due to the correlation of the explanatory variables with the idiosyncratic shock that varies between units and over time. In many applications, the explanatory variables are affected by both of the two sources of endogeneity.
 
Finally, our research extends the literature focused on the comparative analysis of GMM estimation on dynamic panel data models  \citep{Hayakawa2007,Hayakawa2009,Hayakawa2015a,Hayakawa2016,Bun2006,Bun2010,Kiviet2007,Kiviet2017,Kiviet2020,Alvarez2003, Jin2021}. 

The paper is organized as follows. Section \ref{sec:model} introduces the model and Section \ref{sec:CREGMM} presents the Correlated Random Effects GMM-lev estimation (CRE-GMM) and its motivations for the applied researcher. Section \ref{sec:RES} presents the Monte Carlo experiment and the results from our simulations. Section \ref{sec:EA} reports two empirical examples of our approach. Section \ref{CONCLU} concludes. Details and further results of our approach are in the Appendix \ref{app}

\section{The model}\label{sec:model}
We consider a general dynamic model, able to capture many phenomenon like country/regional growth or  unemployment, temporal sequence of students' performance into classes nested within schools, firms' productivity, corporate financial choices, investment in R\&D. In a nutshell, lags of the explanatory variables characterize many economic and sociological behaviours. 
The standard uni-equational linear dynamic model is defined as an ARDL(1,0) or partial adjustment model (PAM): 
\begin{equation}\label{eq:model1}
y_{it}=\alpha+\boldsymbol{\beta}^{\prime}\mathbf{x}_{it}+\boldsymbol{\theta}^{\prime}\mathbf{w}_{i}+\rho y_{it-1}+\nu_{it}
\end{equation}
where $\mathbf{x}_{it}$ is a $1 \times K$ vector of measurable explanatory variables changing with $i$ and $t$, and $\mathbf{w}_{i}$ is a $1 \times D$ vector of measurable time-invariant explanatory variables changing only with $i$. 
 
The lagged dependent variables, $y_{it-1}$, avoids omitted variables and the assumption $|\rho|<1$ implies stability respectively weakly-stationary.

The error term is $\nu_{it}=\mu_{i}+ \upsilon_{it}$ where $\mu_{i} \sim 
i.i.d. (0,\sigma_{\mu}^2)$ represents the individual-specific unobserved effects randomly drawn along with the observed covariates. $\mu_i$ captures unobserved heterogeneity, possibly correlated with $\mathbf{x}_{it}$ and $\mathbf{w}_{i}$ and by construction with $y_{it-1}$. The component $\upsilon_{it} \sim 
i.i.d. (0,\sigma_{\upsilon}^2)$ represents the idiosyncratic errors.
The indvidual heterogeneity is uncorrelated with the random noise, i.e. $Cov(\mu_{i}, \upsilon_{it})=0$; we will however relax this assumption in some of our Monte Carlo simulations.

The error components $\mu_{i}$ and $\upsilon_{it}$ are sometimes referred to as "permanent" and "transitory" components. To be poetic, \citet{Crowder1990} referred to the fixed part of the model as the "\emph{immutable constant of the universe}" (the unique constant term, $\alpha$), to $\mu_{i}$ as the "\emph{lasting characteristic of the individual}" (the individual effects), and to $\upsilon_{it}$ as the "\emph{fleeting aberration of the moment}" (the idiosyncratic shocks). Units are all different from one another in fundamental unmeasured ways. Particularly in behavioural sciences, individual differences are the norm rather than the exception, and the individual effects $\mu_{i}$ capture an additive linear combination of all unit-specific time-invariant omitted variables (such as the institutional and demographic specificities of the country; the cultural background of the people's family of origin; the ability, intelligence, enthusiasm, willingness to take risks, motivation, work ethic of the students; the type of company ownership and management involvement) that can capture the differences of each individual relative to the $\alpha$ benchmark. 

The assumptions regarding the idiosyncratic shock are:
\begin{enumerate}
\item $\mathbb{E}(\upsilon_{it} \upsilon_{jt}) = 0 \;\; \forall j$ and $i=1,..., N$, $t=1,..., T$ with $i\neq j$, errors are uncorrelated across units;
\item $\mathbb{E}(\upsilon_{it} \upsilon_{il}) = 0 \;\; \forall i = 1,..., N, \; l $ and $t=1,..., T$ with $t\neq l$, errors are serially uncorrelated over time;
\end{enumerate}

The easiest way to ensure validity of assumption (1) is to assume that $\upsilon_{it} =\tau_{t}+\varepsilon_{it}$. Adding time dummies $\tau_{t}$ aims to explicitly capture, within the \emph{fleeting aberration of the moment} $\upsilon_{it}$, each period-specific factor of \emph{aggregate influence} on micro units. Examples are the business cycle, macroeconomic events, neighbourhood effects, herd behaviour and social norms. If not accounted for, these unobservable common factors may generate a weak form cross-sectional dependence.\footnote{For an overview over cross-sectional dependence, see \citet{ChudikPesaranTosetti2011}.} Another simple way to account for common correlated effects is to use cross-sectional demeaned data as in \citet{Moral2013,Alvarez2022}. In case of a more complex structure strong cross-sectional dependence can be an outcome. A prominent example is the interactive fixed effects models, $\upsilon_{it} =\varphi_ i \lambda_t +\varepsilon_{it}$, where $\lambda_t$ indicates factors and $\varphi_ i$ individual-specific loadings, strong cross-sectional dependence. To account for strong cross-sectional dependence, regressions are augmented with cross-section averages \citep{Pesaran2006} or principal components \citep{Bai2009}.

To guarantee the assumption 2. of non-autocorrelation, which underlies the appropriate setting of the moment conditions exploited by the GMM, the dynamics of equation \eqref{eq:model1} can be extended. For example, we might consider an ARDL(1,1) and the corresponding equilibrium correction model (ECM):
\begin{equation}\label{eq:model2}
y_{it}=\alpha+\boldsymbol{\beta_1}^{\prime}\mathbf{x}_{it}+\boldsymbol{\beta_2}^{\prime}\mathbf{x}_{it-1}+\boldsymbol{\theta}^{\prime}\mathbf{w}_{i}+\rho y_{it-1}+\nu_{it}
\end{equation}

The absence of autocorrelation of errors also goes hand in hand with the non-rejection of the restrictions $\boldsymbol{\beta_2}=0$ and $\rho=0$. In this case, a static model is obtained, for which the GMM approach is nevertheless interesting as it provides "internal" instruments (information within the model), thus avoiding the difficulties of finding good "external" instruments.

\section{The CRE-GMM estimation}\label{sec:CREGMM}
One of the aims of our Correlated Random Effects GMM-lev estimation method (CRE-GMM) is to keep equation \eqref{eq:model1} in levels as they allow estimating the effects of measurable time-invariant explanatory variables, $\mathbf{w}_{i}$, while also considering the role of unmeasurable individual heterogeneity which, if omitted, would generate upward bias of the autoregressive parameter (and bias of all other parameters through the smearing effect). 
The individual effects $\mu_{i}$ capture an additive and linear combination of all time-invariant unit-specific unobservable variables. 
They capture the differences of each individual with respect to the benchmark $\alpha$. 
The GMM-dif \citep{Anderson1981,Anderson1982,Arellano1991,Arellano1995,Alvarez2003} could solve the upward bias due to endogeneity because of heterogeneity, but it does not allow estimating the parameters associated with the measurable time-invariant variables $\mathbf{w}_i$. This problem also affects the GMM-sys \citep{Arellano1995,Blundell1998}, at least in terms of efficiency, as most of the equations are first-differenced while the level equation is only retained for non-redundant moment conditions, actually for only one more period per panel unit.

Another reason why the GMM-lev is an attractive alternative to the GMM-dif estimator is that its performance does not deteriorate for $\rho$ approaching unity \citep{Binder2005}. The closer the autoregressive parameter is to one, the weaker would be the relation between lagged levels and first difference variables. On the other hand, in GMM-lev, a large autoregressive parameter implies a strong link between lagged first differences and level variables \citep{Bewley1979}. Hence we suppose to get more informative and relevant estimations from the GMM-lev than from the GMM-dif and GMM-sys. 

The addition of Correlated Random Effects to equation \eqref{eq:model1} can be analysed by exploiting the recursive substitutions for $y_{it-1}$ in equation \eqref{eq:model1}  \citep{Bhargava1983,Kiviet2007,Nerlove2008}:
\begin{align}\label{eq:model4}
y_{i\tilde{t}} & =\rho^{\tilde{t}}y_{i0}+\sum_{\tau=0}^{\tilde{t}-1}\rho^{\tau}\alpha_{i}+\sum_{\tau=0}^{\tilde{t}-1}\rho^{\tau}\boldsymbol{\theta}^{\prime}\mathbf{w}_{i}+\sum_{\tau=0}^{\tilde{t}-1}\rho^{\tau} \boldsymbol{\beta}^{\prime}\mathbf{x}_{i\tilde{t}-\tau}+\sum_{\tau=0}^{\tilde{t}-1}\rho^{\tau}\upsilon_{i\tilde{t}-\tau}  \notag \\
& =\rho^{\tilde{t}}y_{i0}+\dfrac{1-\rho^{\tilde{t}-1}}{1-\rho}\alpha_{i}+\dfrac{1-\rho^{\tilde{t}-1}}{1-\rho}\boldsymbol{\theta}^{\prime}\mathbf{w}_{i}+\sum_{\tau=0}^{\tilde{t}-1}\rho^{\tau} \boldsymbol{\beta}^{\prime}\mathbf{x}_{i\tilde{t}-\tau}+\sum_{\tau=0}^{\tilde{t}-1}\rho^{\tau}\upsilon_{i\tilde{t}-\tau}
\end{align}
where $\alpha_i=\alpha+\mu_i$ and $\tilde{t}$ covers the pre-sample and sample periods. The dependent variable can be separated into four components. The first component, $\rho^{t}y_{i0}$ is the term that depends on the initial observations, $y_{i0}$, and influences the behaviour of any estimators as long as $T_i$ is finite; its effect does not vanish, and affects each subsequent period when the time dimension is short. Instead, its relevance decreases when $T_i$ is large, under the condition of weak stationarity $|\rho|<1$. The effect of the initial conditions does not vanish with $T_i$ when $\rho$ is close to unity. 
The starting values may be seen as representing the initial individual endowments. Particularly in longitudinal panels where $T_i$ is rather small and asymptotic concerns $N \rightarrow \infty$, the effects of the initial conditions are not asymptotically diminishing, and hence the assumptions on initial observations play an important role in determining the properties of the various estimators proposed in the literature, like GMM-sys and GMM-lev. \citet{Hahn1999} argues that in estimating an AR(1) model on panel data it is fairly common to disregard the potentially informative role of the distribution of initial conditions $y_{i0}$ for the estimation of the autoregressive parameter $\rho$. This practice is understandable because misspecification of the distribution of $y_{i0}$ would result in the inconsistency of the resultant estimator. Perhaps because of this concern, efficiency in dynamic panel literature has been discussed in the framework where $y_{i0}$ was assumed to be ancillary for the parameter of interest. He shows that the marginal information contained in the initial condition is substantially even when $T_i$ is relatively large, and the efficiency gain tends to be larger for $\rho$ close to one, as the coefficient $\rho^t$ of $y_{i0}$ indicates that the importance of initial condition in $y_{i\tilde{t}}$ is an increasing function of $|\rho|$. 

The second term, $\dfrac{1-\rho^{\tilde{t}-1}}{1-\rho}\alpha_{i}+\dfrac{1-\rho^{\tilde{t}-1}}{1-\rho}\boldsymbol{\theta}^{\prime}\mathbf{w}_{i}$, is a modified intercept that depends on the unit-specific unobservable effects, $\mu_i$ and possibly on observable individual characteristics, $\mathbf{w}_{i}$; individual effects and the autocorrelation coefficient interact to determine the long run means of the $(y_{i1}, \dots, y_{iT_i})$ series, constant for each individual $i$. The third term, $\sum_{\tau=0}^{\tilde{t}-1}\rho^{\tau} \boldsymbol{\beta}^{\prime}\mathbf{x}_{i\tilde{t}-\tau}$, is a component depending on current and past values of  $\mathbf{x}_{i\tilde{t}}$, and related to the dynamics of the model 
Finally, the last term is a moving average of the disturbances $\upsilon_{i\tilde{t}}$.

It is reasonable to assume that the process driving \(y_{i\tilde{t}}\) has been in process sufficiently long in the pre-sample period, such that \(\tilde{t}\rightarrow \infty\). Then Equation \ref{eq:model4} can be re-formulated as:
\begin{align}\label{eq:model4a}
y_{it} & =\sum_{\tau=0}^{\infty}\rho^{\tau}\alpha_{i}+\sum_{\tau=0}^{\infty}\rho^{\tau}\boldsymbol{\theta}^{\prime}\mathbf{w}_{i}+\sum_{\tau=0}^{\infty}\rho^{\tau} \boldsymbol{\beta}^{\prime}\mathbf{x}_{it-\tau}+\sum_{\tau=0}^{\infty}\rho^{\tau}\upsilon_{it-\tau}  \notag \\
& =\dfrac{1}{1-\rho}\alpha_{i}+\dfrac{1}{1-\rho}\boldsymbol{\theta}^{\prime}\mathbf{w}_{i}+\sum_{\tau=0}^{\infty}\rho^{\tau} \boldsymbol{\beta}^{\prime}\mathbf{x}_{it-\tau}+\sum_{\tau=0}^{\infty}\rho^{\tau}\upsilon_{it-\tau}
\end{align}

This implies that it is impossible to estimate $ARDL(\infty, \infty)$ models with finite time dimension $T_i$ (even if $T_i$ were 40 periods). Instead, we truncate the dynamics of explanatory variables at some $(p, q)$ lags, say $(1,0)$ in a Partial Adjustment Model (PAM) or $(1,1)$ in an Error Correction Model (ECM); hence we omit $\sum_{\tau=q+1}^{\infty}\rho^{\tau} \boldsymbol{\beta}_{\tau}^{\prime}\mathbf{x}_{it-\tau}$ in equation \eqref{eq:model1}.

The individual effects can thus be considered as random and interpreted in terms of the past histories of each individual in the panel prior to the time when estimates begin, and these past histories are functions of the stochastic variables, omitted because of the lag truncation in $\mathbf{x}_{it}$ and $y_{it}$.\footnote{Take the example in \citet{Nerlove2008} in which $\mathbf{x}_{it}=\gamma^{\prime}_i \mathbf{x}_{it-1}+\omega_{it}$ with $\omega_{it} \sim i.i.d. (0,\sigma_{\omega_i}^2)$, cross-sectionally and serially unrelated. For $t, j \in \left \{0, \dots, Q \right \}$ (the set of indices for which $\mathbf{x}_{it}$ is observed, with $q$ used to specify the dynamics chosen much less than $Q$), the j-order autocorrelation is $ \mathbb{E}(\mathbf{x}_{it} \mathbf{x}_{it-j}) = \dfrac{\gamma_{i}^{j}}{1-\gamma_{i}^{2}}\sigma_{\omega_i}^2$. It follows that $\mathbf{x}_{it}$ and $\alpha_i$ are correlated, $ \mathbb{E}(\mathbf{x}_{ij} \alpha_{i}) = \sum_{\tau=q+1}^{\infty} \boldsymbol{\beta}_{\tau}^{\prime} \mathbb{E}(\mathbf{x}_{ij} \mathbf{x}_{it-\tau})=\dfrac{\sigma_{\omega_i}^2}{1-\gamma_{i}^{2}} \sum_{\tau=q+1}^{\infty} \boldsymbol{\beta}_{\tau} \gamma_{i}^{\left | j-\tau \right |}$, with a correlation depending on how close to the beginning of the sample period the observation on $\mathbf{x}_{it}$ is taken. This introduces additional ``$\boldsymbol{\beta}_{\tau}$'' parameters in equation (\ref{eq:model1}) capturing the relationship between the individual effects and the observed past values of the explanatory variables $\mathbf{x}_{it}$; the greater $\sigma_{\omega_i}^2$ the greater is the signal to noise ratio on one side, but the greater the dependence between $\mathbf{x}_{it}$ and $\alpha_i$ on the other side (especially for $j$ near the beginning of the observation period). Further useful insights are in footnote 7 of \citet{Bhargava1987}.}

As there is no good reason to treat the initial observation $y_{i0}$ differently from subsequent observations or from unobserved previous values, let us suppose that the temporal observations can be split as:  
\begin{align*}
& \mbox{pre-sample} \qquad \quad \mbox{estimation sample} \notag \\
& \overlinesegment{s=1, 2, 3, \dots, S_{i}} \quad \overlinesegment{t=S_i+1, \dots, T_{i}} \notag \\
& \overlinesegment{s= \dots, -2, -1, 0} \quad \overlinesegment{t=1, 2, 3, \dots, T_{i}} \notag \\
& \overlinesegment{\tau=q+1, \dots, \infty} \quad \overlinesegment{\tau = 0, 1, \dots \dots, q}
\end{align*}

The starting values computed by exploiting the $s = 1, \dots, S_i$ observations may be seen as representing the initial individual endowments on each unit $i$. Hence we can express equation (\ref{eq:model4a}) as:
\begin{align}\label{eq:model4b}
y_{it} & = \dfrac{1}{1-\rho}\alpha_{i}+\dfrac{1}{1-\rho}\boldsymbol{\theta}^{\prime}\mathbf{w}_{i}+\sum_{\tau=0}^{S_{i}-1}\rho^{\tau} \boldsymbol{\beta}^{\prime}\mathbf{x}_{it-\tau}+\sum_{\tau=S_{i}}^{\infty}\rho^{\tau} \boldsymbol{\beta}^{\prime}\mathbf{x}_{it-\tau}+ \notag \\
 & +\sum_{\tau=0}^{S_{i}-1}\rho^{\tau}\upsilon_{it-\tau}+\sum_{\tau=S_{i}}^{\infty}\rho^{\tau}\upsilon_{it-\tau}
\end{align}
where $\sum_{\tau=0}^{S_{i}-1}\rho^{\tau} \boldsymbol{\beta}^{\prime}\mathbf{x}_{it-\tau}$ includes the observations used in the estimation sample, $\mathbf{x}_{it}, \mathbf{x}_{it-1}, \dots, \mathbf{x}_{it-S_{i}+1}$, while $\sum_{\tau=S_{i}}^{\infty}\rho^{\tau} \boldsymbol{\beta}^{\prime}\mathbf{x}_{it-\tau}$ includes the observations of the pre-sample period used to proxy initial conditions, $\mathbf{x}_{it-S_{i}}, \mathbf{x}_{it-S_{i}-1}, \dots, \mathbf{x}_{i1}$.

Lag truncation leads to omission of observations over time. The observations of $y_{it}$ and $\mathbf{x}_{it}$ in those time periods can represent the individual endowments of each unit $i$.\footnote{The lag length of the ARDL(p,q) should be selected in such a way as to imply uncorrelated errors.} The individual effects $\alpha_i$ can be then written as:
\begin{equation}\label{eq:model4c}
\alpha_i=\boldsymbol{\pi}_{x}^{\prime}\breve{\mathbf{x}}_{i.}+\pi_{y} \breve{y}_{i.}^{1}+e_i
\end{equation}
where $\breve{\mathbf{x}}_{i.}=S_{i}^{-1}\sum_{s=1}^{S_{i}} \mathbf{x}_{is}$ and $\breve{y}_{i.}^{1}=S_{i}^{-1}\sum_{s=1}^{S_{i}}y_{is-1}$, and $S_i < T_i$ is the pre-sample period.  
To put it differently, we calculate the averages of the explanatory variables for the periods $s=1, \dots, S_i$ to capture permanent components, and then implement the estimations over the periods $t=S_i+1, \dots, T_i$.  

According to equation (\ref{eq:model4c}), equation (\ref{eq:model4b}) can be written as:
\begin{align}\label{eq:model4d}
y_{it} & =  \dfrac{1}{1-\rho} \left( \boldsymbol{\pi}_{x}^{\prime}\breve{\mathbf{x}}_{i.}+\pi_{y} \breve{y}_{i.}^{1}\right) +\dfrac{1}{1-\rho}\boldsymbol{\theta}^{\prime}\mathbf{w}_{i}+\sum_{\tau=0}^{S_{i}-1}\rho^{\tau} \boldsymbol{\beta}^{\prime}\mathbf{x}_{it-\tau}+ \notag \\
 & +\sum_{\tau=0}^{S_{i}}\rho^{\tau}\upsilon_{it-\tau}+\eta_{it}
\end{align}
where $\eta_{it} \approx e_i+ \sum_{\tau=S_{i}}^{\infty}\rho^{\tau}\upsilon_{it-\tau}$.

In equation \eqref{eq:model4d} we suppose that the unobserved individual effects/initial conditions can be estimated from the history of observations of the explanatory variables entering our equation of interest. Using values dated before the estimation sample to compute proxies for the unobserved heterogeneity avoids correlation with any subsequent shock in the equation of interest. This has the distinct advantage of  producing weakly exogenous (pre-determined) regressors as the measurement of individual effects is based exclusively on pre-sample information. For example, using the average innovative activity carried out by firms in the pre-estimation period makes it possible to capture unobservable differences in accumulated knowledge, returns to scale and cost of R\&D not affected by endogeneity \citep{Bontempi2023}. 
For stationary stochastic processes, such as ARs, the pre-sample mean is a more informed estimate of the steady state solution than the first sample observation. Averages condense past informations and therefore are better suited to represent the initial conditions in comparison to first observations in a sample. A further advantage is that they mitigate possible large variations in the time series and measurement errors (being divided by $S_i$, the statistical averaging effect is achieved). In contrast, the single initial observation can be strongly influenced by the short-term cyclical position of the variables and/or the occurrence of random shocks. Indeed, the use of the $i^{th}$ individual's time series mean (\citet{Mundlak1978}'s approach) is  more restrictive than using each observed variable at all the different time periods for each unit $i$, $\tilde{y}=(y_{i1}, \dots, y_{iS_{i}})$ and $\tilde{\mathbf{x}}=(\mathbf{x}_{i1}, \dots, \mathbf{x}_{iS_{i}})$, as in the \citet{Chamberlain1980}'s approach. However, the simulation results in \citet{Hsiao2018} suggest that to deal with the incidental parameters issues arising from the presence of individual-specific effects,\footnote{In a longitudinal, short $T$, panel, asymptotic theory relies on $N \rightarrow \infty$ and as $N \rightarrow \infty$ so too does the number of individual effects to estimate. This is the \textit{incidental-parameters} problem \citep{Neyman1948,Lancaster2000a,Moon2015}.} following the \citet{Mundlak1978}’s suggestion to condition on the time series average of individual’s observed regressors performs better than conditioning on each observed variable at all different time periods as suggested by \citet{Chamberlain1980}. The averages tend to perform better when the temporal dimension is large (possibly larger than $N$). For small $S$ ($N>S$) the two approaches can lead to asymptotically unbiased inference, with the \citet{Mundlak1978}'s approach yielding the smaller bias and RMSE.

Our idea to capture initial conditions is sufficiently general and encompasses several other specifications of the initial values considered in the literature as special cases. We are close to \citet{Bhargava1983} suggesting to use $y_{i0}=b_0+\sum_{k=1}^{K_1} \sum_{t=0}^{T}b_{1kt} x_{kit} + \sum_{k=1}^{K_2} b_{2k} w_{ki}+ \mu_i + \upsilon_{i0}$ as the equation determining the initial observations. \citet{Anderson1981} show (in their Table 1) the sensitivity of maximum likelihood estimators to four alternative assumptions about initial conditions. At one extreme, it can be assumed that the initial observations $y_{i0}$ are fixed constants specified independently of the parameters of the model; under this specification there are no unit-specific effects at the initial period $t=1$ which considerably simplifies the analysis.\footnote{However, as pointed out by \citet{Nerlove1992}, unless there is a specific argument in favour of treating $y_{i0}$ as fixed, in general such an assumption is not justified and can lead to biased estimates.} Alternatively, the initial values can be assumed to be random draws from a distribution with a common mean: $y_{i0}=c_0+u_{i0}$ where $u_{i0}$ is assumed to be independent of $\alpha_i$ and $\upsilon_{it}$. Once again, the starting value and the equilibrium level are independent. At the other extreme, the initial endowment affects the level, $y_{i0}=c_0+\alpha_{i0}$, and the initial observation is no different from any other observations (an assumption close to our assumption). 

A drawback of our approach is the requirement of a sufficient number of pre-sample observations for each unit $i$. In the case of unbalanced panel data, the initial conditions of the units entering the sample after $S$ (and thus with no observations available in the pre-sample period) could be proxied by the averages computed for other units available in the pre-sample and characterised by ``similar'' initial endowments; for example, we could use individual averages of firms characterised by the same size, within the same industry, localized in the same geographical zone, and with other cluster-level characteristics. Our approach is also valid under the condition that the $\pi_x$ and $\pi_y$ parameters of equation \eqref{eq:model4c} do not vary across $i$ otherwise we will have the incidental parameters problem; however, the correlated random effects approach does not lead to incidental-parameter bias when $T$ and $N$ are of comparable dimension \citep{Bai2009a}. 

\subsection{The CRE-GMM estimation - advantages}\label{sec:CREGMMadv}
Based on \citet{Mundlak1978} idea, the CRE approach models the relationship between $\alpha_{i}$, $\mathbf{x}_{it}$ and the lagged dependent variable $y_{it-1}$. An attractive assumption of \eqref{eq:model4c} is that the individual specific effect is decomposed into a \textit{systematic component} depending on the individual specific mean of the regressors, $\breve{\mathbf{x}}_{i.}$ and $\breve{y}_{i.}^{1}$, and the remaining \textit{unsystematic component} treated as an additional random term, $e_{i}$. In the words of \citet{Bell2015}, the endogeneity due to correlation between lower-level predictors, $\textbf{x}_{it}$, and higher-level part of residuals, $\mu_i$, occurs so regularly, alongside consideration of the \citep{Hausman1978}'s specification test. The fixed effects solution which circumvents this problem by controlling out all differences between higher-level units brings to a very limited model, being unable to estimate the effects of higher-lever variables, $\textbf{w}_i$. The random effects estimation used in a formulation similar to that originally proposed by \citet{Mundlak1978} which partitions the effect of lower-level covariates into two parts, treats endogeneity as a substantive phenomenon which occurs when a given lower-level variable with different within and between processes is assumed to have a single homogeneous effect. A well-specified random effects model can be used to achieve everything that the fixed effects models achieve, and much more besides. 

As an example, firm-specific technical efficiency and economies of scale are between effects, while technological changes over time are within effects. Permanent unemployment is a between measure, while temporary unemployment is a within measure. When studying the birth weight of newborns \citep{Abrevaya2006}, weight may be more related to maternal smoking behaviour than smoking cessation.
When studying the impact of GDP ($x_{it}$) on democracy ($y_{it}$), a fixed-effects estimation yields a single parameter. This leads to the counter-intuitive finding that GDP has no relationship with democratization \citep{Acemoglu2008}. Unobserved factors unique to each country, related to historical, cultural, and political factors, can affect democracy levels. These factors play a role in pushing countries to higher or lower democracy levels. Ignoring the "between" cluster effects can lead to incorrect inferences when "within" cluster data show no GDP-democracy relationship. In psychological research, the \textit{ecological fallacy} arises from confounding within and between group differences \citep{Robinson1950}.

Augmenting the dynamic panel data regression with the systematic part of the individual effects yields a model representation that includes the random and fixed effect specifications as special cases:
\begin{equation}\label{eq:model4e}
y_{it}=\boldsymbol{\beta}^{\prime}\mathbf{x}_{it}+\boldsymbol{\theta}^{\prime}\mathbf{w}_{i}+\rho y_{it-1} + \boldsymbol{\pi}_{x}^{\prime}\breve{\mathbf{x}}_{i.}+\pi_{y}\breve{y}_{i.}^{1}+\upsilon_{it}+e_{i}
\end{equation}

In our random effects framework, initial conditions give rise to a ``between’ equation'' \eqref{eq:model4c}, which captures sample variation across units, while the ``within'' estimated parameters $\boldsymbol{\beta}$ of equation \eqref{eq:model4e} captures the sample variation within each unit over time.\footnote{A similar idea in the maximum likelihood framework is in \citet{Lee2020}. The RE estimator under the hypothesis of exogenous $\mathbf{x}_{it}$ is investigated also by \citet{Hsiao2018,Hsiao2020}.} 

The CRE-GMM approach based on equation \eqref{eq:model4e} has the following advantages:
\begin{enumerate}
\item It avoids the omission/inability to capture individual heterogeneity and the mixture of ``spurious'' (past/state dependence) persistence, due to the serial correlation generated by the unobserved permanent $\mu_i$, and ``true'' (path/behavioural) persistence, measured by $\rho$) which is the causal effect of past values of the dependent variables on its current realization. 
\item The inclusion of individual averages avoids bias from omitted variables and the ``endogeneity bias due to heterogeneity''. In general, it is a rather difficult hypothesis to rule out a statistical dependence between the individual specific effects and the explanatory variables. The GMM-lev estimation applied to equation \eqref{eq:model4e} extended in a \citet{Mundlak1978}'s fashion treats the endogeneity as a \textit{substantive phenomenon} without necessarily requiring instrumental variables estimation for such an endogeneity. 
\item Instead of using (erroneously?) a single weighted average of the within and between effects of the covariates $\mathbf{x}_{it}$, we keep the partition and assess whether the between effect is relevant and thus cannot be neglected as in fixed effects/first-differences estimations. Hence, we simultaneously estimate the within effect (effect of unit-specific temporary deviations from the period averages of $\mathbf{x}_{it}$) and the between effect (effect of permanent differences between individuals in $\mathbf{x}_{it}$). 
\item Obtaining the within estimate of each $\mathbf{x}_{it}$ while controlling for systematic differences in the levels of the covariates across $i$ results in a more convincing analysis. We can estimate coefficients $\boldsymbol{\theta}$ on time-constant variables while reproducing the fixed effects estimates of the time-varying variables, $\boldsymbol{\beta}$ \citep{Wooldridge2019,Wooldridge2021,Schunck2013,Schunck2017}. 
\item A causal interpretation to the coefficients of $\mathbf{w}_i$ can be given only if $\mathbb{E}(\mu_i|\mathbf{x}_{it}, \mathbf{w}_i) = \mathbb{E}(\mu_i|\mathbf{x}_{it} = \boldsymbol{\pi}_{x}^{\prime}\breve{\mathbf{x}}_{i.})$ meaning that $\mathbf{w}_i$ is uncorrelated with $\mu_i$ once the time-varying covariates are controlled for. This assumption is too strong in many applications, but one still might want to include time-constant covariates \citep{Wooldridge2010}. If $\mathbf{w}_i$ is correlated with individual heterogeneity, the effect of time-invariant observable variables can be identified in the spirit of \citet{Hausman1981} who use, as instruments, both the within and between transformations of the components of $\mathbf{x}$ uncorrelated with individual heterogeneity. In our CRE-GMM approach, we directly add to the model the individual averages of $\mathbf{x}$ to capture the individual heterogeneity (see further discussion in Section \ref{sec:GMM}). Equation \eqref{eq:model4c} could also be extended to include measurable time-invariant variables correlated with individual heterogeneity.
\item Indeed, equation \eqref{eq:model4e} allows for the variable addition, or regression based, robust version of the \citet{Hausman1978} test on $H_{0}: \boldsymbol{\pi}_{x}=0$ for individual effects uncorrelated with covariates \citep{Arellano1993}; this test can be implemented also for sub-sets of covariates. This version avoids computational problems and can be robust, avoiding the severe size distortion on inference after using the non-robust Hausman \citep{Guggenberger2010}.    
\item We have greater variability, due to the combination of the variation across cross-sectional units (between) with the variation over time (within), and more informative data. As a results, we could achieve more efficient estimations and mitigate multicollinearity problems, particularly in the weighting matrix used by the GMM to handle the moment conditions.
\end{enumerate}

\subsection{The CRE-GMM estimation - moment conditions}\label{sec:GMM}
The moment conditions exploited to estimate equation \eqref{eq:model4e} in levels must tackle the endogeneity of the lagged dependent variable and the possible \textit{double} endogeneity of the $\mathbf{x}_{it}$ variables:

\begin{align*}
& \mathbb{E}[y_{it-p} (\mu_i + \upsilon_{it})] \neq 0 \quad p \geq 1, \forall t=S_i+1, \dots, T_i  \\
& \mathbb{E}[\mathbf{x}_{it-q} (\mu_i + \upsilon_{it})]=\mbox{\textcolor{red}{?}} \quad q \geq 0, \forall t=S_i+1, \dots, T_i 
\end{align*}

The lagged dependent variable includes the unobservable $\mu_i$, hence, by definition, it is endogenous because of the \textit{individual heterogeneity}; considering the shock component $\upsilon_{it}$, the lagged dependent variable is predetermined, meaning uncorrelated with $\upsilon_{it}, \upsilon_{it+1}, \dots, \upsilon_{iT_i}$. The $\mathbf{x}_{it}$ variables could be (\textcolor{red}{?}) correlated with both the individual heterogeneity $\mu_i$ and the shock $\upsilon$. Indeed, a variable in $\mathbf{x}_{it}$ could be predetermined instead than strictly exogenous, meaning influenced by past values of the dependent variable (correlated with past shocks); for example, in a model explaining wages of workers, a change in $t$ of the dependent variable wages could determine becoming member of a union in $t+1$ hence producing a change in one of the explanatory variables \citep{Vella1998}). Often the dynamic model in equation \eqref{eq:model4e} could be affected by omitted regressors correlated with $\mathbf{x}_{it}$, measurement errors in $\mathbf{x}_{it}$, simultaneity, thus producing endogeneity of the $\mathbf{x}_{it}$ variables.

When the model is first-differenced, as in GMM-dif, the moment conditions are based on the levels of $y_{it-p}$ and $\mathbf{x}_{it-q}$. The comparison of our moment conditions and those of GMM-dif is in Table \ref{tab:MC}.   
When exploiting the equations in levels, as in GMM-sys, the usual approach to tackle correlation with the individual heterogeneity included in the error term, is to exploit the moment conditions based on the first-differenced variables $\Delta y_{it-p}$ and $\Delta \mathbf{x}_{it-q}$. Compared with GMM-sys, where the equations in levels are added only for the non-redundant period with respect to the periods already exploited by the GMM-dif, our approach in principle uses all the moment conditions available for the level-equations, not just the non-redundant ones. However, it has to be considered that panels often have moderate $N$ and long $T$, and that applied econometricians tend to use in practice less than the total number of available instruments when that number (which depends on $T$) is judged to be not small enough relative to the cross-sectional sample size. To combine this practice with a more structured lag selection, we follow the suggestion of \citet{Ziliak1997,Bun2006} to use fewer IVs that potentially available for GMM-lev in the 
$(T-1) \times (T-2)/2$ matrix,\footnote{The instrument matrix has full column rank if $N \geq T/2$ and $T \geq 2$.} i.e. to restrict the moment conditions to lags from $t-1$ to $t-3$. 
The selected lags are valid to deal with predetermined and endogenous explanatory variables in the equation in levels.

\begin{table}[!ht] 
\centering
\caption{Comparison of moment conditions}\label{tab:MC}
\begin{tabular}{lllllllll}
\toprule
\multicolumn{2}{c}{Classification} & \multicolumn{2}{c}{Method} \\
\midrule 
& & GMM-dif & CRE-GMM-CRE-GMM5 \\
\midrule
Correlat. & $\mathbb{E}[y_{it-p} \mu_i] \neq 0, p>0$ & & \\ 
Predet. & $\mathbb{E}[y_{it-p} \upsilon_{it}] =0, p\geq 1$ & $\mathbb{E}[y_{it-m} \Delta \upsilon_{it}] = 0$ & $\mathbb{E}[\Delta y_{it-m} (\mu_i + \upsilon_{it})] = 0$ \\
&  & $m \geq 2; \quad t=3, \dots, T_i$  &  $m=1, 2, 3; \quad t=S_i+2, \dots, T_i$ \\
& & & \scriptsize{CRE-GMM and CRE-GMM3} \\
& & & \scriptsize{additional MCs:} \\
& & & $\mathbb{E}[\breve{y}_{i.}^{1} (\mu_i + \upsilon_{it})] = 0$ \\
\midrule
& & GMM-dif & CRE-GMM-CRE-GMM2 \\
\midrule
Correlat. & $\mathbb{E}[\mathbf{x}_{it-q} \mu_i] \neq 0, q \geq 0$ & & \\
Predet. & $\mathbb{E}[\mathbf{x}_{it-q} \upsilon_{it}]=0, q \geq 1$ & $\mathbb{E}[x_{it-m} \Delta \upsilon_{it}] = 0$ &  \\
&  & $m \geq 1  \quad t=3, \dots, T_i$ & $\mathbb{E}[\Delta x_{it-m} (\mu_i + \upsilon_{it})] = 0$ \\
Endog. & $\mathbb{E}[\mathbf{x}_{it-q} \upsilon_{it}] \neq 0, q = 0$ & $\mathbb{E}[x_{it-m} \Delta \upsilon_{it}] = 0$ &  $m=1, 2, 3  \quad t=S_i + 2, \dots, T_i$ \\
&  & $m \geq 2  \quad t=3, \dots, T_i$ &  \\
& & & \scriptsize{CRE-GMM-CRE-GMM1} \\
& & & \scriptsize{additional MCs:} \\
& & & $\mathbb{E}[\breve{\mathbf{x}}_{i.}^{1} (\mu_i + \upsilon_{it})] = 0$ \\
\midrule
& & GMM-dif & CRE-GMM3-CRE-GMM5 \\
\midrule
Uncorrelat. & $\mathbb{E}[\mathbf{x}_{it-q} \mu_i] = 0, q \geq 0$ & & \\ 
Predet. & $\mathbb{E}[\mathbf{x}_{it-q} \upsilon_{it}]=0, q \geq 1$ & $\mathbb{E}[x_{it-m} \Delta \upsilon_{it}] = 0$ &  \\
&  & $m \geq 1  \quad t=3, \dots, T_i$ & $\mathbb{E}[x_{it-m} (\mu_i + \upsilon_{it})] = 0$ \\
Endog. & $\mathbb{E}[\mathbf{x}_{it-q} \upsilon_{it}] \neq 0, q = 0$ & $\mathbb{E}[x_{it-m} \Delta \upsilon_{it}] = 0$ & $m=1, 2, 3  \quad t=S_i + 2, \dots, T_i$  \\
&  & $m \geq 2  \quad t=3, \dots, T_i$ &  \\
& & & \scriptsize{CRE-GMM3-CRE-GMM4} \\
& & & \scriptsize{additional MCs:} \\
& & & $\mathbb{E}[\breve{\mathbf{x}}_{i.}^{1} (\mu_i + \upsilon_{it})] = 0$ \\
\bottomrule
\end{tabular}
\flushleft
{\scriptsize Note: Under the strict exogeneity assumption of $\mathbf{x}_{it}$ variables, we can exploit $m=0, 1, 2, 3$ moment conditions.}
\end{table}

In all the methods CRE-GMM-CRE-GMM5 we implemented, and in the three CRE-GMM-CRE-GMM2 estimations, we assume that $\mathbf{x}_{it}$, and by definition $y_{it-1}$, are correlated with the individual heterogeneity, $\mu_i$. In the three CRE-GMM3-CRE-GMM5 estimations we suppose that $\mathbf{x}_{it}$ is uncorrelated with the individual heterogeneity, $\mu_i$. GMM-style lags of the explanatory variables in levels are valid instruments for the untransformed equation (in levels) only if they are uncorrelated with the unit-specific error component, while the first set of CRE-GMM-CRE-GMM2 estimations uses first-differenced instruments, as suggested by \citet{Blundell1998}. We explore each of these identification strategies in estimation by comparing the use of lagged first-differenced $\mathbf{x}_{it}$ and lagged levels of $\mathbf{x}_{it}$ as instruments for the equation in levels (respectively, CRE-GMM-CRE-GMM2 and CRE-GMM3-CRE-GMM5). Using IVs in levels could increase the correlation between endogenous variables and IVs, hence the efficiency of the CRE-GMM approach.\footnote{The lagged dependent variable is always estimated with lagged first-differences of $y_{it-1}$ as $y_{it-1}$ is correlated with the individual heterogeneity by definition. Robustness checks about the use of lagged levels of $y_{it-1}$ as instruments do not show any improvement in the results}. 

The inclusion of unit-specific averages in equation \eqref{eq:model4e} allows the comparison of the two sets of estimations, CRE-GMM-CRE-GMM2 and CRE-GMM3-CRE-GMM5, as it explicitly models the individual endogeneity which is hence removed from the error term, making valid the IVs in levels. Additionally, it helps when there is no guarantee that the first-differenced instruments for the untransformed model are uncorrelated with the unit-specific error component. For example, \citet{Macher2021} examine the adoption of fuel-efficient precalciner kilns in the cement industry and have a region-specific term that affects all cement plants in the same geographic region. 
The first-differences are valid instruments only under the case that the region-specific effect is constant over time, a process that could arise in practice due to state-level differences, for example in unionization policies or tax rates. However, the region-specific term could evolve according to a finite moving-average process of order $M$, a possibility arising if construction projects take multiple periods to complete, but there are no spillovers from one construction project to future projects. In such a setting, T-period lags of the endogenous regressors are valid instruments if $T > M$.
The region-specific term could also evolve according to an autoregressive process if construction projects have positive spillovers on future projects, so that the effect of a positive region shock diminishes over time but never fully dies out. Lagged regressors taken from the period $t-T$ are valid instruments if they are orthogonal to the region-specific term in $t-T$ and the entire series of shocks in $t-T, t-T+1, . . . , t$.  

With regard to the assumptions concerning individual averages, let us compare three alternative cases: the pre-sample individual averages $\breve{\mathbf{x}}_{i.}$ and $\breve{y}_{i.}^{1}$ are exogenous (cases CRE-GMM and CRE-GMM3); only individual averages of $\mathbf{x}_{it}$ are exogenous (cases CRE-GMM1 and CRE-GMM4); the pre-sample individual averages of $y_{it-1}$ and $x_{it}$ are endogenous (cases CRE-GMM2 and CRE-GMM5). In Section \ref{sec:RES} we discuss the simulated results with and without instrumenting the proxies for the initial conditions. Individual averages supposed to be endogenous are instrumented by the same suitable lags used to instrument the explanatory variables of the model, $y_{it-1}$ and $\mathbf{x}_{it}$.\footnote{The idea behind these comparisons is to understand what happen when we use suspect moment conditions. \citet{Ditraglia2016} suggests that, in finite samples, the addition of a slightly endogenous but highly relevant instruments can reduce estimator variance by far more than bias is increased.} Instrumenting the averages obtained from the pre-sample with the lags of the variables in the estimating sample resembles the orthogonal forward deviations suggested by \citet{Arellano1995}.\footnote{In \citet{Alvarez2003} for fixed $T$ the IV estimators in orthogonal deviations and first-differences are both consistent, whereas as $T$ increases the former remains consistent but the latter is inconsistent. Using past observations has its antecedent in the work of \citet{Chamberlain1982} considering the use of long lags in regressors.}

An interesting aspect that emerges from comparing CRE-GMM-CRE-GMM1 and CRE-GMM3-CRE-GMM4 estimations with CRE-GMM2 and CRE-GMM5 estimations is whether individual characteristics have evolved over time. If agents maintain their personal characteristics unchanged over time, then it clearly follows that $\mathbb{E}[\breve{y}_{i.}^{1} \mu_i] \neq 0$ and $\mathbb{E}[\breve{\mathbf{x}}_{i.} \mu_i ] \neq 0$ and the moment conditions under CRE-GMM-CRE-GMM1 and CRE-GMM3-CRE-GMM4 and not valid.
If, instead, the behaviour of the agents evolves over time or, even better, if there is a structural and status change in individual characteristics such that $\alpha_i=\boldsymbol{\pi}_{x}^{\prime}\breve{\mathbf{x}}_{i.}+\pi_{y} \breve{y}_{i.}^{1}+e_i$ with $\breve{\mathbf{x}}_{i.}=S_{i}^{-1}\sum_{s=1}^{S_{i}} \mathbf{x}_{is}$ and $\breve{y}_{i.}^{1}=S_{i}^{-1}\sum_{s=1}^{S_{i}}y_{is-1}$ for $S_i < T_i$ but $\alpha_i \neq \boldsymbol{\pi}_{x}^{\prime}\tilde{\mathbf{x}}_{i.}+\pi_{y} \tilde{y}_{i.}^{1}+e_i$ with $\tilde{\mathbf{x}}_{i.}=T_{i}^{-1}\sum_{t=S-i +1}^{T_{i}} \mathbf{x}_{it}$ and $\tilde{y}_{i.}^{1}=T_{i}^{-1}\sum_{t=S_i+1}^{T_{i}}y_{it-1}$ for $t = S_i+1, \dots, T_i$, hence the moment conditions used in CRE-GMM-CRE-GMM1 and CRE-GMM3-CRE-GMM4 are valid.\footnote{We have that $\mathbb{E}[\breve{y}_{i.}^{1} \upsilon_{it}] = 0$ and $\mathbb{E}[\breve{\mathbf{x}}_{i.} \upsilon_{it}] = 0$ by definition, as the individual averages are computed by exploiting the pre-estimation period $s=1, \dots, S_i$.} In other terms, we are explicitly modelling the mean-stationarity assumption implying that the deviations of $y_{it-1}$ and $\mathbf{x}_{it}$ from the initial conditions are uncorrelated with the levels of the initial conditions, and the ``new'' realisations of $y_{it-1}$ and $\mathbf{x}_{it}$ for $t=S_i+1, \dots, T_i$ are not informative for the individual heterogeneity $\mu_i$. 
Comparing estimates CRE-GMM2 and CRE-GMM5 with CRE-GMM-CRE-GMM1 and CRE-GMM3-CRE-GMM4 can add information regarding the behaviour of individuals.

As discussed in \citet{Roodman2009}, the validity of the equations in levels added to GMM-sys is based on whether individuals have achieved their steady state before the study period. Otherwise, the initial distance from the steady state may be greater the higher the individual effect (on which the steady state depends) and this produces, in the early periods of the study, a correlation between the first-differenced $y_{it-1}$ and the error term, $\mathbb{E}[\Delta y_{it-1} (\mu_i + \upsilon_{it})] \neq 0$.
Suppose that we would like to estimate production functions, in which $\mu_i$ may capture the effect of technical efficiency and unobserved managerial practices. 
The mean-stationarity assumption makes it difficult to deal with a mix of companies in which the youngest, still far from their steady state compared to more mature companies, can grow faster at the beginning of the sample period.
Note that an implication of lack of stationarity in mean is that the data in first differences will generally
depend on individual effects, making invalid the IVs for GMM-lev \citep{Alvarez2022}. In short panels, steady state assumptions about initial observations are also critical since estimators that impose them lose consistency if the assumptions fail. Moreover, there are relevant applied situations in which a stable process approximates the dynamics of data well, and yet there are theoretical or empirical grounds to believe that the distribution of initial observations does not coincide with the steady state distribution of the process. 

For example, educational experience has an effect on the earnings structure that is not crudely captured by years of schooling (a between effect), but also depends on on-the-job training (a within effect). During the early stages of their careers, workers may accept lower earnings because they expect that more experience will develop the skills needed to compensate them for higher future earnings \citep{Hause1980}. In these examples, the individual average of working experience added to the equation of interest (our CRE-GMM approach) can help in capturing the heterogeneous different starting points of the workers. 

Another example are growth models \citep{Bond2001G}: assuming that the means of the series, whilst differing across units, are constant through time could be implausible for e.g. per capita GDP.\footnote{The inclusion of the time dummies allows for common long-run growth in per capita GDP, consistent with common technical progress.} However, if the process has been generating the per capita GDP series for long enough, prior to the estimation sample, any influence of the true start-up conditions would be negligible and $\mathbb{E}(\Delta \mathbf{x}_{it} \mu_i )=\mathbb{E}(\Delta y_{it-1} \mu_i )=0 \quad \forall t$. Our CRE-GMM can help in capturing the country-specific initial conditions and the validity of the above assumptions.
 
\citet{Jin2021} generate the dynamic panel data with $m+T$ periods ($m=20$) where the starting value is from $N(0, I_N)$, and then take the last $T$ periods as their sample. By doing so, the initial value in the estimation is close to the steady state. They also use the first period of the simulated data as the initial observation in the estimation sample (so that $m=1$ and the process is away from its steady state). For the $m=1$ case, the GMM-sys has a larger bias; this is consistent with the theoretical prediction because the system GMM estimate requires that initial observations are uncorrelated with the individual effects, which is satisfied if the process has started a long time ago.

\section{Results}\label{sec:RES}
\subsection{Monte Carlo simulations}\label{sec:MC}

To asses the performance of our proposed estimator, we employ a Monte Carlo simulation with the following DGP based on an ARDL(1,1) model:

\begin{align}\label{MC:eq1}
	y_{it} &= \beta_0 + \rho y_{it-1} + \beta_1 x_{it}+ \beta_2 x_{it-1}  + u_{it} \\
	x_{it} &= \gamma_1 \mu_i + \vartheta x_{it-1} + \gamma_2 \epsilon_{it} + \xi_{it} \notag \\
	u_{it} &= \mu_i + e_{it} \notag \\
	e_{it} &= \gamma_3 \mu_i + \epsilon_{it} \notag
\end{align}

An ARDL(1,0) model is obtained from Equation \eqref{MC:eq1} if $\beta_2=0$. The error term \(u_{it}\) is composed into an individual effect \(\mu_i\) and an idiosyncratic shock \(e_{it}\).

The parameters \(\gamma_1\) and \(\gamma_2\) control the degree of endogeneity by specifying the correlation between \(\mathbf{x}_{it}\), the individual effects, and the random noise \(\epsilon_{it}\).
\(\gamma_3\) sets the correlation between the individual effects \(\mu_i\) and the error \(\epsilon_{it}\).\footnote{Note, when $\gamma_3>0$, then the variance of the idiosyncratic shock is lower than the variance of the individual heterogeneity. For a more detailed discussion see Equation \eqref{eq:FixE} in the Appendix \ref{sec:MCset}.} 
Further details on the setup of Monte Carlo simulations are in Appendix \ref{sec:MCset}.

As a baseline we estimate the Pooled Ordinary Least Squares (POLS), the Random Effects (RE) and the Fixed Effects (FE) estimators on model \eqref{MC:eq1}. Under the assumption that $x_{it}$ is uncorrelated with the idiosyncratic shock ($\gamma_2=0$) the POLS, RE and FE estimates serve as benchmarks for the consistent estimate of the $\rho$ parameter. The upper bound is provided by the POLS, affected by omitted heterogeneity bias, and the RE which assumes there is no correlation between the regressors and the individual effects $\mu_i$ ($\gamma_1=0$). The lower bound is the FE which eliminates the influence of any time-invariant variable from the model by exploiting the within-transformation of the data, $x_{it}-x_{i.}$, where $x_{i.}={T_{i}}^{-1}\sum^{T_{i}}_{t=1}x_{it}$. The FE estimator would produce a consistent estimate of the autoregressive parameter for $T_i \rightarrow \infty$, as the \citet{Nickell1981}'s bias disappears.

We then estimate two correlated random effects models. For the first one, named CRE1, we add only the individual average of the lagged dependent variable. For the second one, named CRE2, the individual average of the explanatory variable $x_{it}$ are also added. If the explanatory variable $x_{it}$ is uncorrelated with the idiosyncratic shock, but $x_{it}$ (and obviously $y_{it-1}$) are correlated with $\mu_i$, then the difference in the estimates obtained by CRE1 and CRE2 to the baseline estimators should provide an idea of the bias due to the correlation with the individual heterogeneity.

Finally, we apply the standard GMM-lev (named GL) estimation to equation \eqref{MC:eq1}, and we implement the six CRE-GMM estimations, presented in Table \ref{tab:MC}, on the dynamic model \eqref{MC:eq1} extended to include, as additional explanatory variables, the individual averages of both $y_{it-1}$ and $x_{it}$ computed in the pre-sample period.\footnote{In the ECM specification we add the individual averages of both $x_{it}$ and $x_{it-1}$.} 
The parameters for the simulation are set as follows: we fix $\rho= \theta =0.5$, and $\beta_1=1$. $\gamma_1$ and $\gamma_2$ are varied between 0, 0.25 and 0.8, while $\gamma_3=[0, 0.8]$. With respect to the dimension of the cross-sections and time periods, we focus on three combinations of N and T: longitudinal panel ($N=1000, T=10$), macro panels ($N=25, T=40$), and multilevel panels ($N=100, T=25$).\footnote{Results with $N=[25, 100, 1000]$, $T=[5,10, 20, 40]$, and $\gamma_3=[0,0.25,0.8]$ yield 288 combinations and with different variances as shown in Table \ref{tab:variances} in the Appendix, Section \ref{sec:MCset}, produces 1152 experiments. We present the results for variances defined by equations \eqref{eq:FixE} and \eqref{eq:varXfix}. Equations \eqref{eq:FixEps} and \eqref{eq:varXifix} and alternative combinations of equations in Table \ref{tab:variances} of the Appendix \ref{sec:MCset} produce similar results.} All our simulations are repeated 1000 times and are done in Stata. Estimation are done with the community contributed command \texttt{xtdpdgmm} \citep{Kripfganz2019a}.

\subsection{Monte Carlo results}\label{sec:MCres}
The results for the PAM model are presented in Tables \ref{tab:FIXEPAM05BIGlongi_rho}-\ref{tab:FIXEPAM05BIGmulti_beta} and report the average bias and the root mean square error estimates for the RE, FE, CRE1, CRE2, GL, CRE-GMM-CRE-GMM5 estimators of $\rho$ and $\beta$. The model is specified as an ARDL(1,0) where the $x_{it}$ variable follows an autoregressive process with $\vartheta=0.5$; true $\rho$ and $\beta_1$ are 0.5 and 1, respectively.\footnote{The ECM results, with $\beta_1$ equal to 1 and $\beta_2$ equal to 0.5, are in Section \ref{sec:MCECM} of the Appendix.}

When there is no correlation with the idiosyncratic shock, $\gamma_2=0$, estimates from CRE1 and CRE2 are aligned between those from FE and RE. CRE1 and CRE2 eliminate, from the weighted average of between and within variability produced by RE, the between variability, captured by the individual averages. The correlation with individual effects, measured by individual averages, is unavoidable for the lagged dependent variable, generating the well-known interval for consistent dynamic estimates, with RE providing the upper bound benchmark and FE providing the lower benchmark. Since we do not have problems of standard endogenity, GMM-lev (GL) and the set of CRE-GMM estimators should provide similar estimates. This is in general true for longitudinal panels, while for multilevel and macro panels CRE-GMM is less biased and more efficient, particularly when we use levels of $x_{it}$ as instruments (CRE-GMM3-CRE-GMM5).

The presence of standard endogeneity, $\gamma_2 > 0$, produces a bias in the estimates increasing with the $\gamma_2$ parameter capturing the correlation between $x_{it}$ and $\upsilon_{it}$. The bias is greatly reduced by IVs in CRE-GMM framework: using first-differences or levels as IVs does not alter the results (there is a tendency to prefer levels for $\rho$), since we do not have correlation with individual heterogeneity. Instrumenting individual averages is usually preferable in terms of reduction of bias and standard errors in longitudinal panels. Under the $\gamma_2 = 0.25$ case, the bias in the FE estimates for $\beta_1$ is reduced when we implement IVs approaches: we have 0.231 (FE), -0.010 (GL) and -0.029 (CRE-GMM5) in the longitudinal panel; 0.219 (FE), -0.068 (GL) and 0.082 (CRE-GMM5) in macro panel; 0.226 (FE), -0.051 (GL) and 0.013 (CRE-GMM) in the multilevel panel. The $\rho$ parameter is moderately affected by endogeneity of $x_{it}$: the biases are  -0.110 (FE), 0.006 (GL) and 0.013 (CRE-GMM5) in the longitudinal panel; -0.058 (FE), 0.146 (GL) and 0.009 (CRE-GMM5) in the macro panel; -0.073 (FE), 0.053 (GL) and 0.009 (CRE-GMM5) in the multilevel panel. 

When there is endogeneity due to heterogeneity, $\gamma_1 > 0$, the $\rho$ parameter bias in GL increases from 0.011 (longitudinal) to 0.060 (multilevel) to 0.186 (macro). The best approach seems to be CRE-GMM2, with IVs in first-differences, presenting a bias going from 0.031 (longitudinal) to 0.032 (multilevel) and 0.054 (macro), and an apparently better disentangling of true (path) and spurious (state) dependence.  
The CRE-GMM2 approach is also beneficial for the $\beta_1$ parameter. The bias increases as $\gamma_1$ increases going from -0.009 to 0.019 (GL, longitudinal), -0.027 to 0.078 (GL, multilevel), -0.052 to 0.081 (GL, macro); on the opposite, the increase in the bias is smaller or null for CRE-GMM2, moving from -0.016 to 0.026 (longitudinal), from 0.031 to 0.030 (multilevel), from -0.014 to 0.016 (macro). Also CRE-GMM5, with the IVs in levels, work well, indicating that the inclusion of individual averages is able to capture the impact of individual heterogeneity correlated with $x_{it}$. 

When there is endogeneity from both sources and $0 < \gamma_1 < \gamma_2$, meaning that the correlation with the idiosyncratic shock prevails over the endogeneity due to heterogeneity, CRE-GMM2 works well in longitudinal panels for both $\rho$ and $\beta_1$, while in multilevel/macro panels also CRE-GMM5 has a good performance, particularly for the $\rho$ parameter. Good performance of CRE-GMM5 for $\rho$ in multilevel/macro and $\beta_1$ in longitudinal panels, and of  CRE-GMM2 for $\rho$ in longitudinal and $\beta_1$ in multilevel/macro is obtained when the correlation with individual heterogeneity prevails over the correlation with the idiosyncratic shocks, $0 < \gamma_2 < \gamma_1$. The bias of the $\rho$ parameter is 0.003 (GL) and 0.027 (CRE-GMM2) in longitudinal; 0.033 (GL) and 0.001 (CRE-GMM5) in multilevel; 0.057 (GL) and -0.011 (CRE-GMM5) in macro panel. The bias of the $\beta_1$ parameter is 0.036 (GL) and 0.000 (CRE-GMM5) in longitudinal; 0.194 (GL) and 0.111 (CRE-GMM2) in multilevel; 0.260 (GL) and 0.198 (CRE-GMM2) in macro panel.

The advantage of our approach over GL is even more evident for large $N$ and $T$ (multilevel panels) and $T>N$ (macro panels) for both $\rho$ and $\beta_1$ parameters. Having a longer $T$ improves the possibility to compute better the initial conditions and the CRE-GMM approach is less affected by the spurious persistence (note that for long $T$ the FE tends towards the true $\rho$ parameter for $\gamma_2 = 0$). When individual heterogeneity matters and is correlated with the $x_{it}$ variable, our approach once again outperforms the GL and is in line with FE. Standard endogeneity produces bias in the FE estimates, at least for $\beta_1$, while our method is less affected. With the two sources of endogeneity, the bias of the CRE-GMM approach is in general lower than that of GL. The moment conditions based on the pre-sample individual average of $x_{it}$, if exploited, does not affect the results; the moment conditions based on the pre-sample individual average of  $y_{it-1}$ does the same only as $T$ increases and both $\gamma_1 > 0$ and $\gamma_2 > 0$.

\begin{sidewaystable}
\centering
\caption{MonteCarlo results for $\rho$, ARDL(1,0) with $\rho=0.5$ and $\beta_1=1$}\label{tab:FIXEPAM05BIGlongi_rho} 
\begin{adjustbox}{max width=\textwidth}
{\scriptsize
\begin{tabular}{ll|rrrr|r|rrrrrr}
\toprule
& & \multicolumn{11}{c}{Longitudinal panel $N=1000$, $T=10$. Parameter $\rho$} \\
\midrule
& & RE & FE& CRE1 &CRE2 &GL & CRE-GMM & CRE-GMM1 & CRE-GMM2 &CRE-GMM3 & CRE-GMM4 & CRE-GMM5 \\
\midrule
\multirow{4}*{$\gamma_1=0, \gamma_2=0, \gamma_3=[0, 0.8]$} 
& bias $\rho$ & 0.263 &-0.080 &    0.175 &    0.175 &    0.011 &    0.054 &    0.027 &    0.026  &   0.044 &    0.018 &    0.019 \\
& ese & 0.004  &   0.007   &  0.007   &  0.007 &    0.018  &   0.013  &   0.019 &    0.022  &   0.014  &   0.017  &   0.017  \\
& bias $\rho$ & 0.354  &  -0.038 &    0.161  &   0.195  &   0.050  &   0.108  &   0.070  &   0.070  &   0.081  &   0.039  &   0.042 \\
& ese & 0.005  &   0.005  &   0.009 &    0.009  &   0.031  &   0.013 &    0.023 &    0.029  &   0.014  &   0.022 &    0.020 \\
\midrule
\multirow{4}*{$\gamma_1=0, \gamma_2=0.25, \gamma_3=[0, 0.8]$}
& bias $\rho$ & 0.188  &  -0.110  &   0.108  &  0.131&     0.006     &0.052   &  0.022    & 0.021   &  0.038    & 0.017    & 0.013 \\
& ese &  0.007    & 0.007   &  0.009   &  0.007  &   0.018 &    0.014  &   0.020 &    0.023&     0.014   &  0.018   &  0.018 \\
& bias $\rho$ & 0.323  &  -0.057  &   0.129  &   0.163 &    0.048 &    0.109  &   0.069  &   0.070 &    0.080 &    0.037 &    0.040 \\
& ese &     0.006  &   0.005  &   0.009  &   0.009  &   0.031  &   0.013  &   0.022  &   0.030 &    0.014  &   0.021  &   0.019 \\
\midrule
\multirow{4}*{$\gamma_1=0, \gamma_2=0.8, \gamma_3=[0, 0.8]$} 
& bias $\rho$ &  -0.002   & -0.138 &   -0.043   & -0.018  &  -0.001 &    0.049  &   0.009   &  0.007  &   0.022 &   -0.002   &  0.000 \\
& ese &0.006  &   0.004  &   0.006  &   0.007   &  0.018  &   0.014  &   0.021  &   0.025  &   0.015  &   0.017   &  0.017 \\
& bias $\rho$ & 0.225  &   -0.084  &   0.051  &   0.082 &    0.046 &    0.117 &    0.067 &    0.067  &   0.080  &   0.030  &   0.036 \\
& ese &  0.007    & 0.004   &  0.007 &    0.008 &    0.032  &   0.014 &    0.025  &   0.034 &    0.015  &   0.024  &   0.021 \\
\midrule
\multirow{4}*{$\gamma_1=0.25, \gamma_2=0, \gamma_3=[0, 0.8]$} 
& bias $\rho$ &0.233  &  -0.080    & 0.134    & 0.138  &   0.011&    0.060  &   0.054  &   0.031   &  0.067    & 0.061 &    0.059 \\
& ese &0.004   &  0.007    & 0.006   &  0.006 &    0.019   & 0.012   &  0.014  &   0.020  &   0.013 &    0.013  &   0.013 \\
& bias $\rho$ & 0.326  &   -0.038 &   0.136&   0.168 &   0.056&   0.107&     0.109&    0.077 &    0.095&    0.104  &  0.092 \\
& ese & 0.004   &  0.005    & 0.008   &  0.008 &    0.031   &  0.011 &    0.012   &  0.025 &    0.012   &  0.012 &    0.014 \\
\midrule
\multirow{4}*{$\gamma_1=0.8, \gamma_2=0, \gamma_3=[0, 0.8]$} 
& bias $\rho$ &  0.148  &  -0.078 &   0.077   &  0.076  &   0.010    & 0.043   &  0.042   &  0.030   &  0.047  &   0.040  &   0.040 \\
& ese & 0.004   &  0.007   &  0.006   &  0.006    & 0.017   &  0.012     &0.012    & 0.016    & 0.013   & 0.013   &  0.013 \\
& bias $\rho$ & 0.238  &  -0.038 &    0.099    & 0.118   &  0.052  &   0.085     &0.087  &   0.068 &    0.091    & 0.077 &    0.073 \\
&ese&    0.003    & 0.005  &   0.006&     0.006 &    0.028 &    0.010 &    0.012 &    0.019 &    0.011 &    0.011 &    0.013 \\
\midrule
\multirow{4}*{$\gamma_1=0.25, \gamma_2=0.25, \gamma_3=[0, 0.8]$} 
& bias $\rho$         &       0.176&      -0.110&       0.081&       0.091&       0.006&       0.062&       0.054&       0.028&       0.072&       0.064&       0.062\\
& ese          &       0.005&       0.007&       0.007&       0.006&       0.018&       0.013&       0.014&       0.021&       0.013&       0.013&       0.014\\
& bias $\rho$         &       0.300&      -0.057&       0.105&       0.136&       0.052&       0.110&       0.110&       0.076&       0.097&       0.107&       0.094\\
& ese          &       0.004&       0.005&       0.008&       0.008&       0.032&       0.012&       0.012&       0.027&       0.012&       0.012&       0.015\\
\midrule
\multirow{4}*{$\gamma_1=0.25, \gamma_2=0.8, \gamma_3=[0, 0.8]$} 
& bias $\rho$         &       0.022&      -0.138&      -0.044&      -0.029&      -0.004&       0.067&       0.049&       0.013&       0.076&       0.072&       0.071\\
& ese          &       0.005&       0.004&       0.005&       0.006&       0.019&       0.014&       0.016&       0.025&       0.014&       0.015&       0.015\\
& bias $\rho$         &       0.221&      -0.084&       0.034&       0.062&       0.042&       0.119&       0.116&       0.072&       0.101&       0.117&       0.104\\
& ese          &       0.006&       0.004&       0.006&       0.007&       0.030&       0.013&       0.013&       0.029&       0.013&       0.013&       0.016\\
\midrule
\multirow{4}*{$\gamma_1=0.8, \gamma_2=0.25, \gamma_3=[0, 0.8]$} 
& bias $\rho$         &       0.080&      -0.110&       0.020&       0.018&       0.003&       0.043&       0.043&       0.027&       0.047&       0.040&       0.040\\
& ese          &       0.004&       0.007&       0.005&       0.005&       0.017&       0.013&       0.013&       0.017&       0.013&       0.014&       0.014\\
& bias $\rho$         &       0.212&      -0.057&       0.070&       0.088&       0.046&       0.088&       0.090&       0.068&       0.095&       0.078&       0.075\\
& ese          &       0.003&       0.005&       0.006&       0.006&       0.029&       0.010&       0.012&       0.020&       0.011&       0.011&       0.013\\
\bottomrule
\end{tabular}
}
\end{adjustbox}
\flushleft
{\scriptsize Note: Average bias and empirical standard error reported; 1000 replications for each setting. Estimates are implemented in Stata using \texttt{xtdpdgmm} \citep{Kripfganz2019a}, one-step cluster standard errors. Variances are defined by equations \eqref{eq:FixE} and \eqref{eq:varXfix} in Appendix \ref{sec:MCset}.}
\end{sidewaystable} 

\begin{sidewaystable}
\centering
\caption{MonteCarlo results for $\rho$, ARDL(1,0) with $\rho=0.5$ and $\beta_1=1$}\label{tab:FIXEPAM05BIGmacro_rho} 
\begin{adjustbox}{max width=\textwidth}
{\scriptsize
\begin{tabular}{ll|rrrr|r|rrrrrr}
\toprule
& & \multicolumn{11}{c}{Macro panel $N=25$, $T=40$. Parameter $\rho$} \\
\midrule
& & RE & FE& CRE1 &CRE2 &GL & CRE-GMM & CRE-GMM1 & CRE-GMM2 &CRE-GMM3 & CRE-GMM4 & CRE-GMM5 \\
\midrule
\multirow{4}*{$\gamma_1=0, \gamma_2=0, \gamma_3=[0, 0.8]$} 
& bias $\rho$         &       0.185&      -0.017&       0.118&       0.091&       0.112&       0.049&       0.049&       0.048&       0.042&       0.041&       0.041\\
& ese          &       0.015&       0.018&       0.029&       0.027&       0.028&       0.029&       0.029&       0.029&       0.027&       0.027&       0.027\\
& bias $\rho$         &       0.399&      -0.008&       0.193&       0.122&       0.337&       0.072&       0.071&       0.071&       0.057&       0.056&       0.056\\
& ese          &       0.009&       0.012&       0.037&       0.030&       0.023&       0.029&       0.029&       0.029&       0.026&       0.026&       0.026\\
\midrule
\multirow{4}*{$\gamma_1=0, \gamma_2=0.25, \gamma_3=[0, 0.8]$} 
& bias $\rho$         &       0.229&      -0.058&       0.114&       0.067&       0.146&       0.025&       0.025&       0.024&       0.010&       0.009&       0.009\\
& ese          &       0.013&       0.017&       0.034&       0.030&       0.029&       0.031&       0.031&       0.031&       0.027&       0.027&       0.028\\
& bias $\rho$         &       0.388&      -0.028&       0.180&       0.104&       0.323&       0.056&       0.055&       0.055&       0.039&       0.038&       0.038\\
& ese          &       0.009&       0.012&       0.039&       0.031&       0.022&       0.029&       0.029&       0.029&       0.026&       0.026&       0.026\\
\midrule
\multirow{4}*{$\gamma_1=0, \gamma_2=0.8, \gamma_3=[0, 0.8]$} 
& bias $\rho$         &       0.113&      -0.109&       0.041&      -0.001&       0.047&      -0.038&      -0.039&      -0.039&      -0.050&      -0.050&      -0.050\\
& ese          &       0.016&       0.012&       0.031&       0.027&       0.025&       0.025&       0.025&       0.025&       0.023&       0.023&       0.023\\
& bias $\rho$         &       0.381&      -0.061&       0.153&       0.072&       0.329&       0.026&       0.025&       0.025&       0.009&       0.009&       0.009\\
& ese          &       0.017&       0.010&       0.044&       0.032&       0.023&       0.028&       0.028&       0.027&       0.025&       0.025&       0.025\\
\midrule
\multirow{4}*{$\gamma_1=0.25, \gamma_2=0, \gamma_3=[0, 0.8]$} 
& bias $\rho$         &       0.246&      -0.017&       0.110&       0.097&       0.186&       0.055&       0.054&       0.054&       0.063&       0.060&       0.057\\
& ese          &       0.012&       0.018&       0.028&       0.027&       0.025&       0.030&       0.030&       0.030&       0.029&       0.030&       0.030\\
& bias $\rho$         &       0.355&      -0.008&       0.155&       0.115&       0.323&       0.069&       0.068&       0.068&       0.066&       0.068&       0.068\\
& ese          &       0.009&       0.012&       0.032&       0.028&       0.019&       0.027&       0.027&       0.027&       0.026&       0.027&       0.027\\
\midrule
\multirow{4}*{$\gamma_1=0.8, \gamma_2=0, \gamma_3=[0, 0.8]$} 
& bias $\rho$         &       0.140&      -0.017&       0.059&       0.057&       0.105&       0.031&       0.030&       0.030&       0.045&       0.038&       0.031\\
& ese          &       0.012&       0.019&       0.020&       0.020&       0.022&       0.026&       0.026&       0.026&       0.025&       0.026&       0.027\\
& bias $\rho$         &       0.249&      -0.008&       0.100&       0.089&       0.236&       0.053&       0.053&       0.052&       0.066&       0.058&       0.051\\
& ese          &       0.008&       0.013&       0.022&       0.021&       0.016&       0.023&       0.023&       0.023&       0.022&       0.023&       0.023\\
\midrule
\multirow{4}*{$\gamma_1=0.25, \gamma_2=0.25, \gamma_3=[0, 0.8]$} 
& bias $\rho$         &       0.139&      -0.058&       0.053&       0.041&       0.083&       0.010&       0.010&       0.009&       0.015&       0.013&       0.012\\
& ese          &       0.013&       0.017&       0.025&       0.024&       0.026&       0.027&       0.027&       0.027&       0.026&       0.026&       0.026\\
& bias $\rho$         &       0.345&      -0.028&       0.137&       0.096&       0.314&       0.051&       0.050&       0.050&       0.048&       0.050&       0.049\\
& ese          &       0.009&       0.012&       0.031&       0.027&       0.018&       0.026&       0.026&       0.026&       0.025&       0.026&       0.026\\
\midrule
\multirow{4}*{$\gamma_1=0.25, \gamma_2=0.8, \gamma_3=[0, 0.8]$} 
& bias $\rho$         &       0.103&      -0.108&       0.005&      -0.011&       0.047&      -0.043&      -0.044&      -0.044&      -0.031&      -0.034&      -0.036\\
& ese          &       0.012&       0.012&       0.024&       0.023&       0.022&       0.023&       0.023&       0.023&       0.023&       0.023&       0.023\\
& bias $\rho$         &       0.289&      -0.060&       0.098&       0.059&       0.250&       0.017&       0.017&       0.016&       0.018&       0.019&       0.017\\
& ese          &       0.015&       0.010&       0.033&       0.027&       0.022&       0.025&       0.025&       0.025&       0.024&       0.025&       0.025\\
\midrule
\multirow{4}*{$\gamma_1=0.8, \gamma_2=0.25, \gamma_3=[0, 0.8]$} 
& bias $\rho$         &       0.079&      -0.058&       0.008&       0.006&       0.057&      -0.009&      -0.009&      -0.010&       0.002&      -0.003&      -0.011\\
& ese          &       0.012&       0.017&       0.018&       0.018&       0.022&       0.024&       0.024&       0.024&       0.023&       0.024&       0.025\\
& bias $\rho$         &       0.221&      -0.028&       0.078&       0.067&       0.206&       0.034&       0.034&       0.033&       0.048&       0.040&       0.033\\
& ese          &       0.008&       0.012&       0.021&       0.021&       0.017&       0.022&       0.022&       0.022&       0.022&       0.023&       0.023\\
\bottomrule
\end{tabular}
}
\end{adjustbox}
\flushleft
{\scriptsize Note: Average bias and empirical standard error reported; 1000 replications for each setting. Estimates are implemented in Stata using \texttt{xtdpdgmm} \citep{Kripfganz2019a}, one-step cluster standard errors. Variances are defined by equations \eqref{eq:FixE} and \eqref{eq:varXfix} in Appendix \ref{sec:MCset}.}
\end{sidewaystable} 

\begin{sidewaystable}
\centering
\caption{MonteCarlo results for $\rho$, ARDL(1,0) with $\rho=0.5$ and $\beta_1=1$}\label{tab:FIXEPAM05BIGmulti_rho} 
\begin{adjustbox}{max width=\textwidth}
{\scriptsize
\begin{tabular}{ll|rrrr|r|rrrrrr}
\toprule
& & \multicolumn{11}{c}{Multilevel panel $N=100$, $T=20$. Parameter $\rho$} \\
\midrule
& & RE & FE& CRE1 &CRE2 &GL & CRE-GMM & CRE-GMM1 & CRE-GMM2 &CRE-GMM3 & CRE-GMM4 & CRE-GMM5 \\
\midrule
\multirow{4}*{$\gamma_1=0, \gamma_2=0, \gamma_3=[0, 0.8]$} 
& bias $\rho$         &       0.257&      -0.035&       0.163&       0.147&       0.075&       0.039&       0.038&       0.038&       0.035&       0.034&       0.033\\
& ese          &       0.009&       0.014&       0.018&       0.018&       0.030&       0.025&       0.025&       0.027&       0.025&       0.025&       0.025\\
& bias $\rho$         &       0.395&      -0.016&       0.217&       0.186&       0.231&       0.067&       0.065&       0.066&       0.051&       0.050&       0.049\\
& ese          &       0.010&       0.010&       0.023&       0.019&       0.036&       0.023&       0.023&       0.026&       0.024&       0.024&       0.024\\
\midrule
\multirow{4}*{$\gamma_1=0, \gamma_2=0.25, \gamma_3=[0, 0.8]$} 
& bias $\rho$         &       0.207&      -0.073&       0.123&       0.104&       0.053&       0.021&       0.020&       0.020&       0.011&       0.010&       0.009\\
& ese          &       0.009&       0.013&       0.018&       0.019&       0.029&       0.025&       0.025&       0.026&       0.025&       0.026&       0.026\\
& bias $\rho$         &       0.357&      -0.036&       0.186&       0.165&       0.195&       0.056&       0.054&       0.055&       0.039&       0.038&       0.037\\
& ese          &       0.015&       0.010&       0.026&       0.020&       0.038&       0.024&       0.024&       0.026&       0.024&       0.024&       0.024\\
\midrule
\multirow{4}*{$\gamma_1=0, \gamma_2=0.8, \gamma_3=[0, 0.8]$} 
& bias $\rho$         &       0.054&      -0.117&       0.008&       0.018&      -0.001&      -0.021&      -0.023&      -0.024&      -0.031&      -0.032&      -0.032\\
& ese          &       0.020&       0.009&       0.020&       0.019&       0.027&       0.023&       0.023&       0.025&       0.023&       0.023&       0.023\\
& bias $\rho$         &       0.316&      -0.067&       0.116&       0.110&       0.220&       0.044&       0.042&       0.043&       0.023&       0.022&       0.022\\
& ese          &       0.019&       0.008&       0.026&       0.025&       0.039&       0.025&       0.025&       0.029&       0.025&       0.026&       0.026\\
\midrule
\multirow{4}*{$\gamma_1=0.25, \gamma_2=0, \gamma_3=[0, 0.8]$} 
& bias $\rho$         &       0.205&      -0.035&       0.117&       0.114&       0.060&       0.035&       0.034&       0.032&       0.059&       0.048&       0.046\\
& ese          &       0.009&       0.015&       0.016&       0.016&       0.028&       0.024&       0.024&       0.025&       0.024&       0.025&       0.024\\
& bias $\rho$         &       0.342&      -0.017&       0.173&       0.162&       0.215&       0.062&       0.061&       0.059&       0.077&       0.067&       0.066\\
& ese          &       0.008&       0.010&       0.019&       0.016&       0.033&       0.022&       0.022&       0.024&       0.022&       0.023&       0.023\\
\midrule
\multirow{4}*{$\gamma_1=0.8, \gamma_2=0, \gamma_3=[0, 0.8]$} 
& bias $\rho$         &       0.145&      -0.035&       0.071&       0.070&       0.060&       0.026&       0.025&       0.024&       0.049&       0.029&       0.025\\
& ese          &       0.009&       0.015&       0.013&       0.013&       0.024&       0.023&       0.023&       0.023&       0.023&       0.024&       0.024\\
& bias $\rho$         &       0.250&      -0.016&       0.117&       0.117&       0.197&       0.052&       0.050&       0.048&       0.086&       0.051&       0.048\\
& ese          &       0.006&       0.010&       0.013&       0.013&       0.025&       0.019&       0.020&       0.020&       0.019&       0.021&       0.021\\
\midrule
\multirow{4}*{$\gamma_1=0.25, \gamma_2=0.25, \gamma_3=[0, 0.8]$} 
& bias $\rho$         &       0.167&      -0.073&       0.076&       0.073&       0.044&       0.017&       0.016&       0.015&       0.042&       0.028&       0.026\\
& ese          &       0.009&       0.013&       0.015&       0.015&       0.029&       0.023&       0.023&       0.025&       0.024&       0.024&       0.024\\
& bias $\rho$         &       0.338&      -0.036&       0.152&       0.147&       0.235&       0.058&       0.058&       0.056&       0.073&       0.062&       0.060\\
& ese          &       0.009&       0.010&       0.021&       0.017&       0.035&       0.022&       0.023&       0.025&       0.023&       0.024&       0.024\\
\midrule
\multirow{4}*{$\gamma_1=0.25, \gamma_2=0.8, \gamma_3=[0, 0.8]$} 
& bias $\rho$         &       0.066&      -0.117&      -0.009&       0.003&      -0.002&      -0.019&      -0.020&      -0.024&       0.015&      -0.000&      -0.004\\
& ese          &       0.015&       0.009&       0.016&       0.015&       0.027&       0.020&       0.020&       0.023&       0.021&       0.023&       0.022\\
& bias $\rho$         &       0.259&      -0.067&       0.076&       0.087&       0.182&       0.039&       0.039&       0.036&       0.056&       0.045&       0.043\\
& ese          &       0.014&       0.008&       0.020&       0.021&       0.037&       0.022&       0.022&       0.025&       0.023&       0.024&       0.024\\
\midrule
\multirow{4}*{$\gamma_1=0.8, \gamma_2=0.25, \gamma_3=[0, 0.8]$} 
& bias $\rho$         &       0.083&      -0.073&       0.017&       0.017&       0.033&       0.004&       0.004&       0.003&       0.026&       0.005&       0.001\\
& ese          &       0.008&       0.013&       0.012&       0.012&       0.025&       0.022&       0.022&       0.022&       0.022&       0.024&       0.023\\
& bias $\rho$         &       0.228&      -0.036&       0.093&       0.095&       0.186&       0.042&       0.040&       0.039&       0.079&       0.041&       0.037\\
& ese          &       0.006&       0.010&       0.014&       0.013&       0.026&       0.019&       0.020&       0.020&       0.019&       0.021&       0.021\\
\bottomrule
\end{tabular}
}
\end{adjustbox}
\flushleft
{\scriptsize Note: Average bias and empirical standard error reported; 1000 replications for each setting. Estimates are implemented in Stata using \texttt{xtdpdgmm} \citep{Kripfganz2019a}, one-step cluster standard errors. Variances are defined by equations \eqref{eq:FixE} and \eqref{eq:varXfix} in Appendix \ref{sec:MCset}.}
\end{sidewaystable} 

\begin{sidewaystable}
\centering
\caption{MonteCarlo results for $\beta$, ARDL(1,0) with $\rho=0.5$ and $\beta_1=1$}\label{tab:FIXEPAM05BIGlongi_beta} 
\begin{adjustbox}{max width=\textwidth}
{\scriptsize
\begin{tabular}{ll|rrrr|r|rrrrrr}
\toprule
& & \multicolumn{11}{c}{Longitudinal panel $N=1000$, $T=10$. Parameter $\beta_1$} \\
\midrule
& & RE & FE& CRE1 &CRE2 &GL & CRE-GMM & CRE-GMM1 & CRE-GMM2 &CRE-GMM3 & CRE-GMM4 & CRE-GMM5 \\
\midrule
\multirow{4}*{$\gamma_1=0, \gamma_2=0, \gamma_3=[0, 0.8]$} 
& bias $\beta$        &      -0.174&       0.017&      -0.122&      -0.112&      -0.019&      -0.069&      -0.036&      -0.035&      -0.113&      -0.048&      -0.040\\
& ese          &       0.014&       0.012&       0.013&       0.013&       0.050&       0.044&       0.047&       0.052&       0.041&       0.045&       0.044\\
& bias $\beta$        &      -0.167&       0.008&      -0.083&      -0.108&      -0.097&      -0.145&      -0.096&      -0.090&      -0.208&      -0.103&      -0.088\\
& ese          &       0.010&       0.007&       0.010&       0.011&       0.068&       0.038&       0.048&       0.050&       0.038&       0.047&       0.042\\
\midrule
\multirow{4}*{$\gamma_1=0, \gamma_2=0.25, \gamma_3=[0, 0.8]$}
& bias $\beta$        &       0.046&       0.231&       0.096&       0.078&      -0.010&      -0.082&      -0.031&      -0.028&      -0.111&      -0.033&      -0.029\\
& ese          &       0.015&       0.011&       0.015&       0.013&       0.055&       0.049&       0.054&       0.056&       0.047&       0.052&       0.051\\
& bias $\beta$        &      -0.084&       0.088&       0.002&      -0.021&      -0.098&      -0.162&      -0.102&      -0.095&      -0.217&      -0.100&      -0.088\\
& ese          &       0.011&       0.007&       0.010&       0.011&       0.070&       0.040&       0.050&       0.052&       0.040&       0.051&       0.045\\
\midrule
\multirow{4}*{$\gamma_1=0, \gamma_2=0.8, \gamma_3=[0, 0.8]$} 
& bias $\beta$        &       0.396&       0.465&       0.416&       0.401&       0.016&      -0.089&      -0.002&       0.000&      -0.072&       0.014&       0.011\\
& ese          &       0.009&       0.007&       0.008&       0.009&       0.053&       0.051&       0.056&       0.056&       0.049&       0.052&       0.052\\
& bias $\beta$        &       0.091&       0.223&       0.163&       0.147&      -0.102&      -0.201&      -0.114&      -0.108&      -0.238&      -0.088&      -0.085\\
& ese          &       0.009&       0.006&       0.008&       0.009&       0.072&       0.046&       0.059&       0.060&       0.048&       0.057&       0.052\\
\midrule
\multirow{4}*{$\gamma_1=0.25, \gamma_2=0, \gamma_3=[0, 0.8]$} 
& bias $\beta$        &      -0.034&       0.017&      -0.014&      -0.015&      -0.009&      -0.025&       0.000&      -0.019&       0.005&      -0.033&      -0.035\\
& ese          &       0.013&       0.012&       0.012&       0.012&       0.050&       0.042&       0.048&       0.051&       0.040&       0.043&       0.043\\
& bias $\beta$        &      -0.078&       0.008&      -0.030&      -0.051&      -0.087&      -0.105&      -0.074&      -0.072&      -0.114&      -0.102&      -0.100\\
& ese          &       0.010&       0.007&       0.008&       0.009&       0.068&       0.035&       0.042&       0.045&       0.032&       0.033&       0.032\\
\midrule
\multirow{4}*{$\gamma_1=0.8, \gamma_2=0, \gamma_3=[0, 0.8]$} 
& bias $\beta$        &       0.060&       0.017&       0.042&       0.042&       0.019&       0.014&       0.013&       0.005&       0.079&      -0.003&      -0.002\\
& ese          &       0.012&       0.012&       0.011&       0.011&       0.047&       0.039&       0.043&       0.044&       0.038&       0.041&       0.042\\
& bias $\beta$        &       0.013&       0.008&       0.015&       0.006&      -0.040&      -0.048&      -0.042&      -0.038&      -0.036&      -0.049&      -0.053\\
& ese          &       0.009&       0.007&       0.007&       0.008&       0.056&       0.029&       0.032&       0.034&       0.030&       0.030&       0.030\\
\midrule
\multirow{4}*{$\gamma_1=0.25, \gamma_2=0.25, \gamma_3=[0, 0.8]$} 
& bias $\beta$        &       0.179&       0.231&       0.200&       0.193&       0.003&      -0.041&      -0.008&      -0.016&      -0.017&      -0.052&      -0.053\\
& ese          &       0.013&       0.011&       0.012&       0.012&       0.053&       0.046&       0.051&       0.053&       0.045&       0.047&       0.047\\
& bias $\beta$        &       0.008&       0.088&       0.055&       0.036&      -0.089&      -0.121&      -0.088&      -0.079&      -0.127&      -0.119&      -0.114\\
& ese          &       0.010&       0.007&       0.008&       0.009&       0.070&       0.037&       0.043&       0.046&       0.034&       0.035&       0.035\\
\midrule
\multirow{4}*{$\gamma_1=0.25, \gamma_2=0.8, \gamma_3=[0, 0.8]$} 
& bias $\beta$        &       0.463&       0.464&       0.468&       0.463&       0.032&      -0.074&      -0.018&       0.007&      -0.086&      -0.089&      -0.091\\
& ese          &       0.008&       0.007&       0.008&       0.008&       0.053&       0.050&       0.054&       0.055&       0.052&       0.052&       0.052\\
& bias $\beta$        &       0.171&       0.223&       0.205&       0.194&      -0.080&      -0.160&      -0.116&      -0.089&      -0.157&      -0.157&      -0.150\\
& ese          &       0.008&       0.006&       0.007&       0.007&       0.069&       0.042&       0.048&       0.054&       0.040&       0.041&       0.041\\
\midrule
\multirow{4}*{$\gamma_1=0.8, \gamma_2=0.25, \gamma_3=[0, 0.8]$} 
& bias $\beta$        &       0.273&       0.231&       0.254&       0.254&       0.036&       0.012&       0.016&       0.011&       0.085&      -0.002&       0.000\\
& ese          &       0.011&       0.011&       0.011&       0.011&       0.051&       0.043&       0.046&       0.048&       0.043&       0.046&       0.047\\
& bias $\beta$        &       0.103&       0.088&       0.100&       0.093&      -0.035&      -0.058&      -0.050&      -0.041&      -0.046&      -0.056&      -0.060\\
& ese          &       0.008&       0.007&       0.007&       0.008&       0.062&       0.032&       0.033&       0.037&       0.033&       0.032&       0.032\\
\bottomrule
\end{tabular}
}
\end{adjustbox}
\flushleft
{\scriptsize Note: Average bias and empirical standard error reported; 1000 replications for each setting. Estimates are implemented in Stata using \texttt{xtdpdgmm} \citep{Kripfganz2019a}, one-step cluster standard errors. Variances are defined by equations \eqref{eq:FixE} and \eqref{eq:varXfix} in Appendix \ref{sec:MCset}.}
\end{sidewaystable} 

\begin{sidewaystable}
\centering
\caption{MonteCarlo results for $\beta$, ARDL(1,0) with $\rho=0.5$ and $\beta_1=1$}\label{tab:FIXEPAM05BIGmacro_beta} 
\begin{adjustbox}{max width=\textwidth}
{\scriptsize
\begin{tabular}{ll|rrrr|r|rrrrrr}
\toprule
& & \multicolumn{11}{c}{Macro panel $N=25$, $T=40$. Parameter $\beta_1$} \\
\midrule
& & RE & FE& CRE1 &CRE2 &GL & CRE-GMM & CRE-GMM1 & CRE-GMM2 &CRE-GMM3 & CRE-GMM4 & CRE-GMM5 \\
\midrule
\multirow{4}*{$\gamma_1=0, \gamma_2=0, \gamma_3=[0, 0.8]$} 
& bias $\beta$        &      -0.121&       0.010&      -0.078&      -0.058&      -0.121&      -0.057&      -0.056&      -0.055&      -0.065&      -0.064&      -0.063\\
& ese          &       0.038&       0.031&       0.038&       0.037&       0.069&       0.064&       0.064&       0.065&       0.065&       0.065&       0.066\\
& bias $\beta$        &      -0.262&       0.004&      -0.128&      -0.078&      -0.368&      -0.083&      -0.082&      -0.081&      -0.087&      -0.086&      -0.084\\
& ese          &       0.035&       0.019&       0.036&       0.029&       0.061&       0.049&       0.049&       0.049&       0.051&       0.051&       0.051\\
\midrule
\multirow{4}*{$\gamma_1=0, \gamma_2=0.25, \gamma_3=[0, 0.8]$} 
& bias $\beta$        &      -0.000&       0.219&       0.087&       0.127&      -0.068&       0.086&       0.086&       0.087&       0.080&       0.081&       0.082\\
& ese          &       0.042&       0.030&       0.043&       0.039&       0.077&       0.069&       0.070&       0.070&       0.070&       0.070&       0.070\\
& bias $\beta$        &      -0.206&       0.085&      -0.061&      -0.005&      -0.336&      -0.024&      -0.023&      -0.022&      -0.026&      -0.025&      -0.024\\
& ese          &       0.037&       0.019&       0.039&       0.031&       0.062&       0.051&       0.051&       0.051&       0.053&       0.053&       0.053\\
\midrule
\multirow{4}*{$\gamma_1=0, \gamma_2=0.8, \gamma_3=[0, 0.8]$} 
& bias $\beta$        &       0.265&       0.460&       0.329&       0.371&       0.175&       0.305&       0.306&       0.306&       0.310&       0.310&       0.310\\
& ese          &       0.038&       0.022&       0.040&       0.034&       0.065&       0.057&       0.057&       0.057&       0.057&       0.057&       0.057\\
& bias $\beta$        &      -0.114&       0.220&       0.058&       0.123&      -0.318&       0.079&       0.080&       0.081&       0.082&       0.083&       0.084\\
& ese          &       0.041&       0.016&       0.043&       0.031&       0.068&       0.052&       0.051&       0.051&       0.052&       0.051&       0.051\\
\midrule
\multirow{4}*{$\gamma_1=0.25, \gamma_2=0, \gamma_3=[0, 0.8]$} 
& bias $\beta$        &      -0.033&       0.010&      -0.013&      -0.011&      -0.052&      -0.017&      -0.016&      -0.016&      -0.013&      -0.013&      -0.014\\
& ese          &       0.036&       0.031&       0.033&       0.033&       0.073&       0.063&       0.063&       0.063&       0.062&       0.062&       0.063\\
& bias $\beta$        &      -0.144&       0.005&      -0.062&      -0.045&      -0.242&      -0.053&      -0.053&      -0.052&      -0.054&      -0.053&      -0.052\\
& ese          &       0.031&       0.019&       0.028&       0.025&       0.057&       0.044&       0.045&       0.045&       0.044&       0.044&       0.044\\
\midrule
\multirow{4}*{$\gamma_1=0.8, \gamma_2=0, \gamma_3=[0, 0.8]$} 
& bias $\beta$        &       0.061&       0.010&       0.029&       0.029&       0.081&       0.025&       0.025&       0.026&       0.035&       0.031&       0.026\\
& ese          &       0.034&       0.033&       0.033&       0.033&       0.062&       0.059&       0.060&       0.060&       0.061&       0.061&       0.061\\
& bias $\beta$        &      -0.014&       0.005&      -0.004&      -0.004&      -0.050&      -0.010&      -0.010&      -0.010&      -0.012&      -0.011&      -0.010\\
& ese          &       0.026&       0.020&       0.022&       0.022&       0.052&       0.040&       0.040&       0.041&       0.040&       0.041&       0.041\\
\midrule
\multirow{4}*{$\gamma_1=0.25, \gamma_2=0.25, \gamma_3=[0, 0.8]$} 
& bias $\beta$        &       0.180&       0.220&       0.196&       0.198&       0.122&       0.150&       0.151&       0.151&       0.154&       0.154&       0.155\\
& ese          &       0.036&       0.030&       0.034&       0.033&       0.071&       0.063&       0.064&       0.064&       0.062&       0.062&       0.062\\
& bias $\beta$        &      -0.075&       0.085&       0.014&       0.032&      -0.200&       0.010&       0.011&       0.012&       0.011&       0.011&       0.012\\
& ese          &       0.033&       0.020&       0.030&       0.026&       0.062&       0.046&       0.046&       0.046&       0.046&       0.046&       0.047\\
\midrule
\multirow{4}*{$\gamma_1=0.25, \gamma_2=0.8, \gamma_3=[0, 0.8]$} 
& bias $\beta$        &       0.408&       0.461&       0.433&       0.437&       0.307&       0.366&       0.366&       0.366&       0.363&       0.364&       0.365\\
& ese          &       0.032&       0.022&       0.028&       0.027&       0.060&       0.050&       0.050&       0.050&       0.049&       0.049&       0.049\\
& bias $\beta$        &       0.059&       0.221&       0.148&       0.166&      -0.094&       0.122&       0.123&       0.123&       0.122&       0.122&       0.123\\
& ese          &       0.034&       0.017&       0.030&       0.026&       0.063&       0.045&       0.045&       0.045&       0.044&       0.044&       0.044\\
\midrule
\multirow{4}*{$\gamma_1=0.8, \gamma_2=0.25, \gamma_3=[0, 0.8]$} 
& bias $\beta$        &       0.275&       0.220&       0.241&       0.241&       0.260&       0.197&       0.198&       0.198&       0.207&       0.205&       0.201\\
& ese          &       0.034&       0.033&       0.033&       0.033&       0.064&       0.061&       0.061&       0.061&       0.061&       0.062&       0.061\\
& bias $\beta$        &       0.069&       0.085&       0.078&       0.078&       0.017&       0.056&       0.056&       0.057&       0.054&       0.055&       0.056\\
& ese          &       0.026&       0.020&       0.023&       0.022&       0.052&       0.041&       0.041&       0.041&       0.041&       0.041&       0.041\\
\bottomrule
\end{tabular}
}
\end{adjustbox}
\flushleft
{\scriptsize Note: Average bias and empirical standard error reported; 1000 replications for each setting. Estimates are implemented in Stata using \texttt{xtdpdgmm} \citep{Kripfganz2019a}, one-step cluster standard errors. Variances are defined by equations \eqref{eq:FixE} and \eqref{eq:varXfix} in Appendix \ref{sec:MCset}.}
\end{sidewaystable} 

\begin{sidewaystable}
\centering
\caption{MonteCarlo results for $\beta$, ARDL(1,0) with $\rho=0.5$ and $\beta_1=1$}\label{tab:FIXEPAM05BIGmulti_beta} 
\begin{adjustbox}{max width=\textwidth}
{\scriptsize
\begin{tabular}{ll|rrrr|r|rrrrrr}
\toprule
& & \multicolumn{11}{c}{Multilevel panel $N=100$, $T=20$. Parameter $\beta_1$} \\
\midrule
& & RE & FE& CRE1 &CRE2 &GL & CRE-GMM & CRE-GMM1 & CRE-GMM2 &CRE-GMM3 & CRE-GMM4 & CRE-GMM5 \\
\midrule
\multirow{4}*{$\gamma_1=0, \gamma_2=0, \gamma_3=[0, 0.8]$} 
& bias $\beta$        &      -0.171&       0.015&      -0.111&      -0.095&      -0.119&      -0.064&      -0.062&      -0.060&      -0.075&      -0.073&      -0.070\\
& ese          &       0.030&       0.024&       0.031&       0.030&       0.079&       0.070&       0.070&       0.070&       0.072&       0.072&       0.072\\
& bias $\beta$        &      -0.255&       0.007&      -0.143&      -0.120&      -0.367&      -0.106&      -0.103&      -0.101&      -0.107&      -0.104&      -0.101\\
& ese          &       0.027&       0.015&       0.027&       0.023&       0.084&       0.055&       0.055&       0.055&       0.056&       0.056&       0.055\\
\midrule
\multirow{4}*{$\gamma_1=0, \gamma_2=0.25, \gamma_3=[0, 0.8]$} 
& bias $\beta$        &       0.015&       0.226&       0.078&       0.098&      -0.051&       0.010&       0.012&       0.014&       0.009&       0.011&       0.013\\
& ese          &       0.031&       0.023&       0.031&       0.030&       0.086&       0.075&       0.075&       0.075&       0.077&       0.077&       0.077\\
& bias $\beta$        &      -0.173&       0.087&      -0.062&      -0.047&      -0.322&      -0.078&      -0.074&      -0.072&      -0.074&      -0.071&      -0.068\\
& ese          &       0.030&       0.015&       0.029&       0.025&       0.091&       0.059&       0.059&       0.059&       0.060&       0.060&       0.060\\
\midrule
\multirow{4}*{$\gamma_1=0, \gamma_2=0.8, \gamma_3=[0, 0.8]$} 
& bias $\beta$        &       0.329&       0.463&       0.365&       0.356&       0.106&       0.156&       0.159&       0.160&       0.171&       0.173&       0.174\\
& ese          &       0.027&       0.016&       0.026&       0.027&       0.081&       0.070&       0.070&       0.070&       0.069&       0.069&       0.069\\
& bias $\beta$        &      -0.030&       0.222&       0.100&       0.102&      -0.382&      -0.035&      -0.032&      -0.030&      -0.018&      -0.016&      -0.014\\
& ese          &       0.028&       0.012&       0.026&       0.027&       0.102&       0.065&       0.065&       0.065&       0.065&       0.065&       0.065\\
\midrule
\multirow{4}*{$\gamma_1=0.25, \gamma_2=0, \gamma_3=[0, 0.8]$} 
& bias $\beta$        &      -0.029&       0.015&      -0.014&      -0.014&      -0.027&      -0.016&      -0.015&      -0.014&       0.003&      -0.007&      -0.009\\
& ese          &       0.027&       0.024&       0.026&       0.026&       0.075&       0.068&       0.069&       0.069&       0.069&       0.070&       0.070\\
& bias $\beta$        &      -0.136&       0.007&      -0.068&      -0.064&      -0.259&      -0.068&      -0.066&      -0.064&      -0.074&      -0.070&      -0.068\\
& ese          &       0.023&       0.015&       0.021&       0.020&       0.078&       0.051&       0.051&       0.051&       0.051&       0.051&       0.051\\
\midrule
\multirow{4}*{$\gamma_1=0.8, \gamma_2=0, \gamma_3=[0, 0.8]$} 
& bias $\beta$        &       0.060&       0.015&       0.034&       0.034&       0.078&       0.016&       0.016&       0.016&       0.057&       0.025&       0.018\\
& ese          &       0.025&       0.024&       0.024&       0.024&       0.070&       0.066&       0.067&       0.067&       0.066&       0.069&       0.069\\
& bias $\beta$        &      -0.015&       0.007&      -0.003&      -0.005&      -0.098&      -0.026&      -0.025&      -0.025&      -0.034&      -0.029&      -0.026\\
& ese          &       0.019&       0.015&       0.017&       0.017&       0.066&       0.048&       0.048&       0.048&       0.049&       0.048&       0.048\\
\midrule
\multirow{4}*{$\gamma_1=0.25, \gamma_2=0.25, \gamma_3=[0, 0.8]$} 
& bias $\beta$        &       0.175&       0.226&       0.193&       0.194&       0.048&       0.068&       0.069&       0.069&       0.080&       0.076&       0.073\\
& ese          &       0.027&       0.024&       0.026&       0.026&       0.082&       0.072&       0.073&       0.074&       0.075&       0.075&       0.075\\
& bias $\beta$        &      -0.065&       0.087&       0.011&       0.010&      -0.289&      -0.048&      -0.046&      -0.044&      -0.052&      -0.047&      -0.045\\
& ese          &       0.025&       0.015&       0.023&       0.021&       0.089&       0.056&       0.056&       0.056&       0.056&       0.056&       0.056\\
\midrule
\multirow{4}*{$\gamma_1=0.25, \gamma_2=0.8, \gamma_3=[0, 0.8]$} 
& bias $\beta$        &       0.427&       0.463&       0.444&       0.437&       0.180&       0.207&       0.208&       0.211&       0.182&       0.194&       0.196\\
& ese          &       0.021&       0.015&       0.018&       0.019&       0.077&       0.064&       0.064&       0.064&       0.066&       0.066&       0.065\\
& bias $\beta$        &       0.099&       0.222&       0.169&       0.160&      -0.222&       0.012&       0.014&       0.017&       0.002&       0.011&       0.013\\
& ese          &       0.023&       0.012&       0.018&       0.020&       0.093&       0.057&       0.057&       0.057&       0.056&       0.057&       0.056\\
\midrule
\multirow{4}*{$\gamma_1=0.8, \gamma_2=0.25, \gamma_3=[0, 0.8]$} 
& bias $\beta$        &       0.275&       0.226&       0.247&       0.247&       0.194&       0.112&       0.112&       0.111&       0.154&       0.129&       0.120\\
& ese          &       0.024&       0.023&       0.023&       0.023&       0.076&       0.069&       0.070&       0.070&       0.070&       0.070&       0.070\\
& bias $\beta$        &       0.070&       0.087&       0.080&       0.078&      -0.068&       0.009&       0.009&       0.010&      -0.003&       0.007&       0.011\\
& ese          &       0.019&       0.015&       0.017&       0.017&       0.069&       0.049&       0.049&       0.050&       0.051&       0.050&       0.050\\
\bottomrule
\end{tabular}
}
\end{adjustbox}
\flushleft
{\scriptsize Note: Average bias and empirical standard error reported; 1000 replications for each setting. Estimates are implemented in Stata using \texttt{xtdpdgmm} \citep{Kripfganz2019a}, one-step cluster standard errors. Variances are defined by equations \eqref{eq:FixE} and \eqref{eq:varXfix} in Appendix \ref{sec:MCset}.}
\end{sidewaystable} 

\begin{figure}[!]
\centering
\includegraphics[scale=0.16]{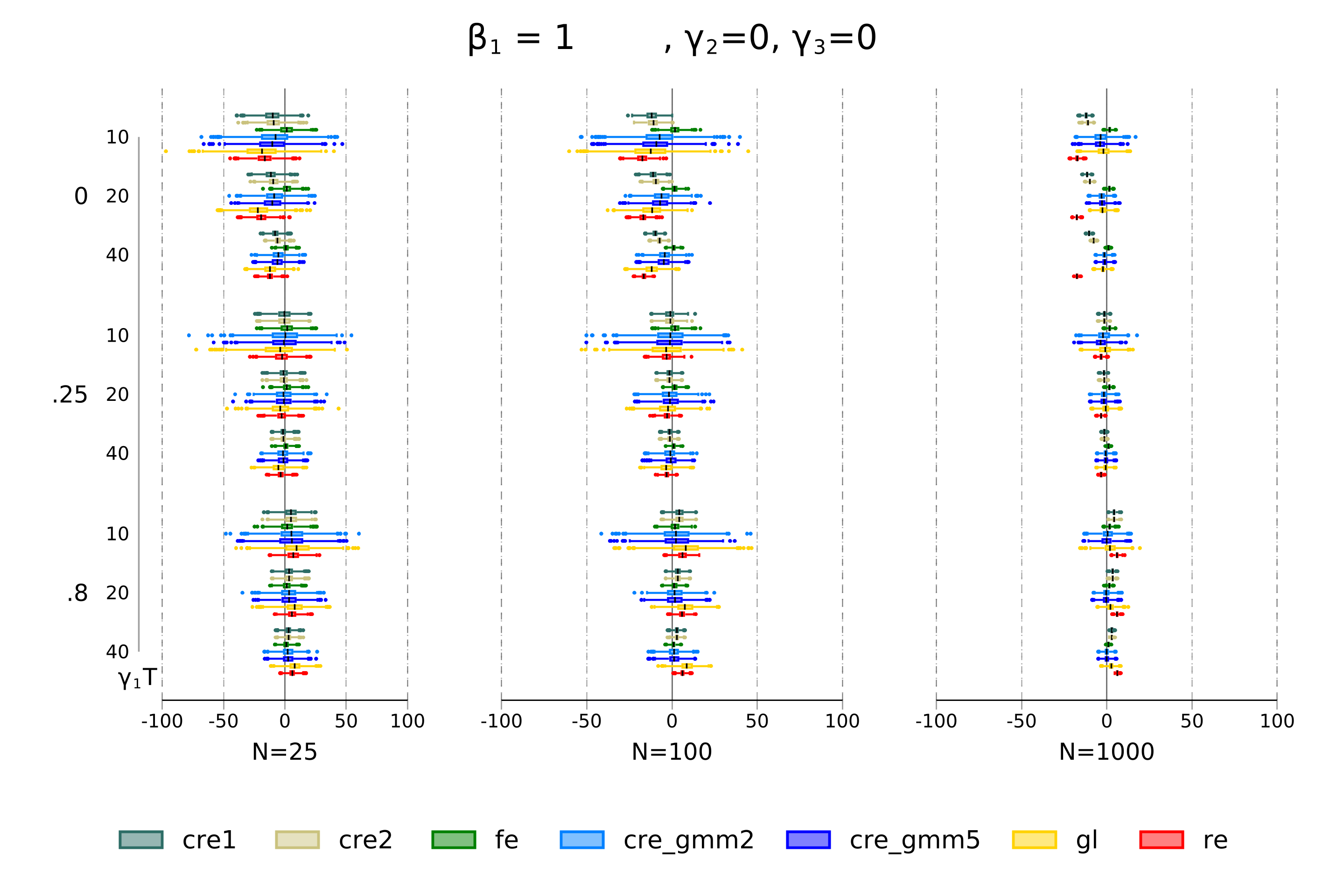}
\caption{\small{Boxplot - No standard endogeneity, PAM ($\rho = 0.5, \beta_1=1$) \\ Boxplot for bias of $\beta_1$ with $\gamma_2=\gamma_3=0$. \\ $\gamma_1 = [0, 0.25, 0.8]$ and $T=10, 20, 40$ are on the horizontal axis.}}
\label{fig:FIXEPAM05BIGbeta}
\end{figure}

\begin{figure}[!]
\centering
\includegraphics[scale=0.14]{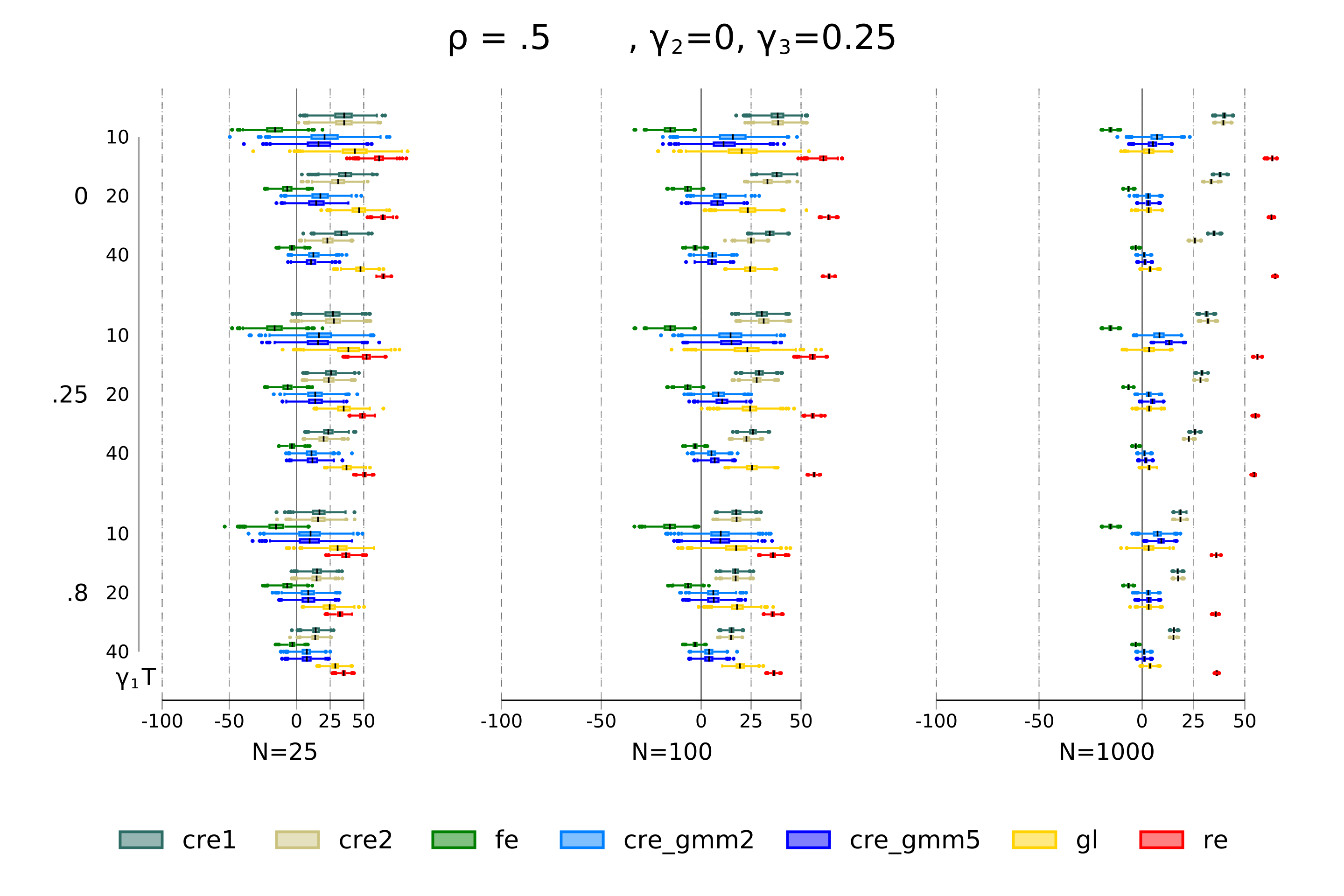}
\caption{\small{Boxplot - No standard endogeneity, PAM ($\rho=0.5, \beta_1= 1$) \\ Boxplot for bias of $\rho$ with $\gamma_2=0, \gamma_3=0.25$ and higher variation in $\mu_i$ than error component $\epsilon_{it}$, $\sigma^2_{\mu}/\sigma^2_{\epsilon}>1$. \\ $\gamma_1 = [0, 0.25, 0.8]$ and $T=10, 20, 40$ are on the horizontal axis.}}
\label{fig:FIXEPAM05BIGrhog3}
\end{figure}

\begin{figure}[!]
\centering
\includegraphics[scale=0.14]{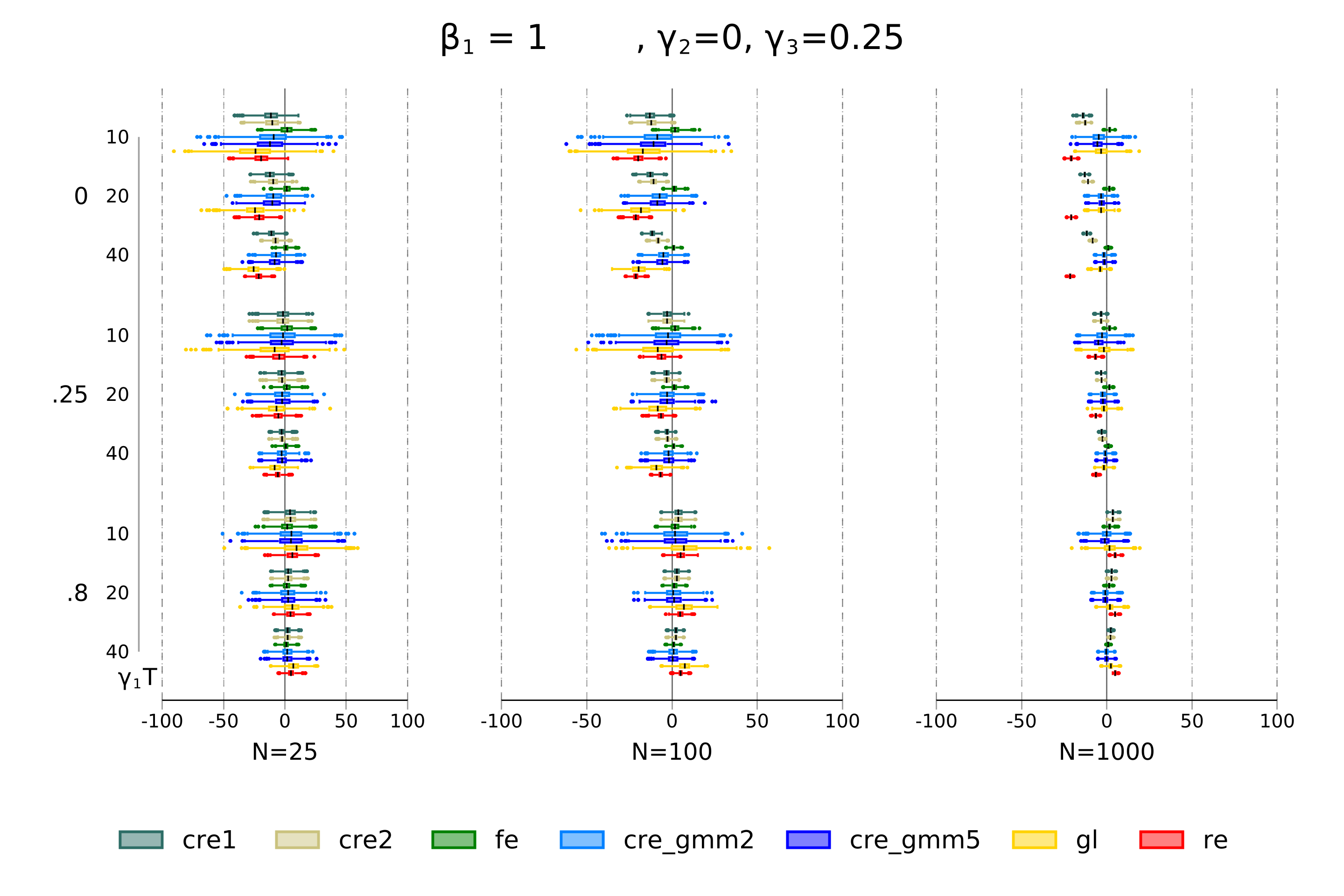}
\caption{\small{Boxplot - No standard endogeneity, PAM ($\rho=0.5, \beta_1 = 1$) \\ Boxplot for bias of $\beta_1$ with $\gamma_2=0, \gamma_3=0.25$ and higher variation in $\mu_i$ than error component $\epsilon_{it}$, $\sigma^2_{\mu}/\sigma^2_{\epsilon}>1$. \\ $\gamma_1 = [0, 0.25, 0.8]$ and $T=10, 20, 40$ are on the horizontal axis.}}
\label{fig:FIXEPAM05BIGbetag3}
\end{figure}

\begin{figure}[!]
\centering
\includegraphics[scale=0.14]{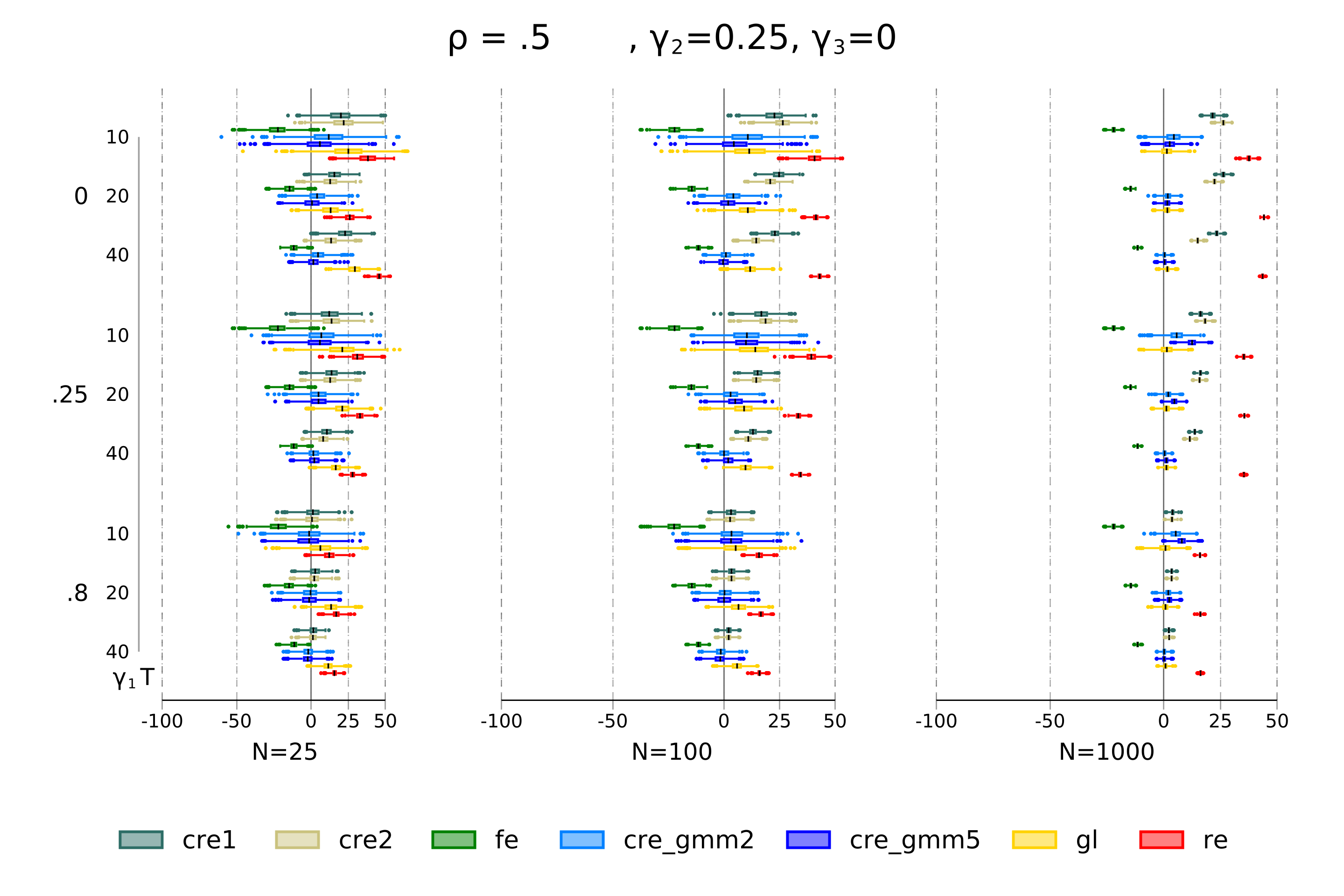}
\caption{\small{Boxplot - Endogeneity from both sources, PAM ($\rho=0.5, \beta_1 = 1$) \\ Boxplot for bias of $\rho$ with correlation between $x_{it}$ and error $\epsilon_{it}$, $\gamma_2=0.25,\gamma_3=0$. \\ $\gamma_1 = [0, 0.25, 0.8]$ and $T=10, 20, 40$ are on the horizontal axis.}}
\label{fig:FIXEPAM05BIGrhog2}
\end{figure}

\begin{figure}[!]
\centering
\includegraphics[scale=0.14]{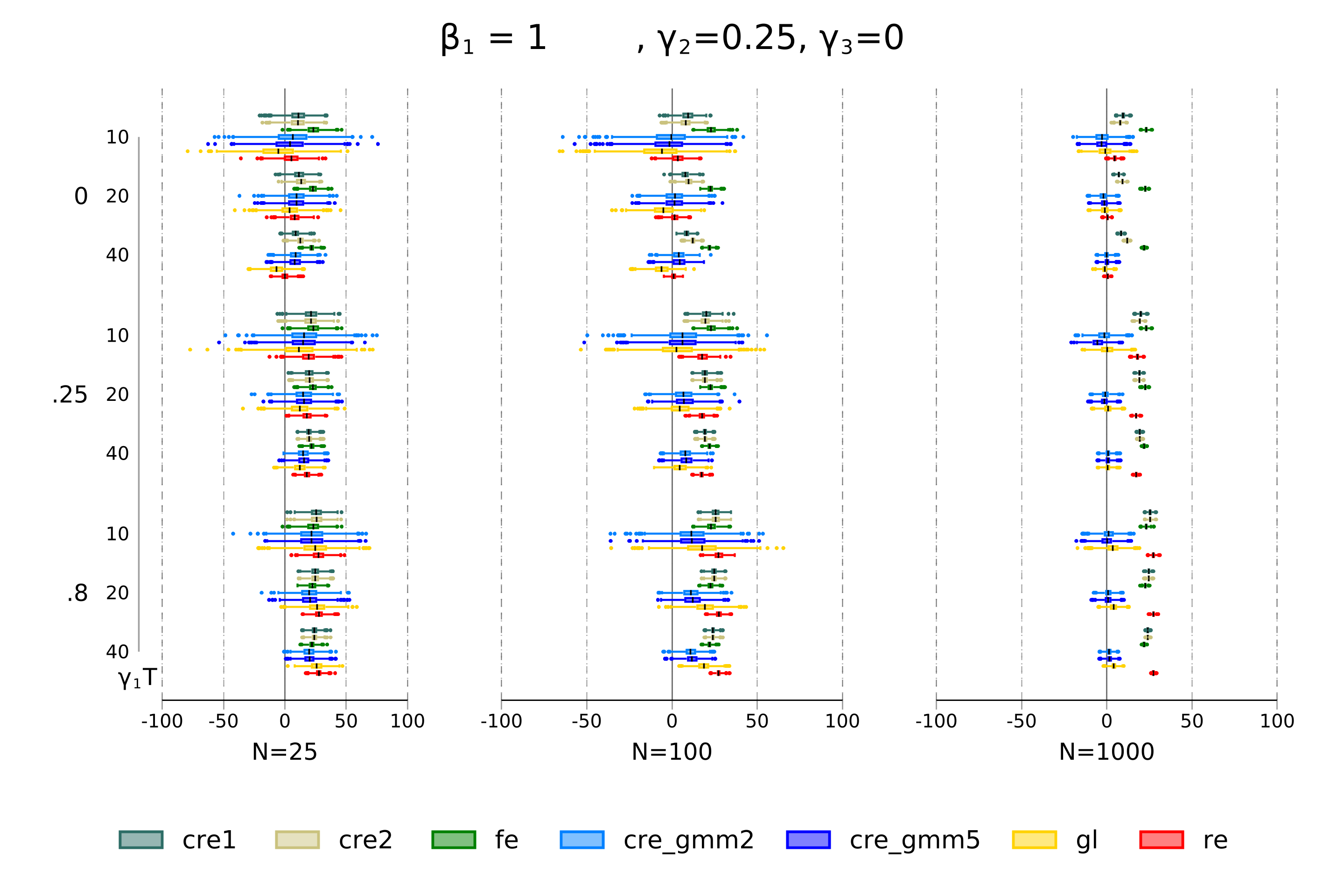}
\caption{\small{Boxplot - Endogeneity from both sources, PAM ($\rho=0.5, \beta_1 = 1$) \\ Boxplot for bias of $\beta_1$ with correlation between $x_{it}$ and error $\epsilon_{it}$, $\gamma_2=0.25,\gamma_3=0$. \\ $\gamma_1 = [0, 0.25, 0.8]$ and $T=10, 20, 40$ are on the horizontal axis.}}
\label{fig:FIXEPAM05BIGbetag2}
\end{figure}

\begin{figure}[!]
\centering
\includegraphics[scale=0.14]{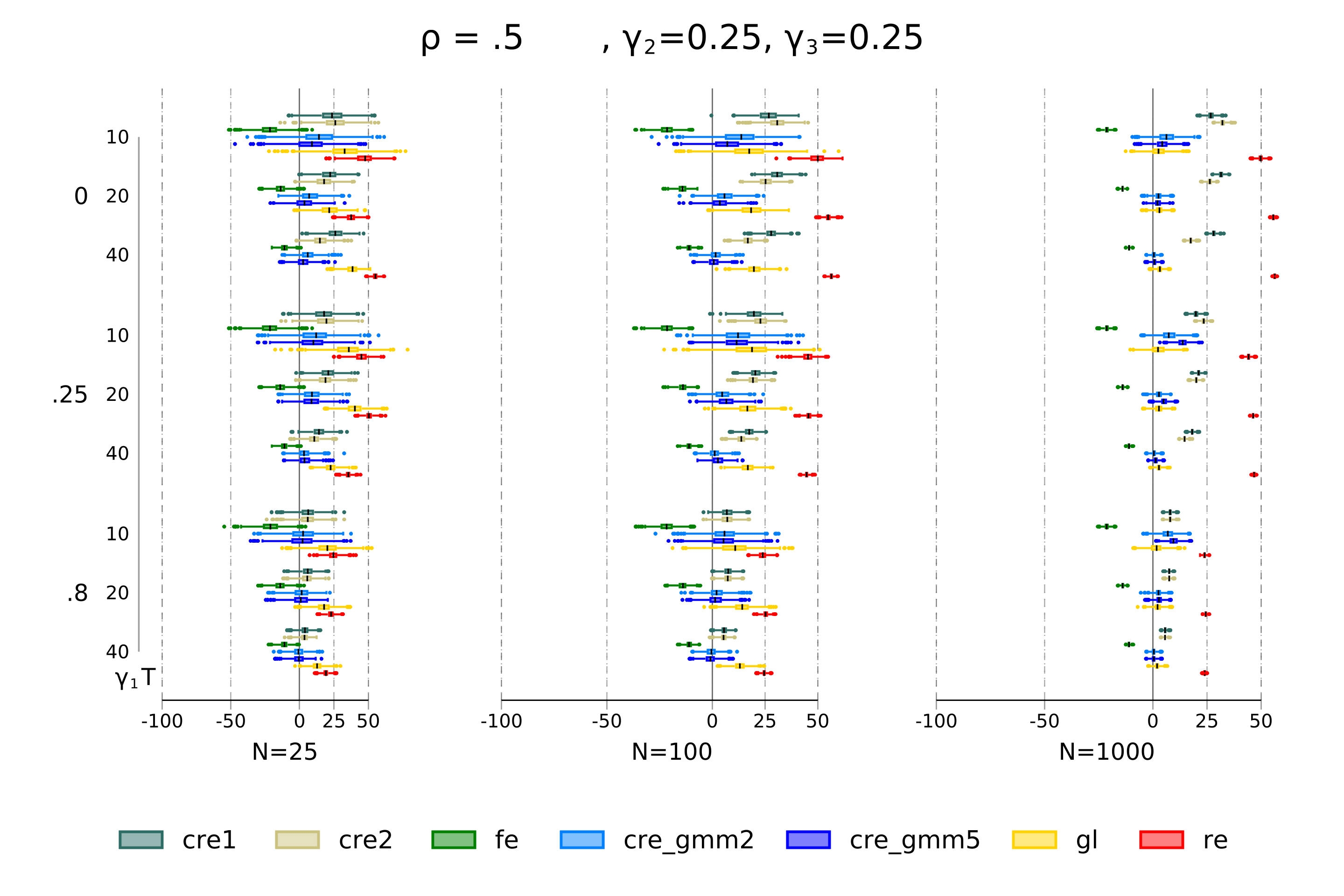}
\caption{\small{Boxplot - Endogeneity from both sources, PAM ($\rho=0.5, \beta_1 = 1$) \\ Boxplot  for bias of $\rho$ with correlation between $x_{it}$ and error $\epsilon_{it}$, $\gamma_2=0.25, \gamma_3=0.25$, and higher variation in $\mu_i$ than error component $\epsilon_{it}$, $\sigma^2_{\mu}/\sigma^2_{\epsilon}>1$. \\ $\gamma_1 = [0, 0.25, 0.8]$ and $T=10, 20, 40$ are on the horizontal axis.}}
\label{fig:FIXEPAM05BIGrhog2g3}
\end{figure}

\begin{figure}[!]
\centering
\includegraphics[scale=0.14]{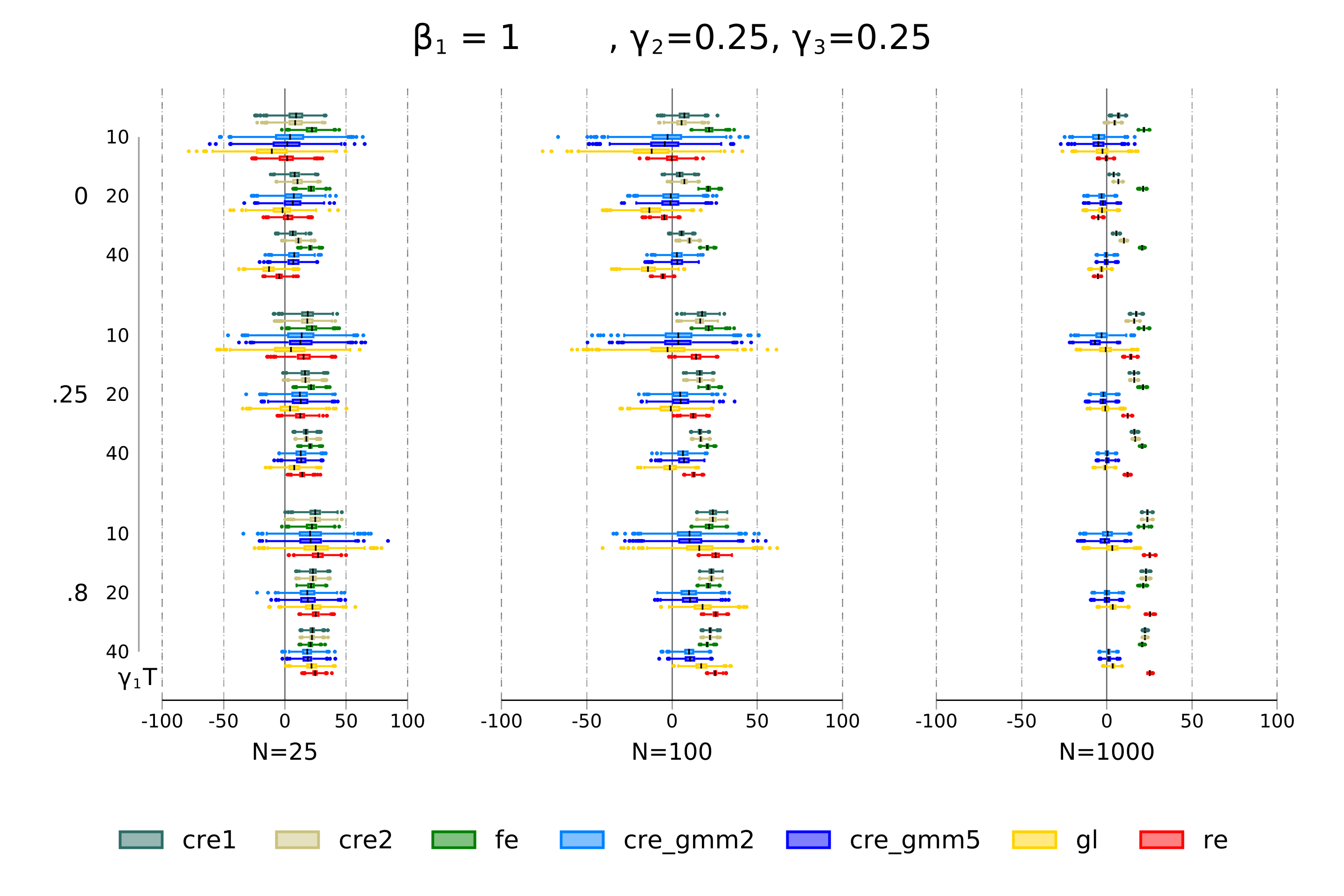}
\caption{\small{Boxplot - Endogeneity from both sources, PAM ($\rho=0.5, \beta_1 = 1$) \\ Boxplot  for bias of $\beta_1$ with correlation between $x_{it}$ and error $\epsilon_{it}$, $\gamma_2=0.25, \gamma_3=0.25$, and higher variation in $\mu_i$ than error component $\epsilon_{it}$, $\sigma^2_{\mu}/\sigma^2_{\epsilon}>1$. \\ $\gamma_1 = [0, 0.25, 0.8]$ and $T=10, 20, 40$ are on the horizontal axis.}}
\label{fig:FIXEPAM05BIGbetag2g3}
\end{figure}

Figures \ref{fig:FIXEPAM05BIGrho}-\ref{fig:FIXEPAM05BIGbeta} shows the role of alternative values for $\gamma_1$, the endogeneity due to heterogeneity, under alternative panel's settings and the $\sigma^2_{\mu}/\sigma^2_{\epsilon}=1$ case, where  $\sigma^2_{\mu}$ is the variance of individual heterogeneity, $\mu_i$, and $\sigma^2_{\epsilon}$ is the variance of the shocks $\epsilon_{it}$. Figures \ref{fig:FIXEPAM05BIGrhog3}-\ref{fig:FIXEPAM05BIGbetag3} shows how results change when the variance of individual heterogeneity is higher than the variance of the shocks, $\sigma^2_{\mu}/\sigma^2_{\epsilon}>1$. Figures \ref{fig:FIXEPAM05BIGrhog2}-\ref{fig:FIXEPAM05BIGbetag2} and Figures \ref{fig:FIXEPAM05BIGrhog2g3}-\ref{fig:FIXEPAM05BIGbetag2g3} compare the cases $\sigma^2_{\mu}/\sigma^2_{\epsilon}=1$ and $\sigma^2_{\mu}/\sigma^2_{\epsilon}>1$ when standard endogeneity, due to the correlation between $x_{it}$ and the idiosyncratic shocks through $\gamma_2=0.25$, is added to the endogeneity due to heterogeneity. To extend the picture of results provided by the Tables, where $\gamma_3=0.8$, in the Figures we present the case of $\gamma_3=0.25$.

\begin{figure}[!]
\centering
\includegraphics[scale=0.16]{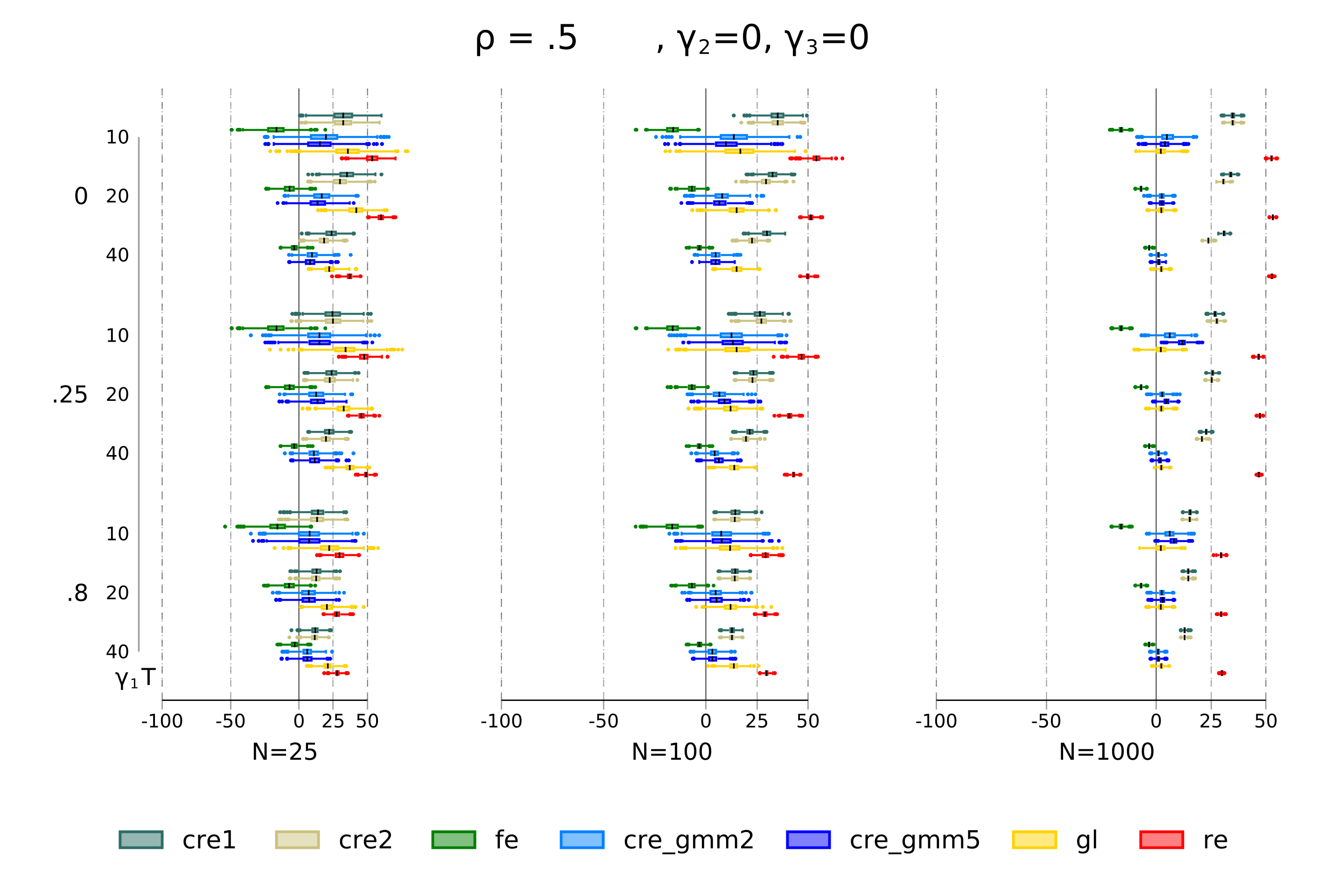}
\caption{\small{Boxplot - No standard endogeneity, PAM ($\rho=0.5, \beta_1=1$) \\ Boxplot for bias of $\rho$ with $\gamma_2=\gamma_3=0$. \\ $\gamma_1 = [0, 0.25, 0.8]$ and $T=10, 20, 40$ are on the horizontal axis.}}
 \label{fig:FIXEPAM05BIGrho} 
\end{figure}

\newpage 

\newpage  

\newpage 

\begin{figure}[!]
\centering
\includegraphics[scale=0.30]{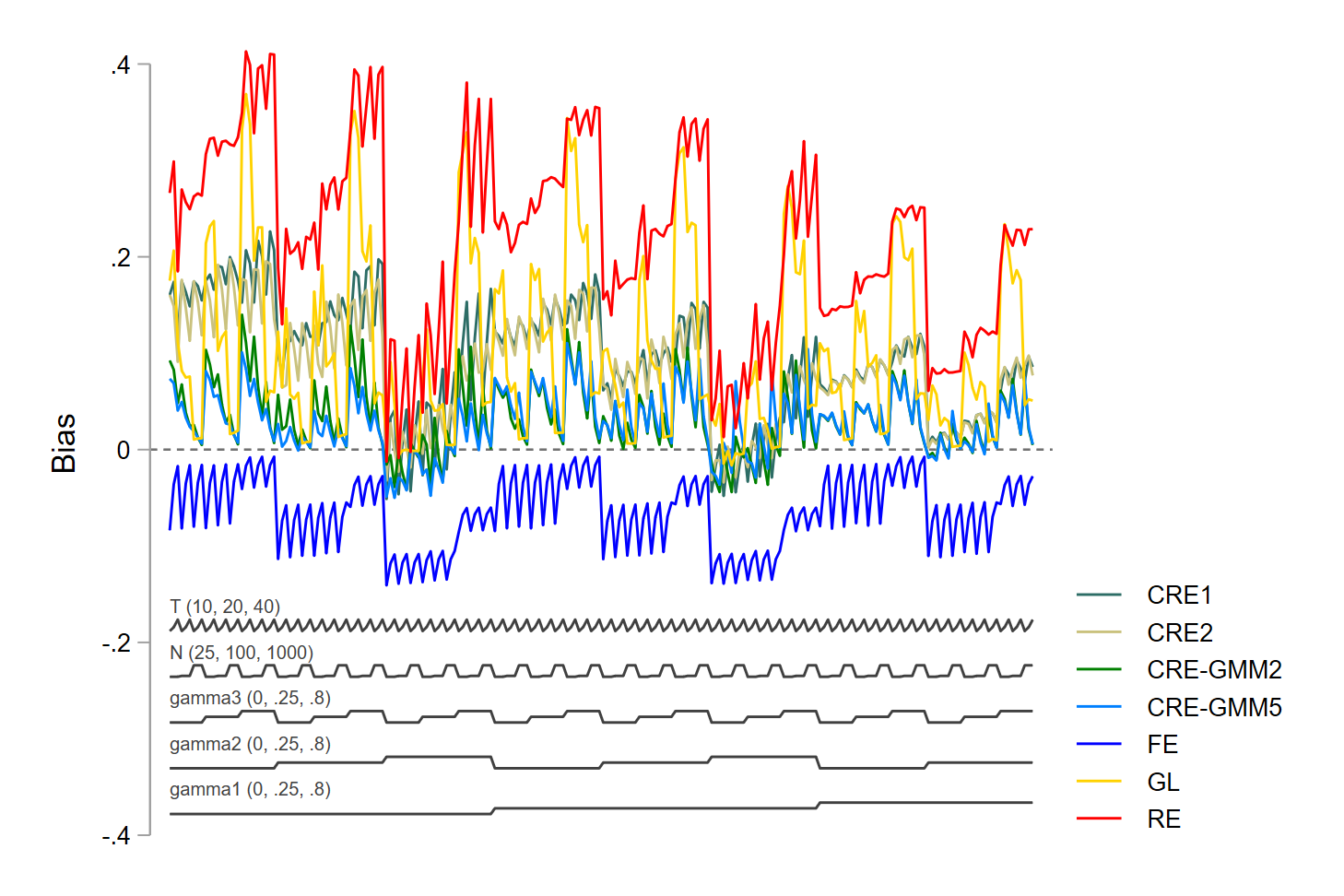}
\caption{\small{Nestedloop Plot for \(\rho\), PAM \\ Bias for $\rho=0.5$ across different specifications.  Parameters shown on horizontal axis.}} 
\label{fig:FIXEPAM05BIGnlooprho}
\end{figure}

\begin{figure}[!]
\centering
\includegraphics[scale=0.30]{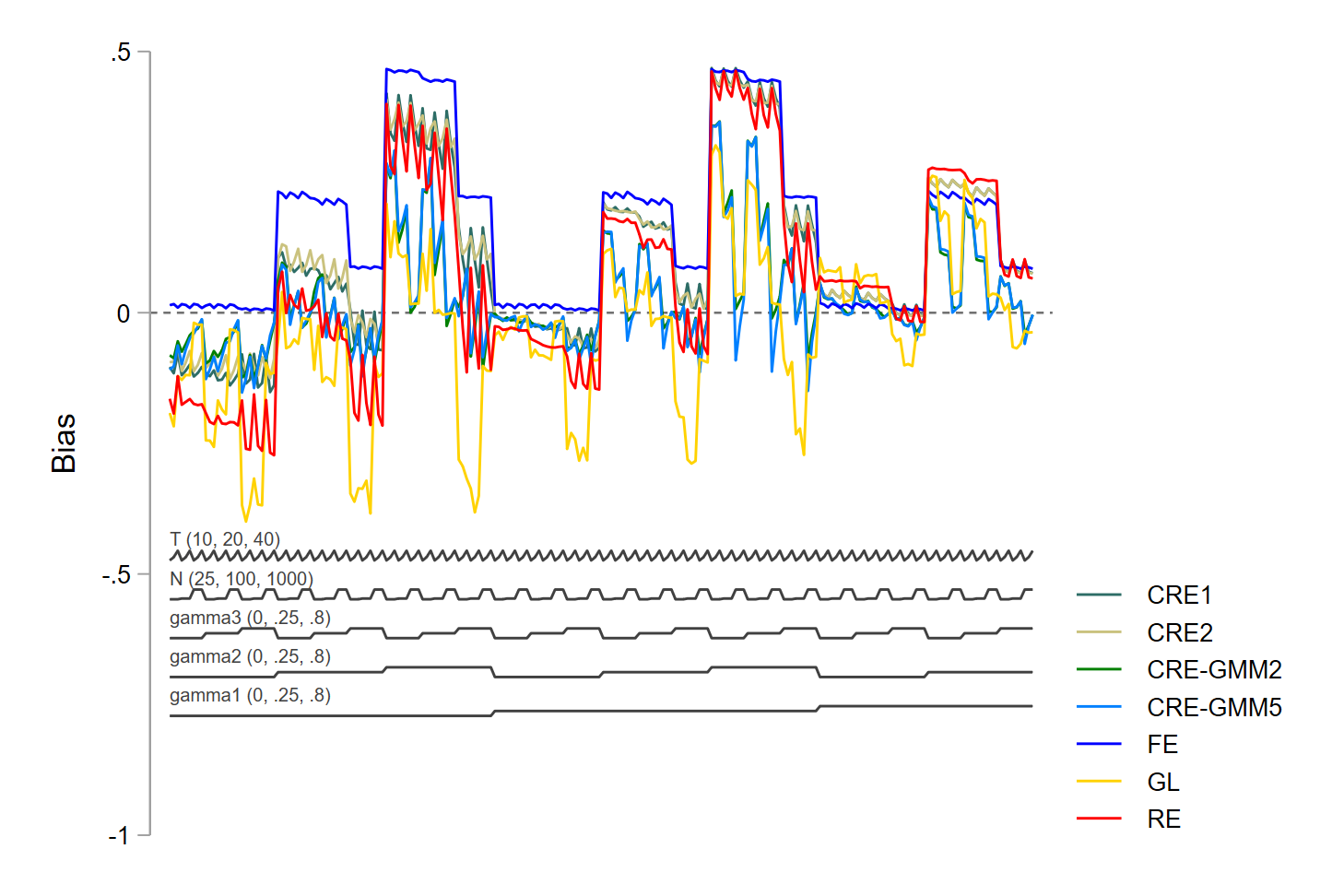}
\caption{\small{Nestedloop Plot for \(\beta_1\), PAM \\ Bias for $\beta_1=1$ across different specifications. Parameters shown on horizontal axis.}} 
\label{fig:FIXEPAM05BIGnloopbeta}
\end{figure}

\begin{figure}[!]
\centering
\includegraphics[scale=0.30]{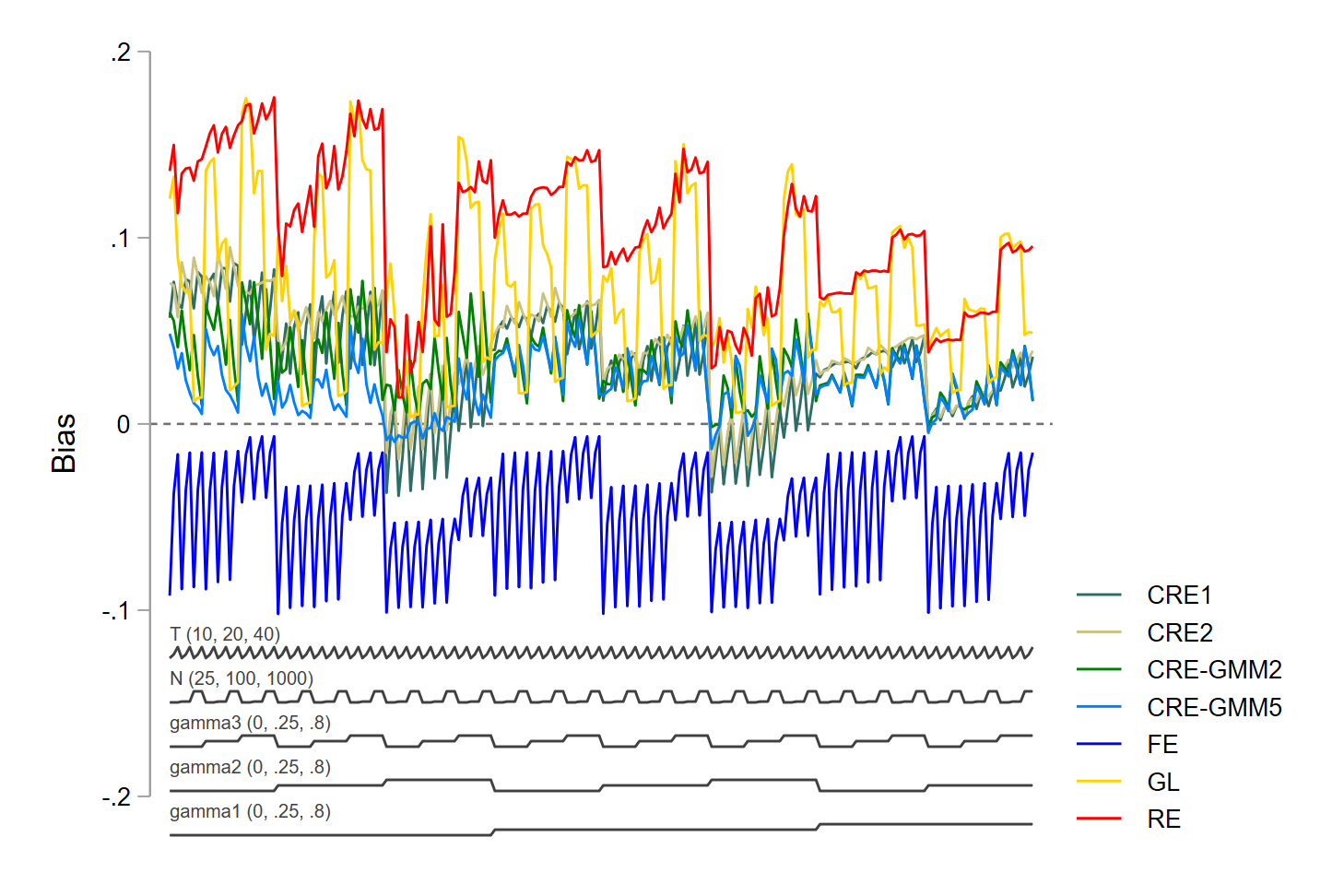}
\caption{\small{Nestedloop Plot for \(\rho\), PAM with $\rho=0.8$ \\ Bias for $\rho=0.8$ across different specifications. Parameters shown on horizontal axis.}}
\label{fig:FIXEPAM08BIGnlooprho}
\end{figure}

\begin{figure}[!]
\centering
\includegraphics[scale=0.30]{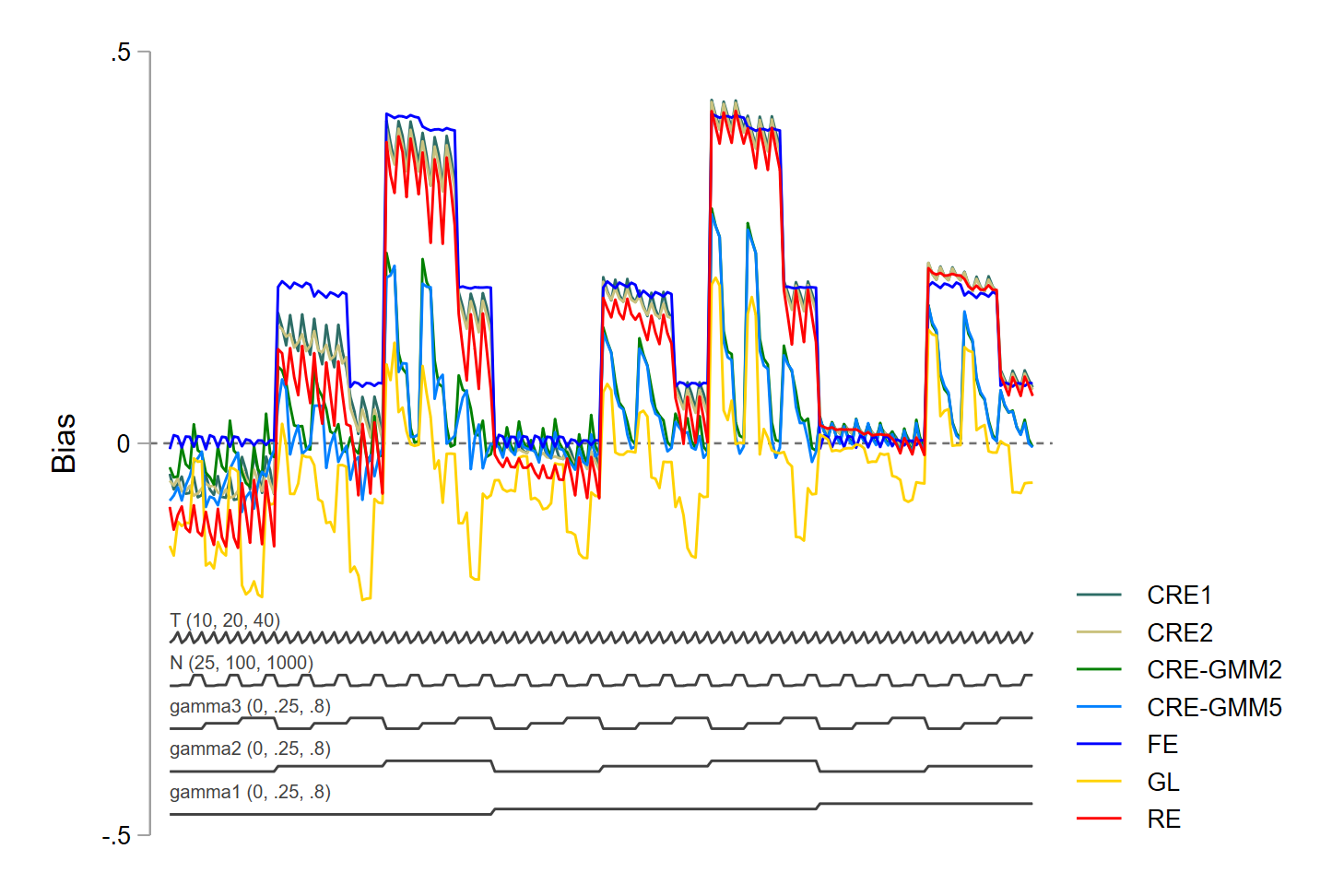}
\caption{\small{Nestedloop Plot for \(\beta_1\), PAM with $\rho=0.8$ \\ Bias for $\beta_1=1$ across different specifications. Parameters shown on horizontal axis.}}
\label{fig:FIXEPAM08BIGnloopbeta}
\end{figure}

\clearpage
Figures \ref{fig:FIXEPAM05BIGnlooprho} and \ref{fig:FIXEPAM05BIGnloopbeta} present the bias for $\rho = 0.5$ and $\beta_1 = 1$ in form of a nestloop plot \citep{Rucker2014} which allows for a straightforward comparison across parametrisations.\footnote{The plots are produced using \texttt{siman} suite in Stata \citep{Marley2022}.} The different parametrisations are on the horizontal axis. The different lines represent the different methods. If in longitudinal panels, CRE-GMM2 is better when $\gamma_1 > 0$ and CRE-GMM5 when $\gamma_1 = 0$, this difference disappears as $T \rightarrow \infty$. Figures \ref{fig:FIXEPAM08BIGnlooprho} and \ref{fig:FIXEPAM08BIGnloopbeta} show the bias when the autoregressive parameter $\rho$ is equal to 0.8.\footnote{The detailed results are in Tables \ref{tab:FIXEPAM08BIGlongi_rho}-\ref{tab:FIXEPAM08BIGmulti_beta} of Appendix \ref{sec:MCPAM08}.} The level equations (CRE-GMM3-CRE-GMM5) deal better with the persistence, and exploiting the additional moment conditions based on the pre-sample average of $y_{it-1}$ and $x_{it}$ reduces the bias particularly in the presence of both sources of endogeneity. 

\section{Empirical applications}\label{sec:EA}
In this section we showcase our estimator on two examples. The first one investigates a specific shape of the production function and the role of net capital and employment on final output on a balanced panel. The second example exploits an unbalanced panel to study an important IO topic concerning the evaluation of the persistence of R\&D activity. The role of accumulated knowledge has to be assessed inside a model that considers, in addition to the unit-and time-varying explanatory variables $x_{it}$, also a list of time-invariant unit-specific measurable explanatory variables $w_i$ which are very important to capture the role of firm-specific market power.

\subsection{Production functions estimation}
We use the dataset in \citet{Blundell2000} to estimate a Cobb-Douglas production function in a sample of 509 US manufacturing firms observed through an eight-year period from 1982 to 1989. This dataset is the equivalent of the UK data used by \citet{Arellano1991} and successively largely analysed by the literature on dynamic panel data models. We estimate the following $ARDL(1,1,1)$ model:

\[ y_{it} = \alpha + \rho y_{it-1} + \beta _1 n_{it} + \beta_2 n_{it-1} + \beta_3 k_{it} + \beta_4 k_{it-1} + \lambda_t + \mu_i + \epsilon_{it} 
\]

where $y_{it}$ is the logarithm of sales, $n_{it}$ is the logarithm of employment, $k_{it}$ is the log of net capital stock, $\lambda_t$ captures time effect and $\alpha_i$ is firms' heterogeneity.. 

According to the ARDL(1,1,1) model, we need the individual averages of the lagged dependent variable $y_{i}^{1}$, the current and lagged employment and net capital, $n_{i.}$, $n_{i.}^{1}$, $k_{i.}$ and $k_{i.}^{1}$; the period $T$ is short, so we selected years 1982-83 as the pre-sample period used to compute the individual averages, and the remaining years 1984-1989 to estimate the model. Table \ref{VaraPROD} compares the characteristics of the dependent variable in terms of variability over the entire period and the two sub-samples.  

\begin{table}[!ht] 
\centering
\caption{Variance decomposition for the dependent variable}
\label{VaraPROD}
    \begin{tabular}{lllllllll}
    \hline
    Period & 1982-1989 & 1982-83 & 1984-89 \\
    \hline
    Between variability & \hspace{1cm}  98.22\% & 99.70\% & 98.87\% \\
    \hline
    Within variability & \hspace{1cm} 1.78\%  & 0.30\% & 1.13\% \\
    \hspace{1cm}  common to all the units & \hspace{1cm} (0.31\%) & (0\%) & (0.17\%) \\
    \hspace{1cm} unit-specific & \hspace{1cm} (1.47\%) & (0.30\%) & (0.95\%) \\
    \hline
\end{tabular}
\flushleft
{\scriptsize Note: Computations implemented by author's user written \texttt{xtsum3} procedure.}
\end{table}

The total variability of the dependent variable is dominated by the variation over the units, with the within variability being mostly firm-specific. Common factors like the business cycle and inflation, which may have the same influence across companies, play a marginal role when explaining within variability. 

\begin{figure}[h!]
\centering
\caption{The log of output, workers and capital over time}\label{fig:PROD}
\includegraphics[scale=0.25]{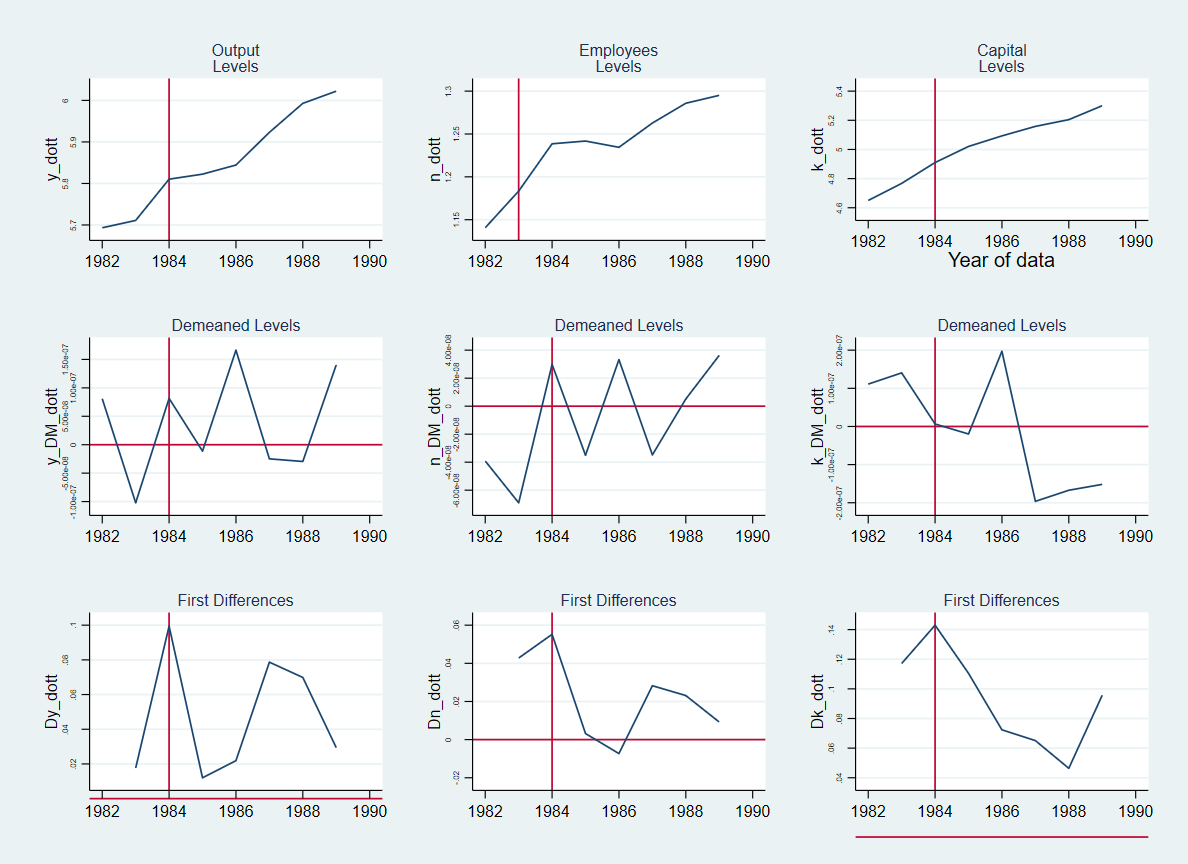}
\end{figure}

Figure \ref{fig:PROD} (and Dickey-Fuller tests not reported) confirms the persistence already noticed by \citet{Blundell2000}. 
Such high persistency leads to the potential weak IVs issue for GMM-dif. Both output and employment present a period of stagnation between 1984 and 1986, which could potentially underline the effects of the oil crisis following the Iranian Revolution in 1979, causing a global recession which influenced the US over the period 1980-1982. The average log output steeply increased over time, especially from 1986. Employment had rapid growth from the beginning period until 1984, starting to rise again in a consistent way from 1986. Besides, capital presents an increasing linear trend from 1982 until the last year considered. 

\newpage 
\begin{sidewaystable}
\centering
{\scriptsize
\caption{Individual means computed over 1982-1983, estimations run over 1984-1989.}
\label{tab:PROD}
\begin{tabular}{lllll|l|lllll}
\toprule
                         &      RE   &     FE   &    CRE1   &    CRE2   &      GL   &  CRE-GMM2   &  CRE-GMM5 & MLtd & ML & QML  \\
                         \midrule
L.y                      &   0.930***&   0.378***&   0.935***&   0.864***&   0.667***&   0.694***&   0.667*** & 0.928*** & 0.604*** & 0.546*** \\
                         &(0.0094)   &(0.0281)   &(0.0114)   &(0.0175)   &(0.0466)   &(0.0514)   &(0.0491)  & (0.0132)& (0.0483) & (0.0757) \\
n                        &   0.454***&   0.460***&   0.453***&   0.438***&   0.459***&   0.688***&   0.779*** & 0.369*** & 0.520*** & 0.411*** \\
                         &(0.0303)   &(0.0325)   &(0.0302)   &(0.0338)   &(0.1249)   &(0.1630)   &(0.1749)  & (0.1127) & (0.0534) & (0.0426 \\
L.n                      &  -0.406***&  -0.004   &  -0.406***&  -0.301***&  -0.232*  &  -0.360***&  -0.453*** & -0.335*** & -0.114** & -0.072 \\
                         &(0.0316)   &(0.0368)   &(0.0315)   &(0.0368)   &(0.1265)   &(0.1355)   &(0.1571)  & (0.0529) & (0.0544) & (0.0602)\\
k                        &   0.238***&   0.200***&   0.235***&   0.177***&   0.573***&   0.400** &   0.383** & 0.219* & 0.093* & 0.182*** \\
                         &(0.0373)   &(0.0377)   &(0.0379)   &(0.0431)   &(0.1309)   &(0.1955)   &(0.1649)  & (0.1205) & (0.0547) & (0.05489) \\
L.k                      &  -0.216***&  -0.125***&  -0.211***&  -0.222***&  -0.464***&  -0.460***&  -0.420*** & -0.204*** & -0.145*** & -0.177*** \\
                         &(0.0361)   &(0.0277)   &(0.0371)   &(0.0345)   &(0.1162)   &(0.1329)   &(0.1252) & (0.0577) & (0.0370) & 0.0530 \\
$y_{i}^{1}$                  &           &           &  -0.007   &   0.079***&           &   0.061   &   0.087   \\
                        &           &           &(0.0084)   &(0.0175)   &           &(0.0630)   &(0.0598)   \\
$n_{i.}$                   &           &           &           &   0.026   &           &  -0.165*  &  -0.090   \\
                         &           &           &           &(0.0278)   &           &(0.0961)   &(0.0748)   \\
$n_{i.}^{1}$                  &           &           &           &  -0.129***&           &  -0.105** &  -0.101** \\
                         &           &           &           &(0.0185)   &           &(0.0446)   &(0.0485)   \\
$k_{i.}$                   &           &           &           &   0.067***&           &   0.215*  &   0.143*  \\
                        &           &           &           &(0.0256)   &           &(0.1213)   &(0.0820)   \\
$k_{i.}^{1}$                  &           &           &           &   0.001   &           &   0.022   &   0.006   \\
                        &           &           &           &(0.0113)   &           &(0.0487)   &(0.0261)   \\
\midrule
NT                       &    3054   &    3054   &    3054   &    3054   &    3054   &    3054   &    3054 & 3054 & 3054 & 3054 \\
N                        &     509   &     509   &     509   &     509   &     509   &     509   &     509  & 509 & 509 & 509 \\
$\bar{T}$                     &       6   &       6   &       6   &       6   &       6   &       6   &       6   & 6 & 6 & 6 \\
ar1 pval.                     &           &   0.00        &    0.00   &    0.00   &    0.00   &    0.00   &    0.00  & & & \\
ar2 pval.                     &           &    0.29       &    0.80   &    0.65   &    0.53   &    0.83   &    0.62  & & & \\
ar3 pval.                     &           &     0.35      &    0.98   &    0.92   &    0.59   &    0.98   &    0.83  & & & \\
Hausman pval.                &           &           &   0.417   &   0.000   &    0.000       &   0.007   &   0.034  & & & \\
Time dummies pval.                     &   0.000   &   0.000   &   0.000   &   0.000   &   0.000   &   0.000   &   0.000  & 0.000 & & \\
\bottomrule
\end{tabular}
}
\flushleft
{\scriptsize Note: Estimates are implemented using \texttt{xtdpdgmm} \citep{Kripfganz2019a} in Stata. ML estimates are from \texttt{xtdpdml} \citep{Williams2018}, QML are from \texttt{xtdpdqml} \citep{Kripfganz2016}. As a robustness check, the results estimated over the whole sample 1982-1989 corroborate our findings.}
\end{sidewaystable} 

Table \ref{tab:PROD} presents the regression results. The RE estimation can be seen as an upper bound, while the FE is a lower bound on the behavioural persistency $\rho$. This means that consistent estimates would be characterized by an autoregressive parameter falling inside $0.378 < \rho < 0.930$. In addition, the discrepancy between RE and FE underlines the presence of firm-specific effects $\mu_i$ correlated with the explanatory variables and corroborated by CRE1 and CRE2. Such heterogeneity is time-invariant and captures firm-specific characteristics, like the CEO's innate abilities or unique features of their products. This sample represents a suitable environment for our CRE-GMM estimator also because individual heterogeneity could be subject to potential changes over time, losing their constancy and intrinsic features. For instance, the overall CEO's ability remains invariant over time until the industry does not hire a new manager. Hence we compare the standard GMM-lev estimation (GL) with the GMM-lev augmented with CRE where instruments are chosen according to the results obtained in our simulations (CRE-GMM2 and CRE-GMM5). 

As far as the second source of endogeneity is concerned, occupation and capital stock are supposed to be correlated with the error term, due to potential simultaneity and omitted variables. For instance, a positive productivity shock could boost output and foster new employment. Also, the capital stock could be linked with the latent financial health of a company, or some omitted financial measures like capital structure or ROE.  
Maximum likelihood estimations (with and without time dummies, MLtd and ML, respectively) and quasi-maximum likelihood estimation (QML) are reported to complete the analysis, despite they are based on the assumption of non endogeneity of capital and workers. The role of this assumption is evident in the not statistical significance of capital in MLtd and ML; the result is different in QML, where capital is again significant, while lagged labour is not significant. Another drawback of maximum likelihood estimations is that they work when panel are strongly balanced, $T$ is relatively small (e.g., less than 10), and there is no missing data which is not the typical situation we face nowadays on panels.  

When exploiting the GMM estimators, \citet{Blundell2000} instrument labour and capital with either lag $t-2$ or lag $t-3$ (preference for $t-3$) to avoid over-fitting with too many IVs. As for the simulations, we preferred to follow \citet{Bun2006} and use both $t-2$ and $t-3$ lags. We do not reject the null hypothesis of valid overidentifying restrictions, according to which the instruments are jointly valid.

The results show that the parameter $\rho$ is consistently estimated by GL, CRE-GMM2 and CRE-GMM5. The estimates differ for $n_{it}$ and $k_{it}$, with CRE-GMM5 producing effects in line with the macroeconomic estimates, 70\% and 30\%, of the shares of GDP to labour and capital. According to the results preferred by \citet{Blundell2000} in Table III for GMM-sys, the estimates are 0.629 (lag $t-2$) and 0.472 (lag $t-3$) for $n_{it}$, and 0.361 (lag $t-2$) and 0.398 (lag $t-3$) for $k_{it}$, with cumulative dynamic multipliers for labour and capital equal to 0.537 and 0.035 (lag $t-2$), and 0.194 and 0.189 (lag $t-3$). Our cumulative dynamic multipliers for labour and capital are more robust, equal to 0.038 and 0.177 from CRE-GMM2, and 0.135 and 0.112 from CRE-GMM5. The decomposition of the within and between effects allowed by the inclusion of individual averages shows that increasing size makes firms able to produce more output, whereas being larger does not mean being more productive; the capital effect is strongest in the long run (the between effect is larger than the within effect). 

%
%
%
%

\subsection{Persistence of R\&D}\label{IO}
As a second example, we exploit the panel of Italian companies performing R\&D \citep{Bontempi2023}. The sample contains 3,971 firms observed over the 1984-2012 period. To obtain the individual averages of time-varying variables, we use the years 1984-95 as the pre-sample period for the computation, and 1996-2012 to estimate following $ARDL(1,1)$ model:

\[ R\&D_{it} = \alpha + \rho R\&D_{it-1} + \boldsymbol{\beta}^{'}\mathbf{x}_{it-1}+\boldsymbol{\theta}^{'}\mathbf{w}_{i} + \lambda_t + \mu_i + \epsilon_{it}. 
\]

The equation is derived from standard IO methods based on the intensive margin where R\&D, the input in levels, is measured as the R\&D investment over employees. \citet{Bontempi2023} argue that the past amount of R\&D investment is not necessarily a meaningful measure, because even after a discovery has been made, companies must continue to invest in R\&D as it may take a long time to convert innovation into economic results. To capture how the cumulative learning (the experience of doing R\&D at any given date) drives the innovation process by making firms progressively better at it, the lag of R\&D is replaced by lagged stocks of R\&D in the second dynamic model we estimate: 

\[ R\&D_{it} = \alpha + \rho_1 R\&D^{stock}_{it-1} + \boldsymbol{\beta}^{'}\mathbf{x}_{it-1}+\boldsymbol{\theta}^{'}\mathbf{w}_{i} + \lambda_t + \mu_i + \epsilon_{it} 
\]

where $R\&D^{stock}$, measured as the logarithm of innovative stock, could capture inter-temporal externalities and temporal spillovers between subsequent R\&D investments. In both specifications, $\mathbf{x}_{it-1}$ includes financial variables like debt and free cash flow, sales from export and other unit-and time-varying explanatory variables, while $\mathbf{w}_{i}$ includes, among the others, the unit-specific market power, a survey-based measure of firm-specific elasticity of demand; $\lambda_t$ is a set of time dummies, as well as an uncertainty index like the EPU from \citet{Baker2013} or EURQ from \citet{Bontempi2021}. 

The CRE-GMM-lev approach is able to tackle the difficulties of the empirical literature on innovation in consistently estimating the causal effect of past R\&D activities on current R\&D investment due to the ``true'' path-dependent nature of technical changes. In fact, spurious persistence emerges as an artefact of the inability to control for the individual heterogeneity producing the past innovation dependence due to  time-invariant unobserved firm characteristics determining the initial conditions. Instead, among the mechanisms that may explain the true path-dependence, three arguments can be seen as complementary \citep{Peters2009}. First, technological knowledge is an economic good characterized by cumulability and non-exhaustibility, and represents at the same time an input and an output of the knowledge-generating process \citep{Antonelli2012}. Hence, firms that have generated new technological knowledge can rely upon such accumulated knowledge to generate new, additional knowledge at a lower cost. Dynamic increasing returns are likely to shape innovative activities: the larger the cumulated size of innovation, the larger is the positive effect on costs (``learning-to-learn'' and ``learning-by-doing'' effects).
A second stream holds that firms need to successfully profit from their innovation so to be able to innovate again. According to this view, commercial success increases the probability of future innovation because it allows for the reallocation of profits to new research projects (``success-breads-success'' effect). Firms that successfully innovate are hence more likely to follow on innovations because of higher permanent market power \citep{LeBas2014}.
Thirdly, according to the ``sunk-costs'' effect theory, innovative activities are characterized by high set up costs for research facilities and the training of personnel, and by long-term commitments in terms of investment. Once research has started, the opportunity cost of interrupting it is quite high. This implies that research and development activities generate high entry and exit barriers as well \citep{Antonelli2013}. 

Among all the measurable individual characteristics, market power is the most relevant one, as the literature still debates between \citet{Schumpeter1943}'s view and \citet{Arrow1962}'s thesis. \citet{Schumpeter1943} argues that firms with greater market power have a higher incentive to innovate because they can better appropriate the returns of their R\&D investment. On the other hand, according to \citet{Arrow1962}, competition positively affects the innovative effort, as a firm that successfully introduces a new product in the market could then become monopolist. 

As shown in Table \ref{VaraRED}, the total variability of the dependent variable is dominated by the variation over the units, with the exception of the pre-sample period in which a temporary tax incentive (``Tremonti Law'') was granted to investments in the years 1994-95.\footnote{The first Tremonti Law granted, for the tax periods 1994 and 1995, a tax benefit consisting in the exclusion from the formation of the company's income of 50\% of the increase in investments made in the current tax period compared to the average of those made in the previous five years.} The within variability is mostly firm-specific. 

\begin{table}[!ht] 
\centering
\caption{Variance decomposition for the dependent variable}
\label{VaraRED}
    \begin{tabular}{lllllllll}
    \hline
    Period & 1984-2012 & 1984-95 & 1996-2012 \\
    \hline
    Between variability & \hspace{1cm}  65.54 \% & 47.49\% & 72.08\% \\
    \hline
    Within variability & \hspace{1cm} 34.46\%  & 52.51\% & 27.92\% \\
    \hspace{1cm}  common to all the units & \hspace{1cm} (0.48\%) & (1.92\%) & (0.15\%) \\
    \hspace{1cm} unit-specific & \hspace{1cm} (33.99\%) & (50.60\%) & (27.77\%) \\
    \hline
\end{tabular}
\flushleft
{\scriptsize Note: Computations implemented by user written \texttt{xtsum3} procedure.}
\end{table}

\begin{figure}[h!]
\centering
\caption{The pattern of R\&D over time}
\label{fig:RED}
\includegraphics[scale=0.3]{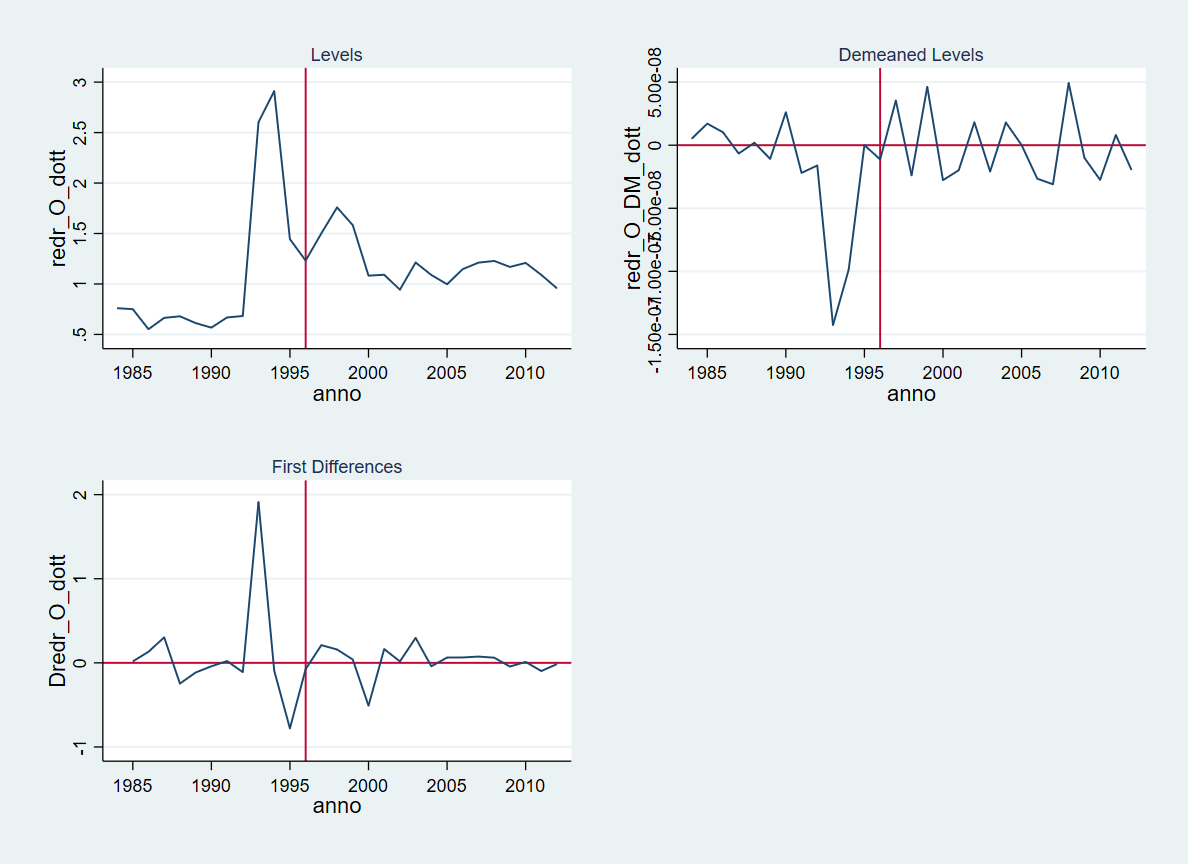}
\end{figure}

From Figure \ref{fig:RED} it is clearly visible the effect of the ``Tremonti Law'' which was temporary and can be smoothed by using the average over 12 years. 

\begin{sidewaystable}
\centering
\caption{Estimation of R\&D for Italian Firms}
\label{tab:RED}
\begin{adjustbox}{max width=\textwidth}
{\scriptsize
\begin{tabular}{l|rrrrrrr|rrrr|rrrr}
\toprule
                         &      RE   &   REcom   &      FE   &   FEcom   &    CRE1   & CRE1com   &    CRE2   &      GL   &   GLcom   &   GLtwo   &GLtwocom   &    CRE-GMM2   & CRE-GMM2two   &    CRE-GMM5   & CRE-GMM5two   \\
\midrule
 \multicolumn{16}{l}{Panel A: Lagged R\&D} \\
\midrule
$R\&D_{it-1}$          &   0.743***&   0.739***&   0.335***&   0.473***&   0.735***&   0.735***&   0.735***&   0.626***&   0.673***&   0.632***&   0.674***&   0.649***&   0.650***&   0.696***&   0.696***\\
                         &(0.0510)   &(0.0641)   &(0.0675)   &(0.0956)   &(0.0647)   &(0.0646)   &(0.0631)   &(0.0616)   &(0.0683)   &(0.0029)   &(0.0001)   &(0.0649)   &(0.0003)   &(0.0764)   &(0.0002)   \\
Market power               &   0.201***&   0.152   &   0.000   &   0.000   &   0.122   &   0.111   &   0.159   &   0.261***&   0.200   &   0.114** &   0.203***&   0.430   &   0.424***&   0.422*  &   0.418***\\
                         &(0.0686)   &(0.2068)   &     (.)   &     (.)   &(0.1993)   &(0.2121)   &(0.1943)   &(0.1006)   &(0.2561)   &(0.0495)   &(0.0031)   &(0.2887)   &(0.0124)   &(0.2446)   &(0.0093)   \\
\midrule
NT                       &    8109   &    1596   &    8109   &    1596   &    1629   &    1596   &    1596   &    8109   &    1596   &    8109   &    1596   &    1596   &    1596   &    1596   &    1596   \\
N                        &    1415   &     284   &    1415   &     284   &     288   &     284   &     284   &    1415   &     284   &    1415   &     284   &     284   &     284   &     284   &     284   \\
Tavg                     &       6   &       6   &       6   &       6   &       6   &       6   &       6   &       6   &       6   &       6   &       6   &       6   &       6   &       6   &       6   \\
ar1 pval.                     &           &           &           &           &           &           &           &    0.00   &    0.00   &    0.00   &    0.00   &    0.00   &    0.00   &    0.00   &    0.00   \\
ar2 pval.                     &           &           &           &           &           &           &           &    0.59   &    0.24   &    0.58   &    0.25   &    0.23   &    0.24   &    0.26   &    0.27   \\
ar3 pval.                     &           &           &           &           &           &           &           &    0.18   &    0.33   &    0.20   &    0.33   &    0.33   &    0.33   &    0.32   &    0.32   \\
Hansen pval.                  &           &           &           &           &           &           &           &    0.00   &    0.97   &    0.08   &    0.97   &    0.95   &    0.94   &    0.95   &    0.95   \\
Hausman pval.                &           &           &           &           &   0.127   &   0.146   &   0.210   &           &           &           &           &   0.058   &   0.000   &   0.355   &   0.000   \\
$R^2$                       &   0.632   &   0.625   &   0.535   &   0.433   &   0.625   &   0.626   &   0.627   &   0.617   &   0.621   &   0.625   &   0.621   &   0.612   &   0.612   &   0.622   &   0.622   \\
\midrule
\multicolumn{16}{l}{Panel B: Accumulated knowledge} \\
\midrule
$R\&D^{stock}_{it-1}$                &   0.199***&   0.086***&   0.023   &  -0.024   &   0.093***&   0.078***&   0.094***&   0.981***&   0.269***&   0.670***&   0.270***&   0.309***&   0.309***&   0.255***&   0.255***\\
                         &(0.0291)   &(0.0289)   &(0.0254)   &(0.0338)   &(0.0221)   &(0.0294)   &(0.0235)   &(0.2044)   &(0.0659)   &(0.0371)   &(0.0010)   &(0.0833)   &(0.0019)   &(0.0815)   &(0.0012)   \\
Market power           &   0.750** &  -0.074   &   0.000   &   0.000   &  -0.144   &  -0.140   &   0.281   &  -0.073   &  -0.574   &   0.016   &  -0.580***&   0.243   &   0.212***&   0.503   &   0.494***\\
                         &(0.3390)   &(0.6183)   &     (.)  &     (.)   &(0.3542)   &(0.6230)   &(0.4175)   &(0.3805)   &(0.4644)   &(0.1992)   &(0.0125)   &(0.6095)   &(0.0315)   &(0.6875)   &(0.0233)   \\
\midrule
NT                       &    6414   &    1203   &    6414   &    1203   &    2068   &    1203   &    1967   &    6414   &    1967   &    6414   &    1967   &    1967   &    1967   &    1967   &    1967   \\
N                        &    1136   &     210   &    1136   &     210   &     349   &     210   &     334   &    1136   &     334   &    1136   &     334   &     334   &     334   &     334   &     334   \\
Tavg                     &       6   &       6   &       6   &       6   &       6   &       6   &       6   &       6   &       6   &       6   &       6   &       6   &       6   &       6   &       6   \\
ar1 pval.                     &           &           &           &           &           &           &           &    0.00   &    0.07   &    0.00   &    0.05   &    0.06   &    0.04   &    0.04   &    0.02   \\
ar2 pval.                     &           &           &           &           &           &           &           &    0.86   &    0.40   &    0.97   &    0.40   &    0.36   &    0.38   &    0.49   &    0.53   \\
ar3 pval.                     &           &           &           &           &           &           &           &    0.50   &    0.52   &    0.41   &    0.50   &    0.43   &    0.41   &    0.39   &    0.40   \\
Hansen pval.                  &           &           &           &           &           &           &           &    0.00   &    0.87   &    0.01   &    0.87   &    0.93   &    0.95   &    0.79   &    0.79   \\
Hausman pval.                &           &           &           &           &   0.050   &   0.147   &   0.232   &           &           &           &           &   0.119   &   0.000   &   0.224   &   0.000   \\
$R^2$                       &   0.199   &   0.394   &   0.018   &   0.001   &   0.371   &   0.395   &   0.381   &   0.174   &   0.362   &   0.189   &   0.362   &   0.344   &   0.344   &   0.369   &   0.369   \\
\midrule
\bottomrule
\end{tabular}
}
\end{adjustbox}
\flushleft
{\scriptsize Note: Estimates are implemented using \texttt{xtdpdgmm} \citep{Kripfganz2019a} in Stata. Individual means computed over 1984-1995, estimations run over 1996-2012. The label's component ``com'' indicates the common estimation sample for which the pre-sample individual averages are available, so that RE, FE, CRE1 and GL are re-estimated over the same sample exploited by CRE2, CRE-GMM2 and CRE-GMM5. Cluster standard errors are used in RE, FE, CRE1-CRE2; cluster one-step standard errors are used in GMM estimates. The label's component ``two'' indicates two-step standard errors in GMM estimates.}
\end{sidewaystable} 

Results are presented in Table \ref{tab:RED}. Estimates for the lagged dependent variable are in a similar range, irrespectively of the estimation method, GL, CRE-GMM2 or CRE-GMM5. The same applies to the use of the common sample resulting from the availability of the pre-sample individual averages or one-step cluster standard errors or two-step standard errors. A completely different situation emerges when we use true accumulated knowledge (the stock of past R\&D); in comparison to the one- or two-step estimators, the GL method greatly overestimates the coefficient. It seems that the decomposition of the effects into the within and between components in the CRE-GMM2 and CRE-GMM5 methods improves the estimation of the weighting matrix used in the second step. 

The most interesting result concerns the estimation of market power, which is obviously not available in the fixed-effects estimation (and therefore not available in the GMM-dif estimation either). The CRE-GMM5 method shows robustness irrespective of the use of one-step or two-step and the way of measuring accumulated knowledge. The Hausman test in general confirms the correlation between the explanatory variables and unobservable individual effects. 

\section{Conclusions}\label{CONCLU}
We present a new approach to dynamic panel data models, also suitable for static models, which merges the GMM applied to equations in levels with the CRE approach. The levels allow estimating the effect of measurable time-invariant covariates. The individual averages computed in the pre-estimation period capture the initial conditions of the units and help to deal with endogeneity due to heterogeneity that may not be removed by the use of instruments in first differences. Our method works well in case of double endogeneity due to correlation with idiosyncratic shocks and individual heterogeneity, and reduces the bias which characterises the GMM-lev when $T$ is large and the variance of individual heterogeneity is greater than the variance of idiosyncratic shocks. It is more efficient than the GMM-lev. The inclusion of individual averages makes level instruments valid, another positive feature of our approach, as instruments in level are preferable when series tend to be persistent, while they produce similar results to using instruments in first differences when the autoregressive parameter is small. When the series are persistent, the estimates are improved by exploiting the additional moment conditions related to the pre-sample individual averages of the lagged dependent variable and the explanatory variables in the case of correlation with individual heterogeneity, and of the explanatory variables alone in the case of non-correlation with individual heterogeneity.

\printbibliography

\newpage
\appendix

\setcounter{table}{0}
\renewcommand{\thetable}{A\arabic{table}}

\setcounter{equation}{0}
\renewcommand{\theequation}{A\arabic{equation}}

\section{For Online Publication - Appendix}\label{app}
\subsection{MonteCarlo Setup}\label{sec:MCset}
The random components of equations \eqref{MC:eq1} are generated such as:
\begin{align}
\mu_i &\sim iid N(\mu_{\mu},\sigma_{\mu}^2) \\
\xi_{it} &\sim iid N(\mu_{\xi},\sigma_{\xi}^2) \\
\epsilon_{it} &\sim iid N(\mu_{\epsilon},\sigma_{\epsilon}^2)
\end{align}

We assume homoskedasticity in all variances, that is $\sigma_{i \mu}^2=\sigma_{\mu}^2, \sigma_{i \xi}^2=\sigma_{\xi}^2$ and $\sigma_{i \epsilon}^2=\sigma_{\epsilon}^2$. The variance of the random component $e_{it}$ can be defined in two ways. First, it can be fixed so that $V(e_{it}) = \sigma_e^2$, where $\sigma_e^2$ is predefined. Then $\epsilon_{it}$ in $e_{it}$ is generated as:

\begin{align}
\sigma_{\epsilon} &= \sqrt{\sigma_{e}^2 - \gamma_3^2 \sigma_{\mu}^2} \label{eq:FixEps}
\end{align}

Alternatively the variance of $\epsilon_{it}$ can be predefined and then the variance of $e_{it}$ becomes:

\begin{align}
V(e_{it}) = \gamma_3^2 V(\mu_i) + V(\epsilon_{it}) \label{eq:FixE}.
\end{align}

There are two ways to control the within variance of $\mathbf{x}_i$. The first is to specify the standard deviation\footnote{Let define $V(\mathbf{x}_i)$ the (empirical) variance of the realized draw of $\mathbf{x}_i$, while $\sigma_x^2$ is the theoretical variance.} of the random term $\xi_{it}$ so that $V(\mathbf{x}_i) = \sigma_x^2$:

\begin{align}
\sigma_{\xi} &= \sqrt{\left(1-\vartheta^2\right) \sigma_x^2 - \gamma_1^2 \sigma_{\mu}^2 - \gamma_2^2 \sigma_{\epsilon}^2} \label{eq:varXifix}
\end{align}

In this case the within variance of the generated $\mathbf{x}_i=(x_{i1},...,x_{iT})$ will be $\sigma_x^2$. The second way is to specify the standard deviation directly and not fix it. Then the variance of the generated $\xi$ will be $\sigma^2_\xi$ and the variance of the generated $\mathbf{x}_i$ will be:

\begin{align}
V(\mathbf{x}_i) = \frac{1}{1-\vartheta^2} \left(\gamma_1^2V(\mu_i) + \gamma_2^2 V(\epsilon_i) + V(\xi)\right) \label{eq:varXfix}
\end{align}

The means $\mu_{\mu}, \mu_{\xi}$ and $\mu_{\epsilon}$ are all set to zero or can be specified. Within the program it is ensured that $\sigma_{\xi}$ remains strictly positive, which rules out certain combinations of $\gamma_1,\gamma_2,\vartheta$.

\begin{table}%
\center
\scriptsize 
\caption{Variances for \(x\) and \(e\) based on \(x_{it} = \gamma_1 \mu_i + \vartheta x_{it-1} + \gamma_2 \epsilon_{it} + \xi_{it}\) and \(e_{it} = \gamma_3 \mu_i + \epsilon_{it}\).}
\label{tab:variances} 
\begin{tabular}{l l l}\hline\hline
\toprule
\multicolumn{3}{l}{Error component \(e_{it} = \gamma_3 \mu_i + \epsilon_{it}\)}\\ \hline
\(V(e)\) is fixed& \(V(e)=\sigma_e^2 (=1)\) & \(\sigma_\epsilon^2=\sigma_{e}^2 - \gamma_3^2 \sigma_{\mu}^2\); Eq. \eqref{eq:FixEps} \\ 
\(V(\epsilon)\) is fixed& \(V(e)=\gamma_3^2 V(\mu_i) + V(\epsilon_{it})\); Eq. \eqref{eq:FixE} & \(V(\epsilon)=\sigma_\epsilon^2 (=1)\)\\ \hline\hline
\multicolumn{3}{l}{Error component \(x_{it} = \gamma_1 \mu_i + \vartheta x_{it-1} + \gamma_2 \epsilon_{it} + \xi_{it}\)}\\ \hline
\(V(x)\) is fixed& \(V(x) = \sigma_x^2(=1)\) & \(\sigma_\xi^2 = \left(1-\vartheta^2\right) \sigma_x^2 - \gamma_1^2 \sigma_{\mu}^2 - \gamma_2^2 \sigma_{\epsilon}^2\); Eq. \eqref{eq:varXifix} \\
\(V(\xi)\) is fixed & \(V(x) = \frac{1}{1-\vartheta^2} \left(\gamma_1^2V(\mu_i) + \gamma_2^2 V(\epsilon_i) + V(\xi)\right)\); Eq. \eqref{eq:varXfix} & \(V(\xi) = \sigma_\xi^2(=1)\)\\ \hline \hline
\end{tabular}
\flushleft
{\scriptsize Note: \(\sigma_e^2, \sigma_\epsilon^2,  \sigma_x^2\) and \( \sigma_\xi^2\) are predefined. Default values in parentheses.}
\end{table}

\subsection{Variances}
The bias of the OLS estimator for \(\beta_1\) is: 

\begin{align*}
\hat{\beta}_1 &= \beta_1 +  \sum_{i=1}^N \frac{cov(x_{i1},u_i)}{var(x_{1i})}\\
var(x_{i1}) &= \gamma_1^2V(\mu_i) + \gamma_2^2 V(\epsilon_i) + V(\xi_i) + \vartheta^2 V(x_{i1}) \\
\rightarrow V(x_{i1})&= \frac{1}{1-\vartheta^2} \left(\gamma_1^2V(\mu_i) + \gamma_2^2 V(\epsilon_i) + V(\xi_i)\right) \\
cov(x_1, u)&= cov(\gamma_1 \mu_i + \vartheta x_{it-1} + \gamma_2 \epsilon_{it} + \xi_{it},\mu_i + e_{it}) \\
&= \gamma_1 (1+\gamma_3) V(\mu) + \gamma_2 V(\epsilon)+ \vartheta cov(x_{it-1},\mu_i + e_{it})
\intertext{with forwarding \(x_{it-1}\):}
&= \gamma_1 (1+\gamma_3) V(\mu) + \gamma_2 V(\epsilon)+  \sum_{j=t-1}^0 \vartheta^j \left[\gamma_1 (1+\gamma_3) V(\mu) + \gamma_2 V(\epsilon)\right] \\
&= \frac{1}{1-\vartheta} \left(\gamma_1 (1+\gamma_3) V(\mu) + \gamma_2 V(\epsilon)\right)
\end{align*}

Then the Bias becomes:
\begin{align*}
 Bias = \frac{cov(x_1, u)}{var(x_1)} &= \frac{\frac{1}{1-\vartheta} \left(\gamma_1 (1+\gamma_3) V(\mu) + \gamma_2 V(\epsilon)\right)}{\frac{1}{1-\vartheta^2} \left(\gamma_1^2V(\mu) + \gamma_2^2 V(\epsilon) + V(\xi)\right)}
\intertext{Assume \(V(\mu) = V(\epsilon) = 1\):}
&= \frac{1-\vartheta^2}{1-\vartheta} \frac{(1+\gamma_3)\gamma_1 + \gamma_2}{\gamma_1^2 + \gamma_2^2 + V(\xi)}
\end{align*}

Clearly there is no bias if \(\gamma_1 = \gamma_2 = 0\). We also have to restrict the variance of \(V(x_1)\), respectively \(\gamma_1^2 + \gamma_2^2 + V(\xi)\). Currently we fix this term to 1, so the variance of \(x_1\) remains constant and the bias only enters due to \(\gamma_1\) and \(\gamma_2\). The resulting bias is then:

\begin{align*}
Bias = \frac{cov(x_1, u)}{var(x_1)} &= \frac{1-\vartheta^2}{1-\vartheta} \left(\gamma_1 (1+\gamma_3)+\gamma_2\right)
\end{align*}

The change in the bias due to a change of \(\gamma_2\) is then:
\begin{align*}
\frac{\partial Bias}{\partial\gamma_2} =  \frac{1-\vartheta^2}{1-\vartheta}
\end{align*}

If we do not restrict the variance of \(x_1\), then the bias is:
\begin{align*}
Bias = \frac{1-\vartheta^2}{1-\vartheta} \frac{(1+\gamma_3)\gamma_1 + \gamma_2}{\gamma_1^2 + \gamma_2^2 + V(\xi)}
\end{align*}
and a change in \(\gamma_2\) changes the bias by:
\begin{align*}
\frac{\partial Bias}{\partial\gamma_2} =-\frac{1-\vartheta^2}{1-\vartheta} \frac{2\gamma_2}{\left(\gamma_1^2 + \gamma_2^2 + V(\xi)\right)^2}
\end{align*}

Therefore depending on the setup we either get a positive change of the bias or a negative change in the coefficient due to a change in \(\gamma_2\).

\subsection{Results for persistent PAM model}\label{sec:MCPAM08}
Tables \ref{tab:FIXEPAM08BIGlongi_rho}-\ref{tab:FIXEPAM08BIGmulti_beta} show the specific results obtained by simulating a persistent ARDL(1,0) specification where $\rho=0.8$. They are condensed in the nestedloop plots in Figures \ref{fig:FIXEPAM08BIGnlooprho} and \ref{fig:FIXEPAM08BIGnloopbeta}.

\begin{sidewaystable}
\centering
\caption{MonteCarlo results for $\rho$, ARDL(1,0) with $\rho=0.8$ and $\beta_1=1$}\label{tab:FIXEPAM08BIGlongi_rho} 
\begin{adjustbox}{max width=\textwidth}
{\scriptsize
\begin{tabular}{ll|rrrr|r|rrrrrr}
\toprule
& & \multicolumn{11}{c}{Longitudinal panel $N=1000$, $T=10$. Parameter $\rho$} \\
\midrule
& & RE & FE& CRE1 &CRE2 &GL & CRE-GMM & CRE-GMM1 & CRE-GMM2 &CRE-GMM3 & CRE-GMM4 & CRE-GMM5 \\
\midrule
\multirow{4}*{$\gamma_1=0, \gamma_2=0, \gamma_3=[0, 0.8]$} 
&bias    $\rho$    &       0.131&      -0.088&       0.062&       0.089&       0.012&       0.057&       0.035&       0.048&       0.055&       0.008&       0.011\\
&ese          &       0.002&       0.007&       0.005&       0.005&       0.015&       0.007&       0.015&       0.019&       0.008&       0.012&       0.012\\
&bias    $\rho$   &       0.164&      -0.040&       0.046&       0.077&       0.040&       0.072&       0.055&       0.072&       0.074&       0.015&       0.022\\
&ese          &       0.002&       0.004&       0.004&       0.005&       0.020&       0.005&       0.013&       0.014&       0.007&       0.016&       0.013\\
\midrule
\multirow{4}*{$\gamma_1=0, \gamma_2=0.25, \gamma_3=[0, 0.8]$}
&bias   $\rho$     &       0.103&      -0.098&       0.027&       0.055&       0.010&       0.055&       0.029&       0.045&       0.048&       0.004&       0.007\\
&ese          &       0.003&       0.006&       0.005&       0.005&       0.013&       0.006&       0.015&       0.019&       0.008&       0.011&       0.011\\
&bias  $\rho$      &       0.158&      -0.049&       0.032&       0.062&       0.040&       0.072&       0.055&       0.073&       0.073&       0.014&       0.021\\
&ese          &       0.002&       0.004&       0.004&       0.005&       0.019&       0.005&       0.013&       0.013&       0.008&       0.016&       0.013\\
\midrule
\multirow{4}*{$\gamma_1=0, \gamma_2=0.8, \gamma_3=[0, 0.8]$} 
&bias   $\rho$     &       0.028&      -0.098&      -0.036&      -0.014&       0.004&       0.047&       0.015&       0.034&       0.029&      -0.002&       0.000\\
&ese          &       0.004&       0.004&       0.004&       0.004&       0.011&       0.006&       0.013&       0.018&       0.007&       0.009&       0.009\\
&bias  $\rho$     &       0.131&      -0.060&      -0.001&       0.026&       0.033&       0.071&       0.050&       0.071&       0.069&       0.009&       0.016\\
&ese          &       0.003&       0.003&       0.004&       0.004&       0.019&       0.005&       0.013&       0.015&       0.008&       0.015&       0.013\\
\midrule
\multirow{4}*{$\gamma_1=0.25, \gamma_2=0, \gamma_3=[0, 0.8]$} 
&bias   $\rho$     &       0.111&      -0.088&       0.052&       0.065&       0.016&       0.047&       0.040&       0.041&       0.042&       0.043&       0.042\\
&ese          &       0.002&       0.006&       0.005&       0.004&       0.013&       0.007&       0.008&       0.013&       0.007&       0.007&       0.007\\
&bias   $\rho$     &       0.141&      -0.040&       0.041&       0.064&       0.047&       0.061&       0.055&       0.061&       0.050&       0.057&       0.056\\
&ese          &       0.001&       0.004&       0.004&       0.004&       0.018&       0.005&       0.006&       0.010&       0.006&       0.006&       0.006\\
\midrule
\multirow{4}*{$\gamma_1=0.8, \gamma_2=0, \gamma_3=[0, 0.8]$} 
&bias  $\rho$      &       0.070&      -0.088&       0.033&       0.035&       0.021&       0.029&       0.028&       0.027&       0.028&       0.028&       0.028\\
&ese          &       0.002&       0.006&       0.003&       0.004&       0.012&       0.007&       0.007&       0.010&       0.008&       0.008&       0.008\\
&bias  $\rho$     &       0.101&      -0.040&       0.033&       0.045&       0.053&       0.043&       0.042&       0.042&       0.035&       0.042&       0.042\\
&ese          &       0.001&       0.004&       0.003&       0.003&       0.015&       0.005&       0.005&       0.006&       0.005&       0.005&       0.005\\
\midrule
\multirow{4}*{$\gamma_1=0.25, \gamma_2=0.25, \gamma_3=[0, 0.8]$} 
&bias   $\rho$     &       0.087&      -0.098&       0.019&       0.035&       0.012&       0.046&       0.038&       0.039&       0.038&       0.041&       0.041\\
&ese          &       0.002&       0.006&       0.004&       0.004&       0.013&       0.006&       0.008&       0.014&       0.007&       0.007&       0.007\\
&bias  $\rho$      &       0.135&      -0.049&       0.027&       0.049&       0.045&       0.061&       0.055&       0.061&       0.048&       0.057&       0.056\\
&ese          &       0.001&       0.004&       0.004&       0.004&       0.018&       0.005&       0.006&       0.010&       0.006&       0.006&       0.006\\
\midrule
\multirow{4}*{$\gamma_1=0.25, \gamma_2=0.8, \gamma_3=[0, 0.8]$} 
&bias  $\rho$      &       0.042&      -0.098&      -0.032&      -0.021&       0.006&       0.043&       0.032&       0.035&       0.027&       0.036&       0.037\\
&ese          &       0.002&       0.004&       0.003&       0.004&       0.011&       0.005&       0.007&       0.013&       0.006&       0.006&       0.006\\
&bias  $\rho$     &       0.114&      -0.060&      -0.003&       0.016&       0.036&       0.060&       0.052&       0.059&       0.041&       0.055&       0.055\\
&ese          &       0.002&       0.003&       0.003&       0.004&       0.016&       0.005&       0.006&       0.010&       0.006&       0.005&       0.005\\
\midrule
\multirow{4}*{$\gamma_1=0.8, \gamma_2=0.25, \gamma_3=[0, 0.8]$} 
&bias  $\rho$      &       0.045&      -0.098&       0.004&       0.006&       0.017&       0.029&       0.027&       0.026&       0.027&       0.027&       0.027\\
&ese          &       0.002&       0.006&       0.003&       0.003&       0.012&       0.006&       0.007&       0.010&       0.007&       0.007&       0.007\\
&bias    $\rho$    &       0.093&      -0.049&       0.020&       0.031&       0.048&       0.043&       0.042&       0.042&       0.034&       0.041&       0.041\\
&ese          &       0.001&       0.004&       0.003&       0.003&       0.015&       0.005&       0.005&       0.007&       0.005&       0.005&       0.005\\
\bottomrule
\end{tabular}
}
\end{adjustbox}
\flushleft
{\scriptsize Note: Average bias and empirical standard error reported; 1000 replications for each setting. Estimates are implemented in Stata using \texttt{xtdpdgmm} \citep{Kripfganz2019a}, one-step cluster standard errors. Variances are defined by equations \eqref{eq:FixE} and \eqref{eq:varXfix} in Appendix \ref{sec:MCset}.}
\end{sidewaystable} 


\begin{sidewaystable}
\centering
\caption{MonteCarlo results for $\rho$, ARDL(1,0) with $\rho=0.8$ and $\beta_1=1$}\label{tab:FIXEPAM08BIGmacro_rho} 
\begin{adjustbox}{max width=\textwidth}
{\scriptsize
\begin{tabular}{ll|rrrr|r|rrrrrr}
\toprule
& & \multicolumn{11}{c}{Macro panel $N=25$, $T=40$. Parameter $\rho$} \\
\midrule
& & RE & FE& CRE1 &CRE2 &GL & CRE-GMM & CRE-GMM1 & CRE-GMM2 &CRE-GMM3 & CRE-GMM4 & CRE-GMM5 \\
\midrule
\multirow{4}*{$\gamma_1=0, \gamma_2=0, \gamma_3=[0, 0.8]$} 
&bias  $\rho$      &       0.113&      -0.016&       0.065&       0.057&       0.089&       0.041&       0.041&       0.041&       0.030&       0.030&       0.030\\
&ese          &       0.007&       0.012&       0.016&       0.016&       0.015&       0.019&       0.019&       0.019&       0.017&       0.017&       0.018\\
&bias  $\rho$     &       0.172&      -0.007&       0.077&       0.070&       0.167&       0.054&       0.053&       0.054&       0.033&       0.033&       0.034\\
&ese          &       0.008&       0.008&       0.018&       0.016&       0.009&       0.017&       0.017&       0.017&       0.015&       0.016&       0.016\\
\midrule
\multirow{4}*{$\gamma_1=0, \gamma_2=0.25, \gamma_3=[0, 0.8]$} 
&bias  $\rho$      &       0.108&      -0.034&       0.054&       0.045&       0.085&       0.030&       0.029&       0.030&       0.012&       0.012&       0.012\\
&ese          &       0.008&       0.011&       0.016&       0.016&       0.015&       0.019&       0.019&       0.019&       0.017&       0.017&       0.017\\
&bias  $\rho$     &       0.174&      -0.016&       0.068&       0.063&       0.172&       0.049&       0.049&       0.049&       0.027&       0.027&       0.027\\
&ese          &       0.008&       0.008&       0.020&       0.017&       0.007&       0.017&       0.017&       0.017&       0.015&       0.015&       0.015\\
\midrule
\multirow{4}*{$\gamma_1=0, \gamma_2=0.8, \gamma_3=[0, 0.8]$} 
&bias   $\rho$     &       0.052&      -0.053&       0.010&       0.013&       0.059&       0.007&       0.007&       0.007&      -0.010&      -0.010&      -0.010\\
&ese          &       0.018&       0.008&       0.019&       0.018&       0.013&       0.017&       0.017&       0.017&       0.015&       0.015&       0.016\\
&bias   $\rho$     &       0.125&      -0.029&       0.036&       0.037&       0.141&       0.035&       0.035&       0.036&       0.012&       0.013&       0.013\\
&ese          &       0.016&       0.006&       0.019&       0.018&       0.010&       0.016&       0.016&       0.016&       0.015&       0.015&       0.015\\
\midrule
\multirow{4}*{$\gamma_1=0.25, \gamma_2=0, \gamma_3=[0, 0.8]$} 
&bias   $\rho$    &       0.120&      -0.016&       0.055&       0.052&       0.113&       0.039&       0.039&       0.039&       0.037&       0.038&       0.038\\
&ese          &       0.005&       0.012&       0.013&       0.013&       0.009&       0.017&       0.017&       0.018&       0.016&       0.017&       0.018\\
&bias   $\rho$     &       0.143&      -0.007&       0.061&       0.060&       0.141&       0.047&       0.046&       0.046&       0.035&       0.040&       0.041\\
&ese          &       0.005&       0.008&       0.015&       0.014&       0.007&       0.015&       0.015&       0.015&       0.014&       0.015&       0.015\\
\midrule
\multirow{4}*{$\gamma_1=0.8, \gamma_2=0, \gamma_3=[0, 0.8]$} 
&bias   $\rho$     &       0.069&      -0.016&       0.030&       0.029&       0.068&       0.022&       0.022&       0.022&       0.025&       0.023&       0.021\\
&ese          &       0.005&       0.012&       0.010&       0.011&       0.009&       0.016&       0.016&       0.016&       0.014&       0.016&       0.016\\
&bias   $\rho$     &       0.104&      -0.007&       0.043&       0.043&       0.106&       0.034&       0.034&       0.034&       0.032&       0.033&       0.032\\
&ese          &       0.003&       0.008&       0.011&       0.010&       0.005&       0.013&       0.013&       0.013&       0.011&       0.013&       0.013\\
\midrule
\multirow{4}*{$\gamma_1=0.25, \gamma_2=0.25, \gamma_3=[0, 0.8]$} 
&bias  $\rho$      &       0.092&      -0.034&       0.034&       0.031&       0.084&       0.021&       0.021&       0.021&       0.017&       0.019&       0.019\\
&ese          &       0.006&       0.011&       0.013&       0.013&       0.011&       0.016&       0.017&       0.017&       0.015&       0.016&       0.016\\
&bias     $\rho$   &       0.148&      -0.016&       0.053&       0.052&       0.150&       0.041&       0.041&       0.041&       0.029&       0.037&       0.038\\
&ese          &       0.004&       0.008&       0.016&       0.014&       0.005&       0.015&       0.015&       0.015&       0.014&       0.015&       0.015\\
\midrule
\multirow{4}*{$\gamma_1=0.25, \gamma_2=0.8, \gamma_3=[0, 0.8]$} 
&bias  $\rho$      &       0.052&      -0.053&      -0.002&       0.002&       0.057&      -0.000&      -0.000&      -0.000&      -0.000&      -0.001&      -0.001\\
&ese          &       0.011&       0.008&       0.014&       0.013&       0.010&       0.014&       0.014&       0.014&       0.013&       0.013&       0.014\\
&bias   $\rho$     &       0.129&      -0.029&       0.026&       0.031&       0.139&       0.030&       0.030&       0.031&       0.019&       0.025&       0.026\\
&ese          &       0.007&       0.006&       0.016&       0.016&       0.006&       0.014&       0.014&       0.015&       0.013&       0.014&       0.014\\
\midrule
\multirow{4}*{$\gamma_1=0.8, \gamma_2=0.25, \gamma_3=[0, 0.8]$} 
&bias   $\rho$     &       0.046&      -0.034&       0.008&       0.007&       0.052&       0.005&       0.005&       0.005&       0.007&       0.005&       0.003\\
&ese          &       0.005&       0.011&       0.010&       0.010&       0.008&       0.015&       0.015&       0.015&       0.013&       0.015&       0.016\\
&bias  $\rho$     &       0.097&      -0.016&       0.033&       0.034&       0.102&       0.027&       0.027&       0.027&       0.026&       0.026&       0.025\\
&ese          &       0.003&       0.008&       0.011&       0.010&       0.005&       0.012&       0.013&       0.013&       0.011&       0.013&       0.013\\
\bottomrule
\end{tabular}
}
\end{adjustbox}
\flushleft
{\scriptsize Note: Average bias and empirical standard error reported; 1000 replications for each setting. Estimates are implemented in Stata using \texttt{xtdpdgmm} \citep{Kripfganz2019a}, one-step cluster standard errors. Variances are defined by equations \eqref{eq:FixE} and \eqref{eq:varXfix} in Appendix \ref{sec:MCset}.}
\end{sidewaystable} 

\begin{sidewaystable}
\centering
\caption{MonteCarlo results for $\rho$, ARDL(1,0) with $\rho=0.8$ and $\beta_1=1$}\label{tab:FIXEPAM08BIGmulti_rho} 
\begin{adjustbox}{max width=\textwidth}
{\scriptsize
\begin{tabular}{ll|rrrr|r|rrrrrr}
\toprule
& & \multicolumn{11}{c}{Multilevel panel $N=100$, $T=20$. Parameter $\rho$} \\
\midrule
& & RE & FE& CRE1 &CRE2 &GL & CRE-GMM & CRE-GMM1 & CRE-GMM2 &CRE-GMM3 & CRE-GMM4 & CRE-GMM5 \\
\midrule
\multirow{4}*{$\gamma_1=0, \gamma_2=0, \gamma_3=[0, 0.8]$} 
&bias $\rho$      &       0.137&      -0.035&       0.078&       0.077&       0.071&       0.043&       0.041&       0.043&       0.026&       0.024&       0.023\\
&ese          &       0.006&       0.011&       0.010&       0.010&       0.018&       0.015&       0.016&       0.017&       0.015&       0.016&       0.016\\
&bias  $\rho$      &       0.163&      -0.016&       0.060&       0.075&       0.136&       0.054&       0.053&       0.056&       0.023&       0.023&       0.025\\
&ese          &       0.005&       0.007&       0.011&       0.011&       0.014&       0.012&       0.012&       0.014&       0.013&       0.013&       0.013\\
\midrule
\multirow{4}*{$\gamma_1=0, \gamma_2=0.25, \gamma_3=[0, 0.8]$} 
&bias   $\rho$     &       0.115&      -0.052&       0.052&       0.060&       0.061&       0.036&       0.034&       0.037&       0.012&       0.010&       0.011\\
&ese          &       0.008&       0.010&       0.013&       0.011&       0.019&       0.014&       0.016&       0.017&       0.015&       0.015&       0.015\\
&bias   $\rho$    &       0.159&      -0.025&       0.048&       0.063&       0.137&       0.052&       0.051&       0.054&       0.018&       0.018&       0.021\\
&ese          &       0.005&       0.007&       0.011&       0.011&       0.014&       0.012&       0.012&       0.014&       0.013&       0.013&       0.014\\
\midrule
\multirow{4}*{$\gamma_1=0, \gamma_2=0.8, \gamma_3=[0, 0.8]$} 
&bias $\rho$       &       0.014&      -0.066&      -0.018&      -0.004&       0.018&       0.015&       0.012&       0.016&      -0.011&      -0.011&      -0.007\\
&ese          &       0.009&       0.007&       0.009&       0.010&       0.016&       0.012&       0.013&       0.014&       0.012&       0.013&       0.012\\
&bias   $\rho$    &       0.124&      -0.038&       0.017&       0.031&       0.119&       0.046&       0.045&       0.048&       0.008&       0.008&       0.013\\
&ese          &       0.008&       0.005&       0.010&       0.010&       0.016&       0.011&       0.012&       0.014&       0.012&       0.013&       0.013\\
\midrule
\multirow{4}*{$\gamma_1=0.25, \gamma_2=0, \gamma_3=[0, 0.8]$} 
&bias   $\rho$     &       0.112&      -0.035&       0.057&       0.058&       0.074&       0.036&       0.035&       0.035&       0.033&       0.036&       0.035\\
&ese          &       0.004&       0.011&       0.009&       0.008&       0.015&       0.014&       0.014&       0.015&       0.013&       0.015&       0.015\\
&bias $\rho$      &       0.142&      -0.016&       0.050&       0.063&       0.128&       0.047&       0.047&       0.048&       0.029&       0.046&       0.044\\
&ese          &       0.003&       0.007&       0.009&       0.009&       0.011&       0.011&       0.011&       0.012&       0.011&       0.012&       0.012\\
\midrule
\multirow{4}*{$\gamma_1=0.8, \gamma_2=0, \gamma_3=[0, 0.8]$} 
&bias   $\rho$     &       0.070&      -0.036&       0.033&       0.033&       0.060&       0.022&       0.022&       0.022&       0.027&       0.022&       0.021\\
&ese          &       0.003&       0.010&       0.007&       0.007&       0.011&       0.013&       0.013&       0.014&       0.012&       0.015&       0.015\\
&bias  $\rho$     &       0.102&      -0.016&       0.038&       0.046&       0.100&       0.034&       0.035&       0.035&       0.031&       0.034&       0.034\\
&ese          &       0.002&       0.007&       0.007&       0.007&       0.008&       0.010&       0.010&       0.010&       0.009&       0.011&       0.011\\
\midrule
\multirow{4}*{$\gamma_1=0.25, \gamma_2=0.25, \gamma_3=[0, 0.8]$} 
&bias  $\rho$      &       0.091&      -0.052&       0.033&       0.037&       0.059&       0.028&       0.027&       0.028&       0.023&       0.026&       0.026\\
&ese          &       0.005&       0.010&       0.009&       0.009&       0.015&       0.013&       0.013&       0.014&       0.012&       0.014&       0.014\\
&bias  $\rho$      &       0.137&      -0.025&       0.038&       0.051&       0.128&       0.044&       0.044&       0.046&       0.025&       0.043&       0.041\\
&ese          &       0.003&       0.007&       0.009&       0.010&       0.012&       0.011&       0.011&       0.012&       0.011&       0.012&       0.012\\
\midrule
\multirow{4}*{$\gamma_1=0.25, \gamma_2=0.8, \gamma_3=[0, 0.8]$} 
&bias  $\rho$      &       0.050&      -0.066&      -0.014&      -0.004&       0.047&       0.018&       0.017&       0.019&       0.014&       0.017&       0.017\\
&ese          &       0.006&       0.007&       0.008&       0.009&       0.014&       0.011&       0.011&       0.012&       0.010&       0.012&       0.012\\
&bias  $\rho$      &       0.111&      -0.038&       0.010&       0.022&       0.111&       0.039&       0.039&       0.041&       0.019&       0.036&       0.035\\
&ese          &       0.005&       0.006&       0.008&       0.009&       0.013&       0.010&       0.010&       0.012&       0.010&       0.011&       0.011\\
\midrule
\multirow{4}*{$\gamma_1=0.8, \gamma_2=0.25, \gamma_3=[0, 0.8]$} 
&bias    $\rho$    &       0.045&      -0.052&       0.009&       0.009&       0.049&       0.014&       0.014&       0.014&       0.018&       0.012&       0.011\\
&ese          &       0.003&       0.010&       0.006&       0.007&       0.011&       0.012&       0.012&       0.013&       0.011&       0.013&       0.013\\
&bias   $\rho$     &       0.093&      -0.025&       0.026&       0.034&       0.097&       0.031&       0.031&       0.031&       0.028&       0.030&       0.030\\
&ese          &       0.002&       0.007&       0.007&       0.007&       0.008&       0.009&       0.010&       0.010&       0.009&       0.011&       0.011\\
\bottomrule
\end{tabular}
}
\end{adjustbox}
\flushleft
{\scriptsize Note: Average bias and empirical standard error reported; 1000 replications for each setting. Estimates are implemented in Stata using \texttt{xtdpdgmm} \citep{Kripfganz2019a}, one-step cluster standard errors. Variances are defined by equations \eqref{eq:FixE} and \eqref{eq:varXfix} in Appendix \ref{sec:MCset}.}
\end{sidewaystable} 

\begin{sidewaystable}
\centering
\caption{MonteCarlo results for $\beta$, ARDL(1,0) with $\rho=0.8$ and $\beta_1=1$}\label{tab:FIXEPAM08BIGlongi_beta} 
\begin{adjustbox}{max width=\textwidth}
{\scriptsize
\begin{tabular}{ll|rrrr|r|rrrrrr}
\toprule
& & \multicolumn{11}{c}{Longitudinal panel $N=1000$, $T=10$. Parameter $\beta_1$} \\
\midrule
& & RE & FE& CRE1 &CRE2 &GL & CRE-GMM & CRE-GMM1 & CRE-GMM2 &CRE-GMM3 & CRE-GMM4 & CRE-GMM5 \\
\midrule
\multirow{4}*{$\gamma_1=0, \gamma_2=0, \gamma_3=[0, 0.8]$} 
&bias  $\beta$     &      -0.079&      -0.004&      -0.043&      -0.067&      -0.019&      -0.021&      -0.013&       0.024&      -0.169&      -0.027&      -0.023\\
&ese          &       0.013&       0.012&       0.012&       0.012&       0.040&       0.034&       0.035&       0.041&       0.034&       0.034&       0.033\\
&bias   $\beta$     &      -0.048&      -0.002&      -0.017&      -0.035&      -0.065&      -0.025&      -0.018&       0.038&      -0.226&      -0.050&      -0.042\\
&ese          &       0.008&       0.007&       0.007&       0.008&       0.046&       0.024&       0.027&       0.027&       0.032&       0.034&       0.029\\
\midrule
\multirow{4}*{$\gamma_1=0, \gamma_2=0.25, \gamma_3=[0, 0.8]$}
&bias  $\beta$     &       0.124&       0.199&       0.165&       0.139&      -0.015&      -0.030&      -0.014&       0.022&      -0.171&      -0.014&      -0.014\\
&ese          &       0.012&       0.011&       0.012&       0.013&       0.040&       0.035&       0.036&       0.039&       0.038&       0.035&       0.034\\
&bias   $\beta$     &       0.027&       0.073&       0.060&       0.043&      -0.071&      -0.031&      -0.022&       0.035&      -0.235&      -0.047&      -0.042\\
&ese          &       0.008&       0.007&       0.007&       0.008&       0.049&       0.025&       0.026&       0.027&       0.034&       0.035&       0.030\\
\midrule
\multirow{4}*{$\gamma_1=0, \gamma_2=0.8, \gamma_3=[0, 0.8]$} 
&bias  $\beta$     &       0.389&       0.419&       0.411&       0.400&      -0.001&      -0.028&      -0.001&       0.020&      -0.115&       0.011&       0.008\\
&ese          &       0.008&       0.007&       0.007&       0.008&       0.032&       0.030&       0.031&       0.031&       0.036&       0.029&       0.028\\
&bias   $\beta$    &       0.166&       0.199&       0.192&       0.182&      -0.064&      -0.038&      -0.024&       0.029&      -0.245&      -0.031&      -0.032\\
&ese          &       0.007&       0.006&       0.006&       0.006&       0.046&       0.024&       0.028&       0.026&       0.039&       0.034&       0.030\\
\midrule
\multirow{4}*{$\gamma_1=0.25, \gamma_2=0, \gamma_3=[0, 0.8]$} 
&bias  $\beta$      &      -0.019&      -0.005&       0.003&      -0.007&      -0.018&       0.005&       0.025&       0.028&      -0.035&       0.015&       0.014\\
&ese          &       0.012&       0.012&       0.011&       0.011&       0.039&       0.033&       0.034&       0.037&       0.031&       0.032&       0.032\\
&bias  $\beta$     &      -0.017&      -0.002&       0.001&      -0.010&      -0.063&      -0.004&       0.025&       0.036&      -0.081&       0.013&       0.011\\
&ese          &       0.008&       0.007&       0.007&       0.007&       0.044&       0.022&       0.024&       0.025&       0.024&       0.021&       0.021\\
\midrule
\multirow{4}*{$\gamma_1=0.8, \gamma_2=0, \gamma_3=[0, 0.8]$} 
&bias   $\beta$     &       0.018&      -0.005&       0.025&       0.023&      -0.007&       0.016&       0.024&       0.024&       0.018&       0.020&       0.022\\
&ese          &       0.011&       0.012&       0.011&       0.011&       0.037&       0.031&       0.034&       0.035&       0.031&       0.034&       0.035\\
&bias  $\beta$     &       0.007&      -0.002&       0.016&       0.011&      -0.050&       0.011&       0.029&       0.030&      -0.003&       0.023&       0.026\\
&ese          &       0.007&       0.007&       0.007&       0.007&       0.038&       0.020&       0.021&       0.022&       0.021&       0.021&       0.022\\
\midrule
\multirow{4}*{$\gamma_1=0.25, \gamma_2=0.25, \gamma_3=[0, 0.8]$} 
&bias   $\beta$     &       0.184&       0.200&       0.210&       0.197&      -0.010&       0.002&       0.024&       0.028&      -0.028&       0.013&       0.011\\
&ese          &       0.011&       0.011&       0.011&       0.011&       0.038&       0.033&       0.034&       0.036&       0.033&       0.032&       0.032\\
&bias   $\beta$    &       0.060&       0.073&       0.078&       0.068&      -0.065&      -0.007&       0.024&       0.035&      -0.077&       0.011&       0.009\\
&ese          &       0.008&       0.007&       0.007&       0.007&       0.047&       0.023&       0.024&       0.024&       0.026&       0.021&       0.021\\
\midrule
\multirow{4}*{$\gamma_1=0.25, \gamma_2=0.8, \gamma_3=[0, 0.8]$} 
&bias    $\beta$    &       0.424&       0.419&       0.437&       0.434&      -0.001&       0.001&       0.024&       0.026&       0.007&       0.009&       0.008\\
&ese          &       0.008&       0.007&       0.007&       0.007&       0.033&       0.028&       0.029&       0.029&       0.028&       0.027&       0.027\\
&bias  $\beta$      &       0.196&       0.199&       0.206&       0.202&      -0.057&      -0.011&       0.022&       0.032&      -0.056&       0.006&       0.004\\
&ese          &       0.007&       0.006&       0.006&       0.006&       0.044&       0.022&       0.023&       0.023&       0.026&       0.020&       0.020\\
\midrule
\multirow{4}*{$\gamma_1=0.8, \gamma_2=0.25, \gamma_3=[0, 0.8]$} 
&bias   $\beta$     &       0.216&       0.199&       0.225&       0.223&      -0.003&       0.018&       0.027&       0.027&       0.024&       0.023&       0.026\\
&ese          &       0.011&       0.011&       0.010&       0.010&       0.037&       0.031&       0.033&       0.034&       0.032&       0.033&       0.034\\
&bias   $\beta$     &       0.085&       0.073&       0.093&       0.089&      -0.051&       0.011&       0.029&       0.030&       0.002&       0.023&       0.026\\
&ese          &       0.007&       0.007&       0.007&       0.007&       0.038&       0.020&       0.021&       0.022&       0.021&       0.021&       0.021\\
\bottomrule
\end{tabular}
}
\end{adjustbox}
\flushleft
{\scriptsize Note: Average bias and empirical standard error reported; 1000 replications for each setting. Estimates are implemented in Stata using \texttt{xtdpdgmm} \citep{Kripfganz2019a}, one-step cluster standard errors. Variances are defined by equations \eqref{eq:FixE} and \eqref{eq:varXfix} in Appendix \ref{sec:MCset}.}
\end{sidewaystable} 

\newpage
\begin{sidewaystable}
\centering
\caption{MonteCarlo results for $\beta$, ARDL(1,0) with $\rho=0.8$ and $\beta_1=1$}\label{tab:FIXEPAM08BIGmacro_beta} 
\begin{adjustbox}{max width=\textwidth}
{\scriptsize
\begin{tabular}{ll|rrrr|r|rrrrrr}
\toprule
& & \multicolumn{11}{c}{Macro panel $N=25$, $T=40$. Parameter $\beta_1$} \\
\midrule
& & RE & FE& CRE1 &CRE2 &GL & CRE-GMM & CRE-GMM1 & CRE-GMM2 &CRE-GMM3 & CRE-GMM4 & CRE-GMM5 \\
\midrule
\multirow{4}*{$\gamma_1=0, \gamma_2=0, \gamma_3=[0, 0.8]$} 
&bias   $\beta$     &      -0.091&       0.009&      -0.052&      -0.044&      -0.100&      -0.044&      -0.043&      -0.042&      -0.058&      -0.057&      -0.056\\
&ese          &       0.036&       0.030&       0.035&       0.034&       0.059&       0.056&       0.057&       0.057&       0.058&       0.058&       0.058\\
&bias  $\beta$      &      -0.127&       0.004&      -0.058&      -0.054&      -0.188&      -0.057&      -0.057&      -0.055&      -0.063&      -0.062&      -0.062\\
&ese          &       0.029&       0.018&       0.027&       0.025&       0.044&       0.039&       0.039&       0.039&       0.042&       0.042&       0.042\\
\midrule
\multirow{4}*{$\gamma_1=0, \gamma_2=0.25, \gamma_3=[0, 0.8]$} 
&bias  $\beta$      &       0.075&       0.202&       0.125&       0.136&      -0.003&       0.075&       0.076&       0.077&       0.067&       0.068&       0.069\\
&ese          &       0.037&       0.029&       0.036&       0.035&       0.060&       0.057&       0.057&       0.057&       0.059&       0.059&       0.059\\
&bias   $\beta$     &      -0.066&       0.077&       0.012&       0.015&      -0.168&      -0.015&      -0.015&      -0.013&      -0.018&      -0.018&      -0.017\\
&ese          &       0.030&       0.018&       0.028&       0.026&       0.046&       0.040&       0.040&       0.040&       0.043&       0.043&       0.043\\
\midrule
\multirow{4}*{$\gamma_1=0, \gamma_2=0.8, \gamma_3=[0, 0.8]$} 
&bias $\beta$       &       0.320&       0.415&       0.359&       0.356&       0.128&       0.220&       0.221&       0.222&       0.227&       0.227&       0.227\\
&ese          &       0.034&       0.021&       0.031&       0.030&       0.050&       0.046&       0.046&       0.046&       0.046&       0.046&       0.046\\
&bias  $\beta$      &       0.080&       0.199&       0.149&       0.146&      -0.088&       0.063&       0.064&       0.065&       0.067&       0.068&       0.067\\
&ese          &       0.027&       0.015&       0.024&       0.025&       0.045&       0.038&       0.038&       0.038&       0.040&       0.040&       0.040\\
\midrule
\multirow{4}*{$\gamma_1=0.25, \gamma_2=0, \gamma_3=[0, 0.8]$} 
&bias  $\beta$     &      -0.030&       0.009&      -0.010&      -0.010&      -0.059&      -0.017&      -0.016&      -0.016&      -0.020&      -0.019&      -0.018\\
&ese          &       0.033&       0.030&       0.031&       0.031&       0.057&       0.055&       0.055&       0.055&       0.055&       0.055&       0.055\\
&bias  $\beta$      &      -0.066&       0.004&      -0.027&      -0.028&      -0.117&      -0.034&      -0.034&      -0.033&      -0.034&      -0.034&      -0.034\\
&ese          &       0.024&       0.018&       0.022&       0.021&       0.039&       0.036&       0.036&       0.036&       0.037&       0.036&       0.037\\
\midrule
\multirow{4}*{$\gamma_1=0.8, \gamma_2=0, \gamma_3=[0, 0.8]$} 
&bias  $\beta$      &       0.019&       0.008&       0.013&       0.013&       0.005&       0.006&       0.006&       0.006&       0.007&       0.006&       0.006\\
&ese          &       0.032&       0.031&       0.031&       0.031&       0.056&       0.054&       0.054&       0.054&       0.054&       0.054&       0.054\\
&bias   $\beta$     &      -0.013&       0.003&      -0.002&      -0.003&      -0.047&      -0.010&      -0.009&      -0.009&      -0.009&      -0.009&      -0.009\\
&ese          &       0.021&       0.019&       0.020&       0.020&       0.037&       0.034&       0.034&       0.034&       0.034&       0.034&       0.034\\
\midrule
\multirow{4}*{$\gamma_1=0.25, \gamma_2=0.25, \gamma_3=[0, 0.8]$} 
&bias $\beta$       &       0.161&       0.202&       0.183&       0.183&       0.068&       0.114&       0.115&       0.116&       0.115&       0.115&       0.116\\
&ese          &       0.033&       0.029&       0.031&       0.031&       0.057&       0.054&       0.055&       0.055&       0.053&       0.053&       0.054\\
&bias  $\beta$      &      -0.000&       0.077&       0.046&       0.043&      -0.092&       0.011&       0.012&       0.013&       0.014&       0.014&       0.013\\
&ese          &       0.024&       0.018&       0.022&       0.022&       0.041&       0.036&       0.036&       0.036&       0.037&       0.037&       0.037\\
\midrule
\multirow{4}*{$\gamma_1=0.25, \gamma_2=0.8, \gamma_3=[0, 0.8]$} 
&bias  $\beta$      &       0.383&       0.416&       0.403&       0.400&       0.201&       0.263&       0.263&       0.263&       0.263&       0.263&       0.263\\
&ese          &       0.025&       0.020&       0.022&       0.023&       0.045&       0.041&       0.041&       0.041&       0.041&       0.041&       0.041\\
&bias    $\beta$    &       0.126&       0.198&       0.175&       0.170&      -0.029&       0.090&       0.091&       0.091&       0.096&       0.095&       0.094\\
&ese          &       0.023&       0.015&       0.019&       0.020&       0.040&       0.034&       0.034&       0.035&       0.034&       0.033&       0.034\\
\midrule
\multirow{4}*{$\gamma_1=0.8, \gamma_2=0.25, \gamma_3=[0, 0.8]$} 
&bias   $\beta$    &       0.215&       0.202&       0.209&       0.209&       0.138&       0.142&       0.142&       0.142&       0.144&       0.144&       0.144\\
&ese          &       0.030&       0.029&       0.030&       0.030&       0.054&       0.052&       0.052&       0.053&       0.052&       0.052&       0.052\\
&bias   $\beta$     &       0.061&       0.076&       0.072&       0.070&      -0.003&       0.039&       0.039&       0.039&       0.039&       0.039&       0.040\\
&ese          &       0.021&       0.018&       0.019&       0.019&       0.037&       0.034&       0.034&       0.034&       0.034&       0.035&       0.035\\
\bottomrule
\end{tabular}
}
\end{adjustbox}
\flushleft
{\scriptsize Note: Average bias and empirical standard error reported; 1000 replications for each setting. Estimates are implemented in Stata using \texttt{xtdpdgmm} \citep{Kripfganz2019a}, one-step cluster standard errors. Variances are defined by equations \eqref{eq:FixE} and \eqref{eq:varXfix} in Appendix \ref{sec:MCset}.}
\end{sidewaystable} 

\newpage
\begin{sidewaystable}
\centering
\caption{MonteCarlo results for $\beta$, ARDL(1,0) with $\rho=0.8$ and $\beta_1=1$}\label{tab:FIXEPAM08BIGmulti_beta} 
\begin{adjustbox}{max width=\textwidth}
{\scriptsize
\begin{tabular}{ll|rrrr|r|rrrrrr}
\toprule
& & \multicolumn{11}{c}{Multilevel panel $N=100$, $T=20$. Parameter $\beta_1$} \\
\midrule
& & RE & FE& CRE1 &CRE2 &GL & CRE-GMM & CRE-GMM1 & CRE-GMM2 &CRE-GMM3 & CRE-GMM4 & CRE-GMM5 \\
\midrule
\multirow{4}*{$\gamma_1=0, \gamma_2=0, \gamma_3=[0, 0.8]$} 
&bias  $\beta$     &      -0.108&       0.009&      -0.064&      -0.061&      -0.101&      -0.033&      -0.031&      -0.027&      -0.069&      -0.063&      -0.053\\
&ese          &       0.028&       0.024&       0.028&       0.027&       0.064&       0.057&       0.057&       0.059&       0.062&       0.061&       0.060\\
&bias  $\beta$      &      -0.084&       0.004&      -0.034&      -0.049&      -0.193&      -0.041&      -0.039&      -0.034&      -0.059&      -0.056&      -0.051\\
&ese          &       0.018&       0.014&       0.017&       0.019&       0.051&       0.038&       0.038&       0.039&       0.044&       0.043&       0.041\\
\midrule
\multirow{4}*{$\gamma_1=0, \gamma_2=0.25, \gamma_3=[0, 0.8]$} 
&bias  $\beta$     &       0.085&       0.205&       0.132&       0.122&      -0.065&       0.015&       0.017&       0.023&      -0.004&       0.002&       0.007\\
&ese          &       0.029&       0.023&       0.029&       0.027&       0.069&       0.059&       0.060&       0.061&       0.064&       0.064&       0.062\\
&bias   $\beta$     &      -0.015&       0.077&       0.041&       0.026&      -0.198&      -0.026&      -0.023&      -0.017&      -0.037&      -0.034&      -0.032\\
&ese          &       0.018&       0.014&       0.017&       0.019&       0.055&       0.040&       0.040&       0.041&       0.046&       0.045&       0.042\\
\midrule
\multirow{4}*{$\gamma_1=0, \gamma_2=0.8, \gamma_3=[0, 0.8]$} 
&bias   $\beta$     &       0.372&       0.417&       0.392&       0.381&       0.046&       0.089&       0.091&       0.095&       0.110&       0.108&       0.102\\
&ese          &       0.018&       0.015&       0.017&       0.019&       0.054&       0.048&       0.048&       0.048&       0.047&       0.047&       0.047\\
&bias   $\beta$     &       0.121&       0.199&       0.174&       0.164&      -0.174&       0.003&       0.005&       0.012&       0.010&       0.011&       0.007\\
&ese          &       0.016&       0.012&       0.014&       0.015&       0.057&       0.039&       0.039&       0.039&       0.043&       0.043&       0.040\\
\midrule
\multirow{4}*{$\gamma_1=0.25, \gamma_2=0, \gamma_3=[0, 0.8]$} 
&bias  $\beta$      &      -0.030&       0.008&      -0.009&      -0.010&      -0.060&      -0.006&      -0.004&      -0.003&      -0.012&      -0.009&      -0.009\\
&ese          &       0.024&       0.024&       0.024&       0.024&       0.061&       0.056&       0.057&       0.058&       0.058&       0.058&       0.058\\
&bias $\beta$      &      -0.041&       0.004&      -0.012&      -0.023&      -0.146&      -0.020&      -0.018&      -0.015&      -0.019&      -0.021&      -0.024\\
&ese          &       0.016&       0.014&       0.016&       0.017&       0.045&       0.036&       0.036&       0.038&       0.037&       0.037&       0.036\\
\midrule
\multirow{4}*{$\gamma_1=0.8, \gamma_2=0, \gamma_3=[0, 0.8]$} 
&bias  $\beta$      &       0.017&       0.008&       0.017&       0.017&      -0.009&       0.008&       0.009&       0.010&       0.012&       0.008&       0.009\\
&ese          &       0.022&       0.024&       0.022&       0.022&       0.056&       0.054&       0.055&       0.056&       0.055&       0.056&       0.057\\
&bias   $\beta$     &      -0.003&       0.003&       0.006&       0.001&      -0.075&      -0.003&      -0.001&       0.000&      -0.001&      -0.002&      -0.002\\
&ese          &       0.015&       0.014&       0.014&       0.015&       0.038&       0.034&       0.034&       0.035&       0.035&       0.035&       0.036\\
\midrule
\multirow{4}*{$\gamma_1=0.25, \gamma_2=0.25, \gamma_3=[0, 0.8]$} 
&bias  $\beta$     &       0.166&       0.205&       0.189&       0.185&      -0.014&       0.048&       0.050&       0.052&       0.048&       0.048&       0.047\\
&ese          &       0.025&       0.023&       0.024&       0.024&       0.061&       0.056&       0.056&       0.056&       0.057&       0.057&       0.057\\
&bias   $\beta$     &       0.032&       0.077&       0.063&       0.053&      -0.144&      -0.003&      -0.000&       0.003&       0.004&      -0.002&      -0.005\\
&ese          &       0.017&       0.014&       0.015&       0.017&       0.047&       0.037&       0.037&       0.038&       0.038&       0.037&       0.037\\
\midrule
\multirow{4}*{$\gamma_1=0.25, \gamma_2=0.8, \gamma_3=[0, 0.8]$} 
&bias  $\beta$      &       0.402&       0.417&       0.418&       0.414&       0.030&       0.113&       0.115&       0.117&       0.117&       0.112&       0.110\\
&ese          &       0.017&       0.015&       0.016&       0.016&       0.054&       0.044&       0.044&       0.044&       0.042&       0.043&       0.043\\
&bias $\beta$       &       0.163&       0.199&       0.192&       0.186&      -0.120&       0.028&       0.031&       0.035&       0.045&       0.031&       0.026\\
&ese          &       0.015&       0.012&       0.013&       0.014&       0.049&       0.035&       0.035&       0.036&       0.034&       0.034&       0.034\\
\midrule
\multirow{4}*{$\gamma_1=0.8, \gamma_2=0.25, \gamma_3=[0, 0.8]$} 
&bias   $\beta$     &       0.214&       0.204&       0.213&       0.213&       0.041&       0.067&       0.068&       0.069&       0.072&       0.072&       0.073\\
&ese          &       0.022&       0.023&       0.022&       0.022&       0.057&       0.054&       0.055&       0.055&       0.056&       0.056&       0.056\\
&bias   $\beta$     &       0.074&       0.077&       0.081&       0.077&      -0.062&       0.018&       0.020&       0.021&       0.021&       0.020&       0.020\\
&ese          &       0.015&       0.014&       0.014&       0.015&       0.040&       0.034&       0.035&       0.035&       0.035&       0.035&       0.035\\
\bottomrule
\end{tabular}
}
\end{adjustbox}
\flushleft
{\scriptsize Note: Average bias and empirical standard error reported; 1000 replications for each setting. Estimates are implemented in Stata using \texttt{xtdpdgmm} \citep{Kripfganz2019a}, one-step cluster standard errors. Variances are defined by equations \eqref{eq:FixE} and \eqref{eq:varXfix} in Appendix \ref{sec:MCset}.}
\end{sidewaystable} 

\clearpage
\subsection{Results for ECM model}\label{sec:MCECM}
\begin{figure}[h!]
\centering
\includegraphics[scale=0.30]{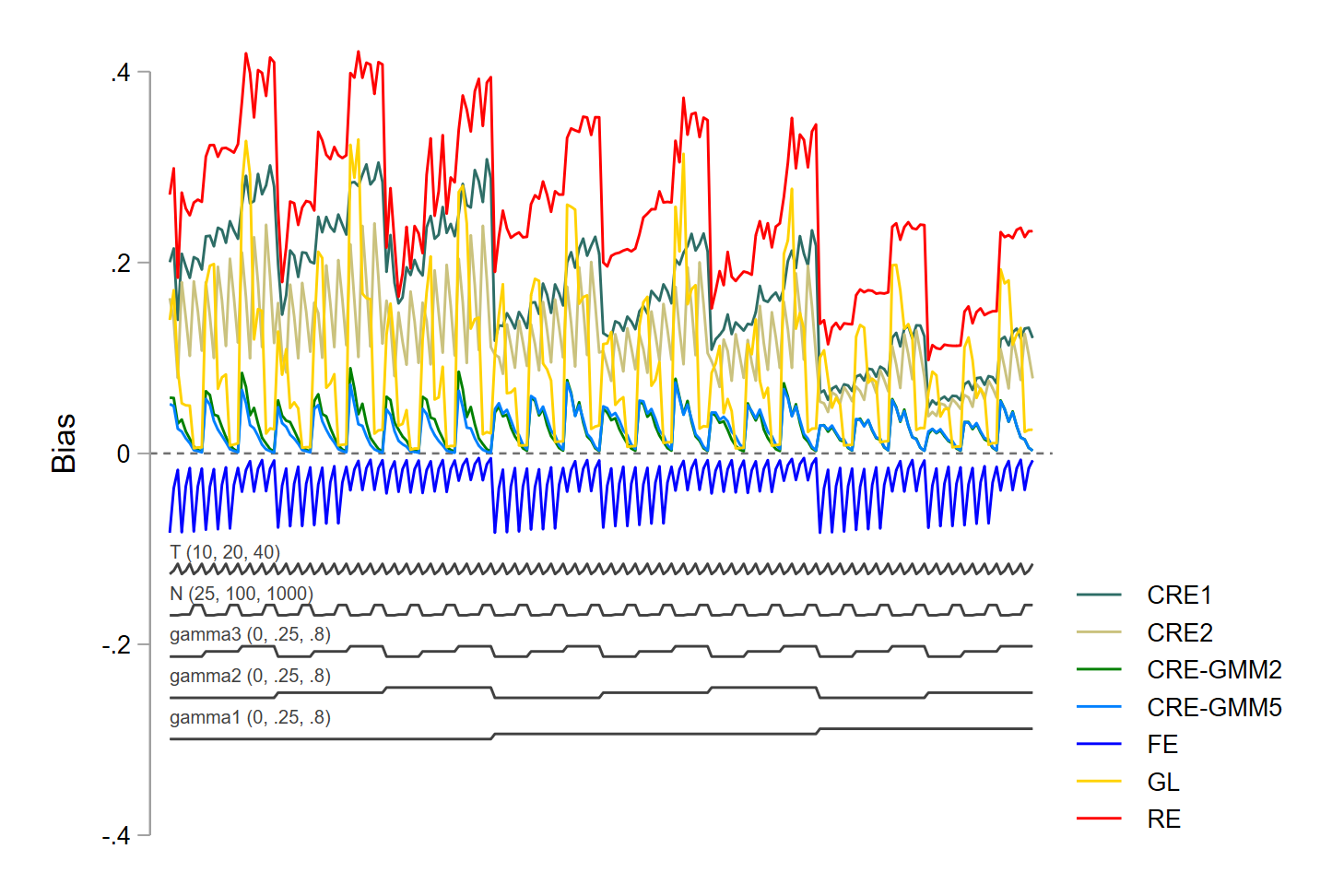}
\caption{\small{Nestedloop Plot for \(\rho\), ECM \\ Bias for $\rho=0.5$ across different specifications. Parameters shown on horizontal axis.}}
\label{fig:FIXEECM05BIG_nloop_L_y}
\end{figure}

\begin{figure}[h!]
\centering
\includegraphics[scale=0.30]{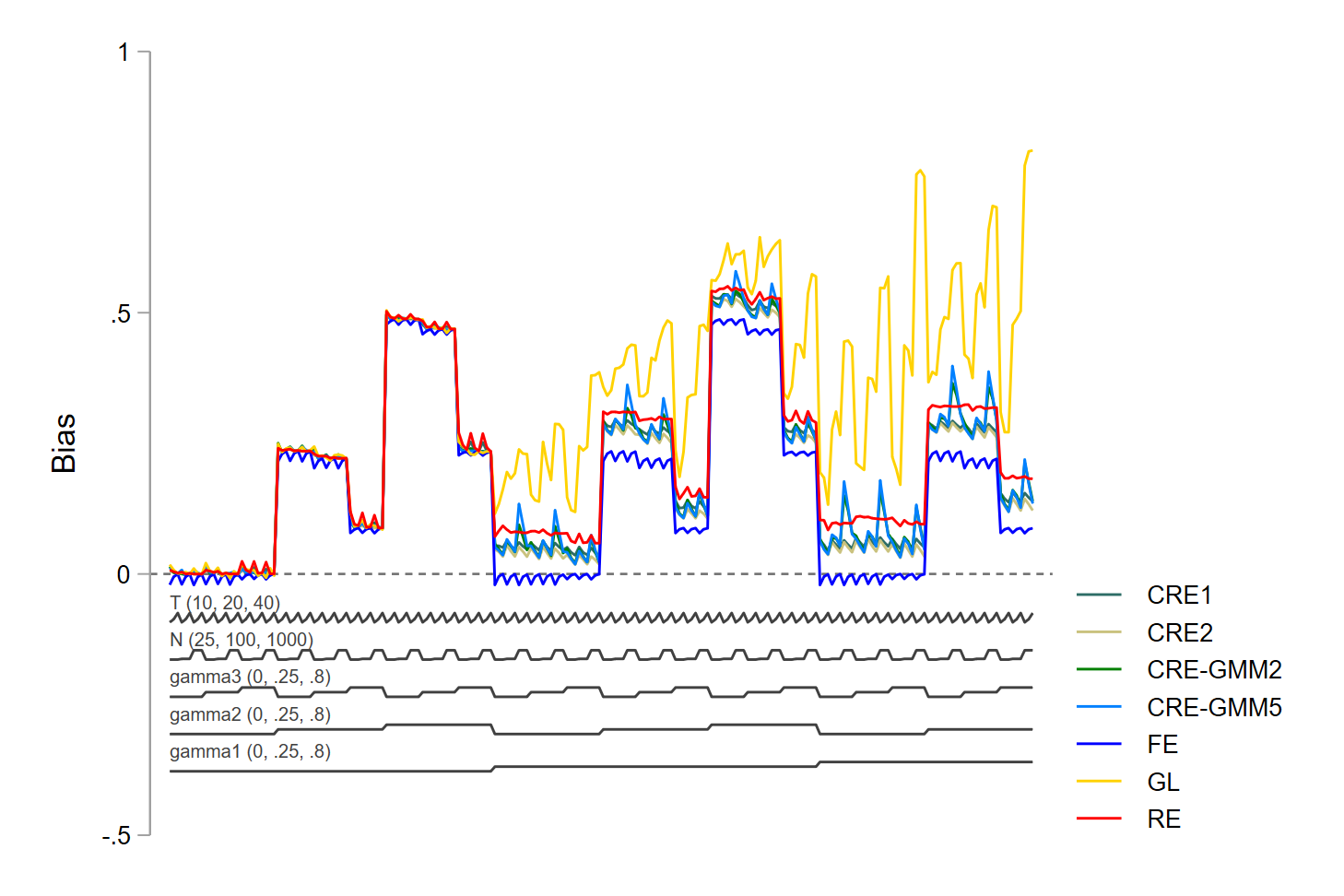}
\caption{\small{Nestedloop Plot for $\beta_1$, ECM \\ Bias for $\beta_1=1$ across different specifications. Parameters shown on horizontal axis.}}
\label{fig:FIXEECM05BIG_nloop_x}
\end{figure}

\begin{figure}[h!]
\centering
\includegraphics[scale=0.30]{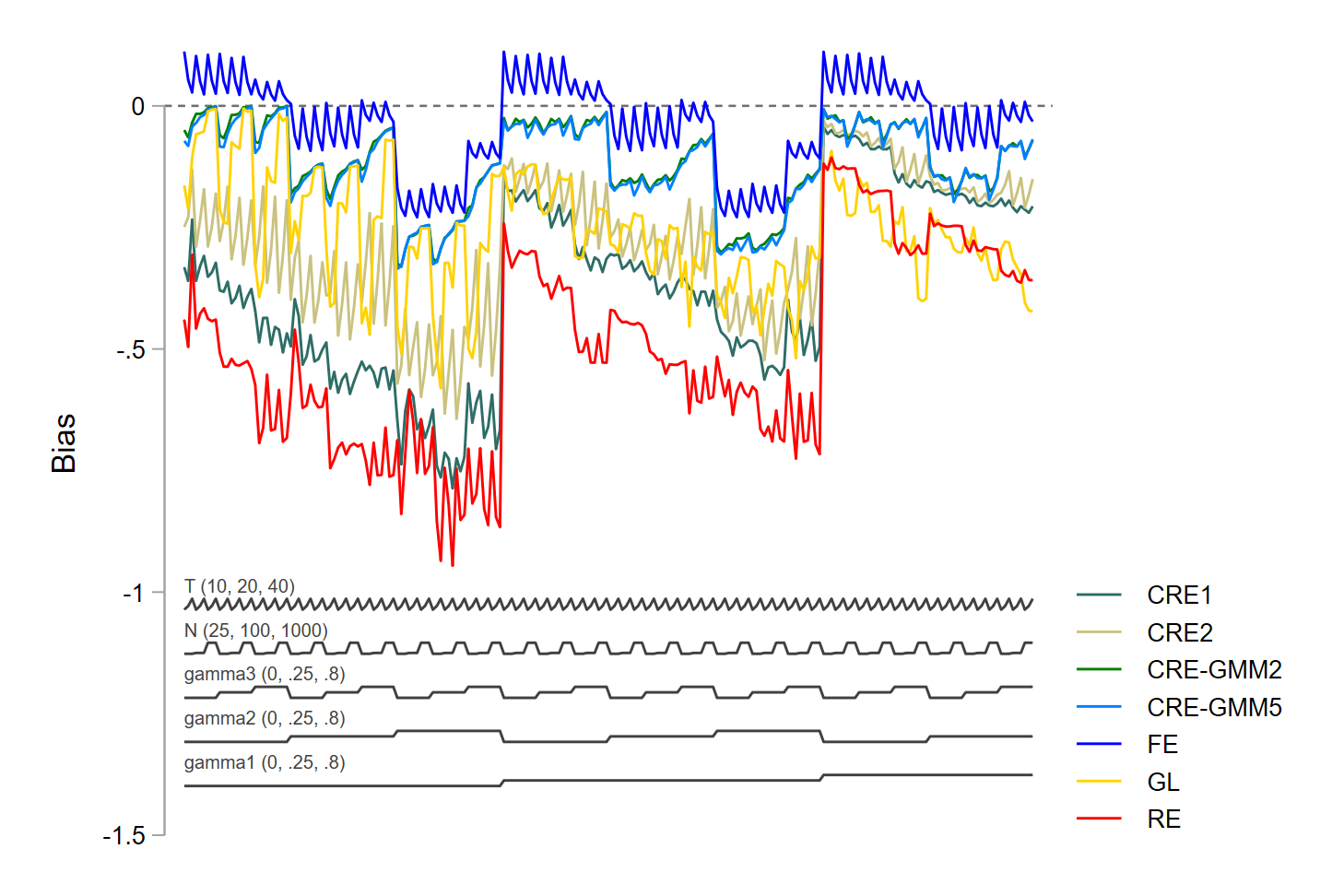}
\caption{\small{Nestedloop Plot for $\beta_2$ , ECM \\ Bias for $\beta_2=0.5$ across different specifications. Parameters shown on horizontal axis.}}
\label{fig:FIXEECM05BIG_nloop_L_x}
\end{figure}

Tables \ref{tab:FIXEECM05BIGlongi_rho}-\ref{tab:FIXEECM05BIGmulti_rho} show how results change for the autoregressive parameter $\rho$ when we simulate an ARDL(1,1) specification. Again, the moment conditions based on lagged levels, CRE-GMM5, work well when $\gamma_1=0$, while, as $\gamma_1>0$, moment conditions based on lagged differences, CRE-GMM2, are better. The GL estimator deteriorates when the variance of individual heterogeneity is higher than the variance of the shock (when $\gamma_3=0.8$) or $T$ is large, while the FE estimator tends to be unbiased in macro panels.  
The Figures \ref{fig:FIXEECM05BIG_nloop_L_y}, \ref{fig:FIXEECM05BIG_nloop_x} and \ref{fig:FIXEECM05BIG_nloop_L_x} compare the overall results for $\rho$, $\beta_1$ and $\beta_2$ according to the alternative values of $\gamma_1=[0, 0.25, 0.8]$, $\gamma_2=[0, 0.25, 0.8]$, $\gamma_3=[0, 0.25, 0.8]$, and various panel data settings ($N=[25, 100, 1000]$ and $T=[10, 20, 40]$).  The bias for $\beta_1$ is almost null when $\gamma_1=\gamma_2=0$, but particularly when $\gamma_2$ increases and tends to 0.8, all the estimates of $\beta_1$ are upward biased, with a tendency towards 1.5 (instead of 1), and those of $\beta_2$ are downward biased, with a tendency towards 0.2 (instead of 0.5). Although the bias is inevitable in the presence of correlation with the idiosyncratic shock, at least the CRE-GMM bias goes from half to more than one-eighth of that of GL, and it decreases as $T$ increases.    

\begin{sidewaystable}
\centering
\caption{MonteCarlo results, ARDL(1,1) with $\rho=0.5$, $\beta_1=1$ and $\beta_2=0.5$}\label{tab:FIXEECM05BIGlongi_rho} 
\begin{adjustbox}{max width=\textwidth}
{\scriptsize
\begin{tabular}{ll|rrrr|r|rrrrrr}
\toprule
& & \multicolumn{11}{c}{Longitudinal panel $N=1000$, $T=10$. Parameter $\rho$} \\
\midrule
& & RE & FE& CRE1 &CRE2 &GL & CRE-GMM & CRE-GMM1 & CRE-GMM2 &CRE-GMM3 & CRE-GMM4 & CRE-GMM5 \\
\midrule
\multirow{4}*{$\gamma_1=0, \gamma_2=0, \gamma_3=[0, 0.8]$} 
& bias $\rho$        &       0.263&      -0.082&       0.206&       0.180&       0.005&       0.028&       0.016&       0.007&       0.016&       0.006&       0.003\\
&ese          &       0.004&       0.008&       0.007&       0.007&       0.014&       0.010&       0.013&       0.017&       0.010&       0.012&       0.012\\
& bias $\rho$        &       0.375&      -0.039&       0.281&       0.240&       0.021&       0.042&       0.032&       0.010&       0.020&       0.010&       0.004\\
&ese          &       0.005&       0.005&       0.009&       0.007&       0.019&       0.008&       0.012&       0.017&       0.009&       0.013&       0.010\\
\midrule
\multirow{4}*{$\gamma_1=0, \gamma_2=0.25, \gamma_3=[0, 0.8]$} 
& bias $\rho$        &       0.258&      -0.076&       0.211&       0.179&       0.006&       0.027&       0.016&       0.006&       0.014&       0.006&       0.004\\
&ese          &       0.004&       0.007&       0.006&       0.007&       0.013&       0.009&       0.012&       0.017&       0.010&       0.011&       0.011\\
& bias $\rho$        &       0.377&      -0.038&       0.287&       0.241&       0.021&       0.042&       0.033&       0.010&       0.019&       0.010&       0.004\\
&ese          &       0.004&       0.005&       0.009&       0.007&       0.020&       0.008&       0.011&       0.017&       0.009&       0.013&       0.011\\
\midrule
\multirow{4}*{$\gamma_1=0, \gamma_2=0.8, \gamma_3=[0, 0.8]$} 
& bias $\rho$        &       0.194&      -0.041&       0.187&       0.170&       0.004&       0.024&       0.013&       0.005&       0.006&       0.002&       0.002\\
&ese          &       0.006&       0.005&       0.008&       0.006&       0.011&       0.007&       0.010&       0.014&       0.008&       0.009&       0.009\\
& bias $\rho$        &       0.343&      -0.028&       0.263&       0.238&       0.020&       0.041&       0.031&       0.010&       0.015&       0.007&       0.004\\
&ese          &       0.005&       0.005&       0.009&       0.007&       0.019&       0.008&       0.011&       0.016&       0.009&       0.013&       0.010\\
\midrule
\multirow{4}*{$\gamma_1=0.25, \gamma_2=0, \gamma_3=[0, 0.8]$} 
& bias $\rho$        &       0.231&      -0.082&       0.148&       0.139&       0.008&       0.042&       0.042&       0.012&       0.047&       0.040&       0.019\\
&ese          &       0.004&       0.008&       0.006&       0.007&       0.013&       0.009&       0.010&       0.016&       0.009&       0.010&       0.013\\
& bias $\rho$        &       0.334&      -0.039&       0.218&       0.201&       0.026&       0.053&       0.049&       0.014&       0.047&       0.054&       0.016\\
&ese          &       0.004&       0.005&       0.008&       0.006&       0.019&       0.007&       0.013&       0.015&       0.008&       0.008&       0.011\\
\midrule
\multirow{4}*{$\gamma_1=0.8, \gamma_2=0, \gamma_3=[0, 0.8]$} 
& bias $\rho$        &       0.136&      -0.082&       0.072&       0.069&       0.008&       0.028&       0.024&       0.012&       0.030&       0.023&       0.015\\
&ese          &       0.004&       0.008&       0.006&       0.006&       0.011&       0.009&       0.011&       0.013&       0.009&       0.010&       0.012\\
& bias $\rho$        &       0.235&      -0.039&       0.134&       0.129&       0.025&       0.045&       0.030&       0.015&       0.048&       0.037&       0.015\\
&ese          &       0.003&       0.005&       0.006&       0.005&       0.014&       0.007&       0.012&       0.013&       0.007&       0.008&       0.011\\
\midrule
\multirow{4}*{$\gamma_1=0.25, \gamma_2=0.25, \gamma_3=[0, 0.8]$} 
& bias $\rho$        &       0.214&      -0.076&       0.143&       0.131&       0.007&       0.039&       0.040&       0.010&       0.045&       0.038&       0.019\\
&ese          &       0.004&       0.007&       0.006&       0.007&       0.012&       0.008&       0.009&       0.015&       0.009&       0.009&       0.013\\
& bias $\rho$        &       0.331&      -0.038&       0.220&       0.200&       0.026&       0.053&       0.048&       0.013&       0.047&       0.053&       0.015\\
&ese          &       0.004&       0.005&       0.008&       0.006&       0.019&       0.007&       0.013&       0.015&       0.008&       0.008&       0.011\\
\midrule
\multirow{4}*{$\gamma_1=0.25, \gamma_2=0.8, \gamma_3=[0, 0.8]$} 
& bias $\rho$        &       0.181&      -0.041&       0.137&       0.125&       0.005&       0.036&       0.037&       0.008&       0.039&       0.037&       0.018\\
&ese          &       0.005&       0.005&       0.006&       0.005&       0.009&       0.006&       0.007&       0.013&       0.007&       0.007&       0.010\\
& bias $\rho$        &       0.300&      -0.028&       0.197&       0.197&       0.022&       0.051&       0.049&       0.012&       0.044&       0.052&       0.015\\
&ese          &       0.004&       0.005&       0.007&       0.006&       0.017&       0.007&       0.011&       0.015&       0.008&       0.008&       0.010\\
 \midrule
\multirow{4}*{$\gamma_1=0.8, \gamma_2=0.25, \gamma_3=[0, 0.8]$} 
& bias $\rho$        &       0.113&      -0.076&       0.060&       0.057&       0.006&       0.024&       0.021&       0.011&       0.025&       0.019&       0.013\\
&ese          &       0.004&       0.007&       0.005&       0.006&       0.010&       0.009&       0.010&       0.013&       0.009&       0.009&       0.011\\
& bias $\rho$        &       0.227&      -0.038&       0.131&       0.125&       0.023&       0.044&       0.030&       0.015&       0.048&       0.036&       0.014\\
&ese          &       0.003&       0.005&       0.006&       0.005&       0.014&       0.007&       0.011&       0.012&       0.007&       0.007&       0.010\\
\bottomrule
\end{tabular}
}
\end{adjustbox}
\flushleft
{\scriptsize Note: Average bias and empirical standard error reported; 1000 replications for each setting. Estimates are implemented in Stata using \texttt{xtdpdgmm} \citep{Kripfganz2019a}, one-step cluster standard errors. Variances are defined by equations \eqref{eq:FixEps} and \eqref{eq:varXfix} in Appendix \ref{sec:MCset}.}
\end{sidewaystable} 
\begin{sidewaystable}
\centering
\caption{MonteCarlo results, ARDL(1,1) with $\rho=0.5$ and $\beta_1=1$ and $\beta_2=0.5$}\label{tab:FIXEECM05BIGmacro_rho} 
\begin{adjustbox}{max width=\textwidth}
{\scriptsize
\begin{tabular}{ll|rrrr|r|rrrrrr}
\toprule
& & \multicolumn{11}{c}{Macro panel $N=25$, $T=40$. Parameter $\rho$} \\
\midrule
& & RE & FE& CRE1 &CRE2 &GL & CRE-GMM & CRE-GMM1 & CRE-GMM2 &CRE-GMM3 & CRE-GMM4 & CRE-GMM5 \\
\midrule
\multirow{4}*{$\gamma_1=0, \gamma_2=0, \gamma_3=[0, 0.8]$} 
& bias $\rho$        &       0.184&      -0.017&       0.140&       0.079&       0.088&       0.033&       0.032&       0.031&       0.028&       0.027&       0.026\\
&ese          &       0.016&       0.019&       0.029&       0.028&       0.026&       0.026&       0.026&       0.026&       0.025&       0.025&       0.025\\
& bias $\rho$        &       0.399&      -0.008&       0.262&       0.100&       0.292&       0.041&       0.041&       0.040&       0.031&       0.030&       0.030\\
&ese          &       0.009&       0.013&       0.037&       0.030&       0.023&       0.022&       0.022&       0.022&       0.021&       0.021&       0.021\\
\midrule
\multirow{4}*{$\gamma_1=0, \gamma_2=0.25, \gamma_3=[0, 0.8]$} 
& bias $\rho$        &       0.216&      -0.016&       0.166&       0.085&       0.109&       0.036&       0.035&       0.034&       0.029&       0.029&       0.028\\
&ese          &       0.013&       0.018&       0.029&       0.027&       0.025&       0.025&       0.025&       0.025&       0.024&       0.024&       0.024\\
& bias $\rho$        &       0.421&      -0.008&       0.280&       0.101&       0.329&       0.043&       0.042&       0.041&       0.031&       0.031&       0.031\\
&ese          &       0.008&       0.012&       0.037&       0.030&       0.021&       0.022&       0.022&       0.022&       0.021&       0.021&       0.021\\
\midrule
\multirow{4}*{$\gamma_1=0, \gamma_2=0.8, \gamma_3=[0, 0.8]$} 
& bias $\rho$        &       0.221&      -0.008&       0.179&       0.081&       0.111&       0.034&       0.033&       0.033&       0.024&       0.024&       0.023\\
&ese          &       0.011&       0.013&       0.026&       0.025&       0.021&       0.020&       0.020&       0.020&       0.019&       0.019&       0.019\\
& bias $\rho$        &       0.360&      -0.006&       0.260&       0.095&       0.241&       0.039&       0.039&       0.038&       0.028&       0.027&       0.027\\
&ese          &       0.009&       0.011&       0.036&       0.028&       0.022&       0.020&       0.020&       0.020&       0.019&       0.019&       0.019\\
\midrule
\multirow{4}*{$\gamma_1=0.25, \gamma_2=0, \gamma_3=[0, 0.8]$} 
& bias $\rho$        &       0.254&      -0.017&       0.134&       0.083&       0.177&       0.040&       0.040&       0.039&       0.047&       0.044&       0.042\\
&ese          &       0.011&       0.019&       0.028&       0.026&       0.019&       0.025&       0.025&       0.025&       0.025&       0.026&       0.026\\
& bias $\rho$        &       0.338&      -0.008&       0.195&       0.091&       0.256&       0.041&       0.041&       0.040&       0.040&       0.040&       0.039\\
&ese          &       0.009&       0.013&       0.033&       0.026&       0.019&       0.021&       0.021&       0.021&       0.020&       0.021&       0.021\\
\midrule
\multirow{4}*{$\gamma_1=0.8, \gamma_2=0, \gamma_3=[0, 0.8]$} 
& bias $\rho$        &       0.114&      -0.017&       0.056&       0.043&       0.078&       0.024&       0.024&       0.023&       0.033&       0.029&       0.025\\
&ese          &       0.013&       0.019&       0.019&       0.020&       0.019&       0.023&       0.023&       0.023&       0.022&       0.023&       0.023\\
& bias $\rho$        &       0.224&      -0.008&       0.112&       0.069&       0.171&       0.034&       0.034&       0.033&       0.042&       0.037&       0.034\\
&ese          &       0.009&       0.013&       0.022&       0.021&       0.013&       0.019&       0.019&       0.019&       0.018&       0.019&       0.019\\
\midrule
\multirow{4}*{$\gamma_1=0.25, \gamma_2=0.25, \gamma_3=[0, 0.8]$} 
& bias $\rho$        &       0.207&      -0.016&       0.122&       0.076&       0.127&       0.036&       0.036&       0.035&       0.043&       0.041&       0.039\\
&ese          &       0.012&       0.018&       0.026&       0.024&       0.021&       0.024&       0.024&       0.024&       0.023&       0.024&       0.024\\
& bias $\rho$        &       0.373&      -0.008&       0.210&       0.093&       0.314&       0.042&       0.042&       0.041&       0.040&       0.041&       0.041\\
&ese          &       0.008&       0.012&       0.032&       0.027&       0.015&       0.021&       0.021&       0.021&       0.020&       0.021&       0.021\\
\midrule
\multirow{4}*{$\gamma_1=0.25, \gamma_2=0.8, \gamma_3=[0, 0.8]$} 
& bias $\rho$        &       0.191&      -0.008&       0.124&       0.070&       0.113&       0.033&       0.033&       0.032&       0.037&       0.036&       0.035\\
&ese          &       0.010&       0.013&       0.023&       0.021&       0.017&       0.019&       0.019&       0.019&       0.018&       0.018&       0.019\\
& bias $\rho$        &       0.351&      -0.006&       0.212&       0.090&       0.277&       0.041&       0.040&       0.040&       0.038&       0.039&       0.039\\
&ese          &       0.009&       0.011&       0.033&       0.025&       0.016&       0.019&       0.019&       0.019&       0.019&       0.019&       0.019\\
 \midrule
\multirow{4}*{$\gamma_1=0.8, \gamma_2=0.25, \gamma_3=[0, 0.8]$} 
& bias $\rho$        &       0.110&      -0.016&       0.051&       0.039&       0.082&       0.022&       0.022&       0.021&       0.032&       0.027&       0.022\\
&ese          &       0.012&       0.018&       0.018&       0.018&       0.017&       0.021&       0.021&       0.021&       0.020&       0.021&       0.022\\
& bias $\rho$        &       0.229&      -0.008&       0.113&       0.068&       0.181&       0.034&       0.034&       0.033&       0.042&       0.037&       0.033\\
&ese          &       0.009&       0.013&       0.021&       0.020&       0.012&       0.019&       0.019&       0.019&       0.018&       0.019&       0.019\\
\bottomrule
\end{tabular}
}
\end{adjustbox}
\flushleft
{\scriptsize Note: Average bias and empirical standard error reported; 1000 replications for each setting. Estimates are implemented in Stata using \texttt{xtdpdgmm} \citep{Kripfganz2019a}, one-step cluster standard errors. Variances are defined by equations \eqref{eq:FixEps} and \eqref{eq:varXfix} in Appendix \ref{sec:MCset}.}
\end{sidewaystable} 

\begin{sidewaystable}
\centering
\caption{MonteCarlo results, ARDL(1,1) with $\rho=0.5$ and $\beta_1=1$ and $\beta_2=0.5$}\label{tab:FIXEECM05BIGmulti_rho} 
\begin{adjustbox}{max width=\textwidth}
{\scriptsize
\begin{tabular}{ll|rrrr|r|rrrrrr}
\toprule
& & \multicolumn{11}{c}{Multilevel panel $N=100$, $T=20$. Parameter $\rho$} \\
\midrule
& & RE & FE& CRE1 &CRE2 &GL & CRE-GMM & CRE-GMM1 & CRE-GMM2 &CRE-GMM3 & CRE-GMM4 & CRE-GMM5 \\
\midrule
\multirow{4}*{$\gamma_1=0, \gamma_2=0, \gamma_3=[0, 0.8]$} 
& bias $\rho$        &       0.257&      -0.034&       0.196&       0.142&       0.050&       0.027&       0.026&       0.023&       0.017&       0.017&       0.016\\
&ese          &       0.009&       0.014&       0.018&       0.018&       0.024&       0.020&       0.020&       0.021&       0.019&       0.019&       0.020\\
& bias $\rho$        &       0.402&      -0.016&       0.293&       0.178&       0.151&       0.036&       0.035&       0.031&       0.017&       0.017&       0.018\\
&ese          &       0.006&       0.010&       0.019&       0.018&       0.032&       0.015&       0.015&       0.018&       0.017&       0.017&       0.018\\
\midrule
\multirow{4}*{$\gamma_1=0, \gamma_2=0.25, \gamma_3=[0, 0.8]$} 
& bias $\rho$        &       0.262&      -0.032&       0.207&       0.144&       0.054&       0.027&       0.027&       0.024&       0.017&       0.016&       0.016\\
&ese          &       0.009&       0.014&       0.017&       0.019&       0.023&       0.019&       0.019&       0.021&       0.019&       0.019&       0.020\\
& bias $\rho$        &       0.409&      -0.015&       0.303&       0.180&       0.163&       0.036&       0.035&       0.032&       0.017&       0.017&       0.018\\
&ese          &       0.006&       0.010&       0.019&       0.019&       0.033&       0.016&       0.016&       0.018&       0.018&       0.018&       0.018\\
\midrule
\multirow{4}*{$\gamma_1=0, \gamma_2=0.8, \gamma_3=[0, 0.8]$} 
& bias $\rho$        &       0.188&      -0.016&       0.163&       0.117&       0.031&       0.021&       0.020&       0.018&       0.009&       0.009&       0.010\\
&ese          &       0.009&       0.010&       0.013&       0.015&       0.017&       0.014&       0.015&       0.016&       0.016&       0.016&       0.016\\
& bias $\rho$        &       0.379&      -0.011&       0.297&       0.176&       0.141&       0.035&       0.034&       0.031&       0.014&       0.014&       0.016\\
&ese          &       0.013&       0.008&       0.021&       0.019&       0.030&       0.014&       0.014&       0.016&       0.017&       0.017&       0.017\\
\midrule
\multirow{4}*{$\gamma_1=0.25, \gamma_2=0, \gamma_3=[0, 0.8]$} 
& bias $\rho$        &       0.226&      -0.034&       0.140&       0.113&       0.064&       0.032&       0.031&       0.027&       0.048&       0.037&       0.035\\
&ese          &       0.009&       0.014&       0.016&       0.016&       0.021&       0.018&       0.018&       0.020&       0.018&       0.019&       0.020\\
& bias $\rho$        &       0.353&      -0.016&       0.225&       0.153&       0.164&       0.039&       0.037&       0.033&       0.044&       0.037&       0.036\\
&ese          &       0.007&       0.010&       0.018&       0.017&       0.027&       0.015&       0.015&       0.016&       0.016&       0.017&       0.017\\
\midrule
\multirow{4}*{$\gamma_1=0.8, \gamma_2=0, \gamma_3=[0, 0.8]$} 
& bias $\rho$        &       0.137&      -0.034&       0.070&       0.061&       0.058&       0.024&       0.022&       0.021&       0.041&       0.024&       0.022\\
&ese          &       0.009&       0.014&       0.013&       0.013&       0.017&       0.017&       0.017&       0.018&       0.016&       0.018&       0.018\\
& bias $\rho$        &       0.242&      -0.016&       0.133&       0.104&       0.135&       0.035&       0.031&       0.029&       0.052&       0.030&       0.029\\
&ese          &       0.006&       0.010&       0.013&       0.013&       0.016&       0.013&       0.014&       0.014&       0.013&       0.014&       0.014\\
\midrule
\multirow{4}*{$\gamma_1=0.25, \gamma_2=0.25, \gamma_3=[0, 0.8]$} 
& bias $\rho$        &       0.210&      -0.032&       0.136&       0.108&       0.056&       0.030&       0.030&       0.026&       0.047&       0.036&       0.035\\
&ese          &       0.009&       0.014&       0.016&       0.016&       0.020&       0.018&       0.018&       0.019&       0.017&       0.019&       0.019\\
& bias $\rho$        &       0.355&      -0.015&       0.230&       0.154&       0.172&       0.040&       0.038&       0.033&       0.044&       0.037&       0.036\\
&ese          &       0.006&       0.010&       0.018&       0.017&       0.026&       0.015&       0.015&       0.016&       0.016&       0.016&       0.016\\
\midrule
\multirow{4}*{$\gamma_1=0.25, \gamma_2=0.8, \gamma_3=[0, 0.8]$} 
& bias $\rho$        &       0.211&      -0.016&       0.146&       0.107&       0.057&       0.030&       0.029&       0.025&       0.044&       0.035&       0.034\\
&ese          &       0.008&       0.010&       0.013&       0.013&       0.018&       0.013&       0.013&       0.014&       0.013&       0.015&       0.015\\
& bias $\rho$        &       0.334&      -0.011&       0.228&       0.150&       0.147&       0.038&       0.037&       0.032&       0.041&       0.036&       0.035\\
&ese          &       0.010&       0.008&       0.018&       0.016&       0.025&       0.013&       0.013&       0.015&       0.014&       0.015&       0.015\\
 \midrule
\multirow{4}*{$\gamma_1=0.8, \gamma_2=0.25, \gamma_3=[0, 0.8]$} 
& bias $\rho$        &       0.114&      -0.032&       0.060&       0.051&       0.046&       0.021&       0.020&       0.018&       0.037&       0.021&       0.019\\
&ese          &       0.009&       0.014&       0.012&       0.012&       0.016&       0.016&       0.016&       0.017&       0.015&       0.018&       0.018\\
& bias $\rho$        &       0.234&      -0.015&       0.131&       0.101&       0.125&       0.034&       0.031&       0.029&       0.052&       0.030&       0.029\\
&ese          &       0.006&       0.010&       0.012&       0.012&       0.015&       0.013&       0.013&       0.014&       0.012&       0.014&       0.014\\
\bottomrule
\end{tabular}
}
\end{adjustbox}
\flushleft
{\scriptsize Note: Average bias and empirical standard error reported; 1000 replications for each setting. Estimates are implemented in Stata using \texttt{xtdpdgmm} \citep{Kripfganz2019a}, one-step cluster standard errors. Variances are defined by equations \eqref{eq:FixEps} and \eqref{eq:varXfix} in Appendix \ref{sec:MCset}.}
\end{sidewaystable} 

\newpage

\end{document}